\documentclass[11pt,a4paper]{article}
\usepackage{amsmath,scalerel, amsfonts,amssymb,mathtools,nicefrac,nccmath,cases}
\usepackage{mathrsfs}
\usepackage{microtype,booktabs,multirow}
\usepackage[usenames,dvipsnames,table]{xcolor}
\usepackage{dsfont}
\usepackage[shortlabels]{enumitem}
\usepackage[normalem]{ulem}
\usepackage{colortbl}
\usepackage{tikz}
\usepackage{verbatim}
\usetikzlibrary{shapes,arrows}
\usetikzlibrary{calc,shapes.callouts,shapes.arrows,bending,intersections}
\usetikzlibrary{shapes.geometric}
\usetikzlibrary{arrows.meta,arrows}
\usetikzlibrary{decorations.pathmorphing}
\usetikzlibrary{decorations.markings}
\usetikzlibrary{shapes.multipart}
\usetikzlibrary{tikzmark}
\usepackage{arydshln}
\usepackage{caption}
\usepackage{subcaption}
\usepackage{bm}
\usepackage[nosort]{cite}
\usepackage{colortbl}
\usepackage{amsthm}
\usepackage{soul}
\usepackage{bbm}
\usepackage[normalem]{ulem}
\usepackage{tikz}
\usetikzlibrary{3d,arrows.meta,decorations.pathmorphing,positioning}

\usepackage{xcolor}
\usepackage{lipsum}

\usepackage[percent]{overpic}

\usepackage[linktocpage=true]{hyperref}
\hypersetup{colorlinks=true,linkcolor=nicecolor,citecolor=nicecolor,urlcolor=nicecolor}
\definecolor{nicecolor}{rgb}{0.1, 0.3, 0.4}

\definecolor{blue}{rgb}{0.06, 0.3, 0.57}
\definecolor{Gray}{gray}{0.4}

\definecolor{nicecolor}{rgb}{0.1, 0.3, 0.4}
\definecolor{blue}{rgb}{0.06, 0.3, 0.57}
\definecolor{Gray}{gray}{0.4}
\colorlet{tableheadcolor}{gray!15} % Table header colour = 25% gray
\colorlet{tablerowcolor}{gray!7} % Table row separator colour = 10% gray

\def\hybrid{\topmargin -20pt    \oddsidemargin 0pt
	\headheight 0pt \headsep 0pt
	\textwidth 6.5in        % US paper
	\textheight 9in         % US paper
	\textwidth 6.25in       % A4 paper
	\textheight 9 in       % A4 paper
	\marginparwidth .875in
	\parskip 5pt plus 1pt 
	\jot = 1.5ex
}
\hybrid
\numberwithin{equation}{section}
\numberwithin{table}{section}
\usepackage{float}
\usepackage{tabularx}
\usepackage{multirow}
\usepackage{hhline}
\usepackage{stackengine}

\newcolumntype{D}{>{\centering\arraybackslash}X}
\newcolumntype{L}{>{$}l<{$}}
\newcolumntype{R}{>{$}r<{$}}
\newcolumntype{C}{>{$}c<{$}}
\usepackage{IEEEtrantools}
\newcommand{\beq}{\begin{equation}}  \newcommand{\eeq}{\end{equation}}
\newcommand{\bal}{\begin{aligned}}   \newcommand{\eal}{\end{aligned}}
\newcommand{\bea}{\begin{eqnarray}}  \newcommand{\eea}{\end{eqnarray}}
\def\beqa{\begin{eqnarray}}
\def\eeqa{\end{eqnarray}}

\newcommand{\bmat}{\left(\begin{array}}
\newcommand{\emat}{\end{array}\right)}

\newcommand{\cD}{\mathcal{D}}

\newcommand{\cW}{\mathcal{W}}

\newcommand{\cU}{\mathcal{U}}

\newcommand{\R}{\text{Re}\,}

\usepackage{cancel}

\usepackage[section]{placeins} % TO FORCE FIGURES IN ASSOCIATED SECTIONS

\newcommand{\be}{\begin{equation}}
\newcommand{\ee}{\end{equation}}

\definecolor{Gray}{gray}{0.95}
\definecolor{darkspringgreen}{rgb}{0.09, 0.45, 0.27}
\definecolor{darkseagreen}{rgb}{0.56, 0.74, 0.56}
\definecolor{darkmouthgreen}{rgb}{0.05, 0.5, 0.06}
\definecolor{darkcyan}{rgb}{0.0, 0.55, 0.55}

\def\d {{\rm d}}
\def\del          {\partial}

\def\ii           {{\rm i}}

\def\Re           {{\rm Re\hskip0.1em}}
\def\Im           {{\rm Im\hskip0.1em}}

\def\cala         {{\cal A}}

\def\calb         {{\cal B}}
\def\calc         {{\cal C}}
\def\cald         {{\cal D}}

\def\calf        {{\cal F}}
\def\calg         {{\cal G}}

\def\cali         {{\cal I}}
\def\calj         {{\cal J}}

\def\call         {{\cal L}}
\def\calm         {{\cal M}}
\def\caln         {{\cal N}}
\def\calo         {{\cal O}}

\def\calr         {{\cal R}}
\def\cals         {{\cal S}}
\def\calt         {{\cal T}}
\def\calu         {{\cal U}}

\def\calw         {{\cal W}}
\def\calz         {{\cal Z}}

\def\sfA         {\mathsf{A}}
\def\sfB         {\mathsf{B}}
\def\sfC         {\mathsf{C}}
\def\sfD         {\mathsf{D}}
\def\sfE         {\mathsf{E}}

\def\sfK         {\mathsf{K}}

\def\sfM         {\mathsf{M}}
\def\sfN         {\mathsf{N}}

\def\sfP         {\mathsf{P}}
\def\sfQ         {\mathsf{Q}}
\def\sfR         {\mathsf{R}}

\def\sfT         {\mathsf{T}}

\def\sfW         {\mathsf{W}}
\def\sfX        {\mathsf{X}}
\def\sfY        {\mathsf{Y}}
\def\sfZ         {\mathsf{Z}}

\def\sfa         {\mathsf{a}}
\def\sfb         {\mathsf{b}}
\def\sfc         {\mathsf{c}}
\def\sfd         {\mathsf{d}}
\def\sfe         {\mathsf{e}}
\def\sff         {\mathsf{f}}

\def\sfk         {\mathsf{k}}
\def\sfm         {\mathsf{m}}
\def\sfn         {\mathsf{n}}

\def\sfp         {\mathsf{p}}

\def\sfu         {\mathsf{u}}

\def\sfv         {\mathsf{v}}
\def\sfw         {\mathsf{w}}

\def\tR      {{\text{\tiny R}}}
\def\tL      {{\text{\tiny L}}}

\usepackage{xcolor}
\definecolor{colorloc1}{RGB}{0,0,102}  
\definecolor{colorloc2}{RGB}{0,125,253} 
\usepackage[framemethod=default]{mdframed}
\newmdenv[skipabove=10pt,
skipbelow=7pt,
rightline=false,
leftline=true,
topline=false,
bottomline=false,
linecolor=colorloc1,
backgroundcolor=colorloc2!5,
innerleftmargin=4pt,
innerrightmargin=0pt,
innertopmargin=0pt,
leftmargin=2pt,
rightmargin=0pt,
linewidth=2pt,
innerbottommargin=0pt]{lbBox}

\begin{document}

\baselineskip=14pt
\parskip 5pt plus 1pt

\vspace*{-1.5cm}
\begin{flushright}    % Publication numbers
  {\small 
 % IFT-UAM/CSIC-21-36
  }
\end{flushright}

\vspace{2cm}
\begin{center}        % Main title

\textbf{\textsc{\huge 
Gaillard-Zumino non-invertible symmetries
%\\\smallskip and the species scale
%Wormholes and the species scale \\\smallskip in the axiverse
}}\\[.3cm]

\end{center}

\vspace{0.5cm}
\begin{center}        % Authors
%{\large   Luca Martucci}
{\large  Fabio Apruzzi and  Luca Martucci}
\end{center}

\vspace{0.15cm}
\begin{center}  
 \emph{Dipartimento di Fisica e Astronomia ``Galileo Galilei",  Universit\`a degli Studi di Padova} \\ 
\emph{\& INFN Sezione di Padova, Via F. Marzolo 8, 35131 Padova, Italy}
%\\${}^3$\emph{Department of Physics, University of Basel, Klingelbergstrasse 82, CH-4056 Basel, Switzerland}
\\
%[2mm]${}^4$\emph{Hamburg}\\[.3cm]
\end{center}

\vspace{2cm}

%%%%%%%%%%%%%%%%%%%%%%%%%%%%%%%%%%%%%%%%%%%%%%%
%%%%%%%%%%%%%%%%%%%%%%%%%%%%%%%%%%%%%%%%%%%%%%%
%%%%%%%%%%%%%%%%%%%%%%%%%%%%%%%%%%%%%%%%%%%%%%%
%%%%%%%%%%%%%%%%%%%%%%%%%%%%%%%%%%%%%%%%%%%%%%%
%%%%%%%%%%%%%%%%%%%%%%%%%%%%%%%%%%%%%%%%%%%%%%%
%%%%%%%%%%%%%%%%%%%%%%%%%%%%%%%%%%%%%%%%%%%%%%%
%%%%%%%%%%%%%%%%%%%%%%%%%%%%%%%%%%%%%%%%%%%%%%%
%%%%%%%%%%%%%%%%%%%%%%%%%%%%%%%%%%%%%%%%%%%%%%%

\begin{abstract}

\noindent 
We uncover an infinite class of novel zero-form non-invertible symmetries in a broad family of four-dimensional models, studied years ago by Gaillard and Zumino (GZ), which includes several extended supergravities as particular subcases. The GZ models consist of abelian gauge fields coupled to a neutral  sector, typically including a set of scalars, whose equations of motion are classically invariant under a continuous group $\mathscr{G}$ acting on the electric and magnetic field strengths via symplectic transformations. The standard lore holds that, at the quantum level,  these symmetries are broken to an integral subgroup $\mathscr{G}_\mathbb{Z}$. We show that, in fact, a much larger subgroup $\mathscr{G}_\mathbb{Q}$ survives, albeit through non-invertible topological defects. We explicitly construct these defects and compute some of their fusion rules. As illustrative examples, we consider the axion-dilaton-Maxwell model and the bosonic sector of a class of  $\mathcal{N}=2$ supergravities of the kind that appear in type II Calabi-Yau compactifications. Finally, we comment on how (part of) these non-invertible zero-form symmetries can be broken by gauging the $\mathscr{G}_\mathbb{Z}$ subgroup of invertible symmetries.

\end{abstract}

\thispagestyle{empty}
\clearpage

\setcounter{page}{1}

%%%%%%%%%%%%%%%%%%%%%%%%%%%%%%%%%%%%%%%%%%%%%%%
%%%%%%%%%%%%%%%%%%%%%%%%%%%%%%%%%%%%%%%%%%%%%%%
%%%%%%%%%%%                 %%%%%%%%%%%%%%%%%%%
%%%%%%%%%%%  DOCUMENT BODY  %%%%%%%%%%%%%%%%%%%
%%%%%%%%%%%                 %%%%%%%%%%%%%%%%%%%
%%%%%%%%%%%%%%%%%%%%%%%%%%%%%%%%%%%%%%%%%%%%%%%
%%%%%%%%%%%%%%%%%%%%%%%%%%%%%%%%%%%%%%%%%%%%%%%
%%%%%%%%%%%%%%%%%%%%%%%%%%%%%%%%%%%%%%%%%%%%%%%

\newpage

  \tableofcontents

%\newpage

\section{Introduction}

Understanding the full symmetry structure of a given physical system provides fundamental information on its dynamical properties. In quantum field theory, it is natural to adopt the generalized notion of symmetry proposed in \cite{Gaiotto:2014kfa}, which identifies global symmetries with extended topological defects. In this framework, codimension-$(p+1)$ operators generate $p$-form symmetries, with ordinary global symmetries corresponding to the case $p = 0$. This viewpoint has led to a wealth of developments that have revealed a much richer landscape of symmetries than previously recognized. Among the most intriguing are the so-called non-invertible symmetries associated with topological defects that lack inverses under fusion -- see e.g.\  \cite{Cordova:2022ruw,McGreevy:2022oyu,Freed:2022iao,Gomes:2023ahz,Schafer-Nameki:2023jdn,Brennan:2023mmt,Bhardwaj:2023kri,Shao:2023gho,Luo:2023ive, Carqueville:2023jhb,Costa:2024wks} for reviews. Non-invertible symmetries have been first discussed in two-dimensional quantum field theories via the study of topological line defects -- see \cite{Frohlich:2004ef, Frohlich:2006ch, Chang:2018iay, Thorngren:2019iar, Thorngren:2021yso, Komargodski:2020mxz} for a partial list of references in this direction. More recently, thanks to the pioneering work of \cite{Rudelius:2020orz, Heidenreich:2021xpr,Kaidi:2021xfk}, it has been discovered that non-invertible symmetries proliferate also in higher dimensions, which can be of various type such us discrete ones \cite{Bhardwaj:2022yxj,Cordova:2022ieu,Bhardwaj:2022lsg,
Bartsch:2022mpm, Kaidi:2022uux, Apruzzi:2022rei,
GarciaEtxebarria:2022vzq, Heckman:2022muc, Heckman:2022xgu, Apruzzi:2023uma,Bhardwaj:2022kot,
Bhardwaj:2022maz,
Bartsch:2022ytj,
Hsin:2022heo,
Kaidi:2023maf,
Carta:2023bqn,
Bhardwaj:2023ayw,
Damia:2023ses,Lawrie:2023tdz,
Mekareeya:2022spm, Antinucci:2022cdi, Antinucci:2022vyk,
Bah:2023ymy,Bhardwaj:2023bbf,Cordova:2024ypu,Choi:2024rjm,Cordova:2025eim, Nardoni:2024sos,Cordova:2023her,Apruzzi:2024cty} or continuous ones \cite{Antinucci:2022eat,Hayashi:2022fkw,Damia:2022seq,
Damia:2022bcd,Niro:2022ctq,Copetti:2023mcq,Argurio:2023lwl,Honda:2024xmk,Chen:2023czk,Choi:2023pdp,Honda:2024yte,Sela:2024okz,Arbalestrier:2024oqg}. 

In this paper we will investigate  the non-invertible symmetries characterizing the broad class of models identified and studied in 1981
by Gaillard and Zumino (GZ) \cite{Gaillard:1981rj}. These models have a Lagrangian of the form 
$\call(F,\phi,\del\phi)$,  depending 
on $n$ $U(1)$ gauge fields $A^I$ ($I=1,\ldots,n$) only through their field strengths $F^I=\d A^I$  and on a quite general neutral sector  collectively represented by the fields $\phi^i$, and satisfying  the following property. Its  classical equations of motion are invariant under a group $\mathscr{G}$ of continuous transformations generated by infinitesimal transformations that act on the gauge sector by infinitesimal symplectic transformations of the form 
\be\label{Tsplit} 
T=\left(\begin{array}{cc} U & V \\
W & Y\end{array}\right)\in {\rm sp}(2n,\mathbb{R})\,.
\ee
   More precisely, \eqref{Tsplit} acts linearly on  the field strengths   $F^I$ and on their electromagnetic duals, $G_I$, mixing them up -- see Section \ref{sec:GZ} for more details. We will refer to these classical symmetries as {\em GZ symmetries}, to the corresponding group $\mathscr{G}$ as {\em GZ group}, and to the models characterized by such symmetries as {\em GZ models}.
 In \cite{Gaillard:1981rj} it was also shown that, even if the equations of motion are invariant, the Lagrangian is not, though in  a very specific way. In particular, its non-invariance is due only to the gauge sector. Indeed, one can construct  a three-form current $\calj_T$, whose closure is obstructed only by the electric and magnetic field strengths as follows \cite{Gaillard:1981rj}: 
\be\label{dJnon0} 
\d\calj_T=\frac1{4\pi}\left(W_{IJ}F^I\wedge F^J-V^{IJ}G_I\wedge G_J-U^I{}_JF^J\wedge G_I+Y_J{}^IF^J\wedge G_I\right)\,.
\ee
Note that we use the same notation as in \cite{Gaiotto:2014kfa}, where a conserved $(d-p-1)$-form current of a $d$-dimensional theory is associated to a $p$-form symmetry. 
 
 It is clear that the classical GZ symmetries cannot be promoted to standard quantum zero-form symmetries. The main issue is the incompatibility with the flux quantization conditions $\frac1{2\pi}\oint F^I\in \mathbb{Z}$ and  $\frac1{2\pi}\oint G_I\in \mathbb{Z}$, since these  are not preserved by the generic symplectic transformation generated by $T\in {\rm sp}(2n,\mathbb{R})$. This seems to imply  that,  using for simplicity the same symbol  $\mathscr{G}$ to denote the abstract continuous group of  GZ symmetries and  its symplectic realization, only the integral discrete
subgroup  $\mathscr{G}_\mathbb{Z}\equiv \mathscr{G}\cap {\rm Sp}(2n,\mathbb{Z})$ can actually survive at the quantum level \cite{Hull:1994ys}. On the other hand, even if apparently relevant  only at classical level, the conditions imposed by the existence of the  GZ symmetries are considerably stronger than just imposing invariance under $\mathscr{G}_\mathbb{Z}$. So, the natural  question that will be addressed in this paper is: can at least  part of the classical non-integral GZ transformations survive as proper quantum symmetries? We will argue that the answer to this question is yes.  

The field strengths $F^I$ and $G_I$ are closed and, in sufficiently simple spacetime topologies with only local operator insertions, the right-hand side of \eqref{dJnon0} can be reabsorbed in a redefinition of the current $\calj_T$, leading to a conserved but not gauge invariant Page charge. This suggests that, at least in some cases, the  GZ symmetries should have some non-trivial implication also at the quantum level. The main goal of this paper is to make this vague idea concrete: when gravity is decoupled, we will show how the full subgroup   $\mathscr{G}_\mathbb{Q}\equiv \mathscr{G}\cap {\rm Sp}(2n,\mathbb{Q})$ of rational GZ symmetries is indeed not broken, and that the rational but non-integral GZ symmetries are  realized by {\em non}-invertible topological defects.

In order to better illustrate our general claims, we will  consider   some concrete examples.
First, we will revisit  the axion-Maxwell theory, which probably provides  the simplest example of GZ model, reproducing some known results \cite{Choi:2022jqy,Cordova:2022ieu}. We will then discuss in some detail a simple two-derivative model with a single gauge field and a dynamical complexified axion-dilaton coupling $\tau=\frac{\vartheta}{2\pi}+\ii e^{-\phi}$, parametrizing the coset ${\rm SL}(2,\mathbb{R})/{\rm U}(1)$. This model has GZ symmetry group $\mathscr{G}={\rm SL}(2,\mathbb{R})$ and we will refer to it as the $\tau$-Maxwell model.  Furthermore, we will consider the bosonic sector of a class of  $\caln=2$ two-derivative supergravities with GZ axionic shift symmetries, of the kind that for instance  arise
in perturbative  type IIA Calabi-Yau compactifications.

Other concrete models  can be treated likewise. In particular, the $\tau$-Maxwell model is the simplest instance  of a large class of  GZ models in which the scalar sector parametrizes cosets $\mathscr{G}/\mathscr{K}$, where  $\mathscr{K}$ is the maximal compact subgroup of   $\mathscr{G}$. These include several extended supergravities, many of which can be obtained through string and M-theory compactifications. The prototypical examples are provided  by the maximal ungauged supergravity  \cite{Cremmer:1979up,deWit:1982bul}, with  $\mathscr{G}={\rm E}_{7(7)}$ and $\mathscr{K}={\rm SU}(8)$, and  half-maximal supergravities \cite{deRoo:1984zyh}, with $\mathscr{G}={\rm SL}(2,\mathbb{R})\times {\rm SO}(6,6+n)$ and $\mathscr{K}={\rm U}(1)\times {\rm SO}(6)\times {\rm SO}(6+n)$. In this context, and in more general $d$-dimensional models, the classical GZ symmetries are often called ``duality symmetries" -- see for instance \cite{Hull:2025jpv} for a recent review and references to the original literature. Our results   provide a novel viewpoint on these symmetry structures, and also on their role in the study of UV divergences of extended supergravities  \cite{Kallosh:2008ic,Bossard:2010dq,Kallosh:2011dp}. Furthermore, from a stringy viewpoint, our results may  also encode the low-energy target space manifestation of the O$(d,d;\mathbb{Q})$ world-sheet topological interfaces pointed out in   \cite{Bachas:2012bj}, see also some more recent work in this direction \cite{Bharadwaj:2024gpj, Damia:2024xju, Angius:2024evd, Cordova:2023qei, Arias-Tamargo:2025fhv}. We postpone a detailed investigation of these interesting connections to future work. 
 We  also note that the appearance of non-invertible symmetries in higher-dimensional supergravities has been already investigated in e.g. \cite{Garcia-Valdecasas:2023mis,Fernandez-Melgarejo:2024ffg}.

It is important to emphasize that most of the GZ models  must be interpreted as low-energy effective field theories. 
As long as gravity is decoupled one may assume that they admit a UV completion characterized, at all energy scales, by a $\mathscr{G}_\mathbb{Q}$ group of invertible and non-invertible zero-form global symmetries. However,  this is expected not to be true once the gravitational interaction is turned on, as for instance in  the already mentioned string/M-theory realizations. Indeed, various arguments based on black hole quantum physics and holography, as well as compelling evidence from a plethora of concrete string theory models, strongly indicate that quantum gravity forbids exact global symmetries \cite{Misner:1957mt,Banks:2010zn,Harlow:2018tng}. In the Swampland program \cite{Vafa:2005ui,Ooguri:2006in} -- see \cite{Brennan:2017rbf,Palti:2019pca,vanBeest:2021lhn,Grana:2021zvf,Agmon:2022thq,VanRiet:2023pnx} for reviews -- this is known as the  ``No Global Symmetry Conjecture". Hence, in a  gravitational effective field theory admitting a UV completion, any symmetry  should either be  `accidental', and hence broken at some UV energy scale, or gauged. The  role of non-invertible symmetries in the realization of this paradigm, and more generically in the Swampland program, has already been investigated e.g.\ in \cite{Rudelius:2020orz,Heidenreich:2021xpr,McNamara:2021cuo,Cordova:2022rer,Heckman:2024obe,Rudelius:2024vmc,Basile:2025zjc,Gagliano:2025oqv}. 

In our context, a natural option is the gauging of (part of) the invertible subgroup $\mathscr{G}_\mathbb{Z}\subset \mathscr{G}_\mathbb{Q}$, and indeed  this is precisely what happens in several string/M-theory models in which $\mathscr{G}_\mathbb{Z}$ coincides with the so-called U-duality group \cite{Hull:1994ys} and identifies different regions in the moduli space. Once a subset of the  $\mathscr{G}_\mathbb{Q}$ zero-form symmetries  are gauged, the other ones must be broken. The specific symmetry-breaking mechanism, and its interplay with U-dualities, can encode important information on the UV completion of the gravitational model. Hence our results appear naturally relevant for the understanding of the ``marked moduli space" \cite{Raman:2024fcv,Delgado:2024skw} of quantum gravity models with symmetric moduli spaces  -- see in particular the recent \cite{Baines:2025upi}.   
Furthermore, the IR emergence of the $\mathscr{G}_\mathbb{Q}$ topological defects  can make the smallness of symmetry breaking couplings  technically natural \cite{Hooft1980}, as for instance in \cite{Cordova:2022ieu}. The existence of string theory realizations of GZ models provide concrete frameworks in which these qualitative ideas can be developed and tested.  While these aspects constitute a significant part of our motivation, we will only briefly touch upon them,  postponing a more detailed discussion to future work. 
%forthcoming paper \cite{AGMV}.  

The structure of the paper is the following. In Section \ref{sec:GZ} we review the basic aspects of the GZ models. In Sections \ref{sec:GZdefects} and \ref{sec:GZdefects2} we construct the topological GZ defects, which realize the GZ non-invertible zero-form symmetries. In Section \ref{sec:AactWH} we analyze the non-invertible action of the GZ defects on line operators. We then consider some explicit examples in details: the axion-Maxwell and $\tau$-Maxwell models in Section \ref{sec:Maxaxdil}, and the bosonic sector of $\caln=2$ models with axionic symmetries in Section \ref{sec:CYdefects}.  In Section \ref{sec:gaugingSB} we comment on the gauging of the invertible integral subgroups of GZ zero-form symmetries, and in Section \ref{sec:conclusion} we present our conclusions and outlook. Finally, the appendices contain useful background material and several complementary results.

\section{Gaillard-Zumino  models}
\label{sec:GZ}

In this section, we will review the main properties of the Gaillard-Zumino (GZ) models \cite{Gaillard:1981rj} and the implications of the flux quantization condition. This  will lay the groundwork for the subsequent discussions. As already mentioned, the Lagrangian  has the following general form
\be\label{Lansatz} 
\call(F,\phi,\del\phi)\,,
\ee
defined on a  four-dimensional spacetime $X$.  Note that in our notation $\call$ is a four-form and, hence, the corresponding action is just given by $\int\call$. It can depend on $n$ field strength two-forms $F^I=\d A^I$, $I=1,\ldots,n$, and on a  generic neutral sector collectively represented by $\phi^i$. For concreteness, we will mostly focus  on a set of real scalars $\phi^i$ parametrizing a scalar field space $\calm $,
but our discussion applies to more general sectors, as for instance the supergravity completion  of the bosonic theories that we will discuss in Section \ref{sec:CYdefects}. We also notice that, as in \cite{Gaillard:1981rj}, the Lagrangian can be put in a form that depends on the elementary fields and on their first derivatives. While from an effective field theory perspective it is natural to restrict  to two-derivative models, there may exist interesting higher derivative Lagrangians of the form   \eqref{Lansatz}, as for instance the  DBI extension of the $\tau$-Maxwell model discussed in Section \ref{sec:Maxaxdil}, which can reliably describe the low-energy dynamics around semiclassical configurations with large but slowly varying derivatives of the fundamental fields.  

Being quite generic, the following discussion may appear too abstract. The reader who prefers to handle more concrete quantities is invited to consult Appendix \ref{app:twoderiv}, in which some of the following general formulas are explicitly applied to  two-derivative models with a scalar neutral sector.

%%%%%%%%%%%%%%%%%%%%%%%%%%%%%

\subsection{Classical continuous GZ symmetries}
\label{sec:GZ1}

The key  property of the GZ models is the  existence of a continuous group $\mathscr{G}$ of transformations, generated by the corresponding algebra $\mathfrak{g}$,  which act on both the neutral and the gauge sector while preserving their classical equations of motion. More precisely an element $T\in \mathfrak{g}$ acts on the neutral sector by a transformation of the form
\be\label{delTphi} 
\delta_T\phi^i=\xi^i_T(\phi)\,.
\ee
Restricting ourselves  to a scalar neutral sector,  we can consider $\xi_T(\phi)\equiv \xi^i_T(\phi)\del_i$ as a globally defined vector field on the field space $\calm$.  

In order to describe the action on the gauge sector,  let us introduce the dual field strength  $G_I$ defined by the following identity
\be\label{G_I} 
\delta_F\call=\frac1{2\pi}\, \delta F^I\wedge G_I\,,
\ee
where  $\delta F^I$ is an arbitrary variation  of $F^I$ (not necessarily of the form $\d\delta A^I$). 
The $A^I$  equations of motion require that 
\be\label{Feom} 
\d G_I=0\,.
\ee
If we organize $F^I$ and $G_J$ in the $2n$-dimensional vector of two-forms
\be \label{bF}
\mathbbm{F}\equiv\left(\begin{array}{c} F^I \\
G_J\end{array}\right)\,,
\ee
the element $T\in\mathfrak{g}$ acts on $\mathbbm{F}$ as an ${\rm sp}(2n,\mathbb{R})$ matrix.
Using for notational simplicity the same symbol $T$ for this matrix and the more abstract element of $\mathfrak{g}$ -- the distinction should be clear from the context -- we can then write
\be\label{delTbF} 
\delta_T \mathbbm{F}=T\, \mathbbm{F}\,,
\ee
with $T$ such that 
\be\label{tcond} 
T^{\rm t}\Omega+\Omega T=0\,,
\ee
where $\Omega$ is the symplectic matrix
\be 
\Omega=\left(\begin{array}{rr} 0 &  \mathds{1} \\
-\mathds{1} & 0\end{array}\right)\,.
\ee

The GZ models are then characterized by the invariance of the classical equations of motion under the combined action  of  \eqref{delTphi} and \eqref{delTbF}. Finite transformations can be obtained by exponentiation. 
For concreteness we will focus on connected Lie groups $\mathscr{G}$, whose elements $\cals$ can be written as products of exponentials $\exp T$, with   $T\in\mathfrak{g}$. Hence the invariance of the equations of motion under \eqref{delTphi} and \eqref{delTbF} implies their invariance under  finite transformations
\begin{subequations}\label{gentr}
\begin{align}
\phi^i\quad &\mapsto\quad\tilde\phi^i=f_\cals^i(\phi)\,,\label{gentra}\\
\mathbbm{F}\quad &\mapsto\quad \tilde{\mathbbm{F}}= \cals\, \mathbbm{F}\,, \label{gentrb}
\end{align}
\end{subequations}
for any $\cals\in \mathscr{G}$. 
In the case of a scalar neutral sector, $\phi^i\mapsto f_\cals^i(\phi)$ defines a diffeomorphism of the field space $\calm$. Furthermore, we are using the same symbol $\cals$ for the abstract element of $\mathscr{G}$ as well as  for the corresponding ${\rm Sp}(2n,\mathbb{R})$ matrix appearing in \eqref{gentrb},
which hence satisfies
\be\label{calScond} 
\cals^{\rm t}\Omega\cals=\Omega\,.
\ee 
Strictly speaking, this notation  makes sense only if $\mathscr{G}$ can be regarded as a subgroup of  $\text{Sp}(2n,\mathbb R)$. However, for simplicity, we will loosely  adopt it also for cases in which   the symplectic action  of $\cals\in\mathscr{G}$ is specified  by a more general homomorphism\footnote{
That is, in these cases,  $\cals$  will  denote both the element in $\mathscr{G}$ and its image through the map $\varphi$. Similarly, we will write $\mathscr{G}\cap{\rm Sp}(2n,\mathbb{Z})$  and $\mathscr{G}\cap{\rm Sp}(2n,\mathbb{Q})$  instead of the more accurate $\varphi(\mathscr{G})\cap {\rm Sp}(2n,\mathbb{Z})$ and $\varphi(\mathscr{G})\cap {\rm Sp}(2n,\mathbb{Q})$, respectively.}  
\begin{equation} \label{symprep}
    \varphi: \mathscr{G} \rightarrow \text{Sp}(2n,\mathbb R)\,.
\end{equation}

As emphasized in \cite{Gaillard:1981rj}, even if the equations of motion are invariant under these transformations, the Lagrangian \eqref{Lansatz} is not, though in a specific way. Namely, under \eqref{delTphi} and \eqref{delTbF} the Lagrangian must transform so that the combination $\call-\frac1{4\pi}F^I\wedge G_I$ remains invariant.  
In terms of the finite transformations \eqref{gentr}, this means that
\be\label{Lvariation} 
\call(\tilde F,\tilde\phi)-\frac1{4\pi}\tilde F^I\wedge \tilde G_I=\call(F,\phi)-\frac1{4\pi}F^I\wedge G_I\,.
\ee
Here and in the following,  we suppress the explicit $\call$  dependence on $\del\phi$ for notational simplicity.

For our purposes, it is convenient to rephrase this condition by disentangling the effect of the two transformations in \eqref{gentr}. First of all, the GZ analysis implies that, given any   Lagrangian $\call$, for {\em any} $\cals\in{\rm Sp}(2n,\mathbb{R})$ we can construct  a new Lagrangian $\call_\cals$ such that
\be
\label{Lvariation2} 
\call_\cals(\tilde F,\phi)-\frac1{4\pi}\tilde F^I\wedge \tilde G_I=\call(F,\phi)-\frac1{4\pi}F^I\wedge G_I\,,
\ee
where the fields strengths are related as in \eqref{gentrb}.    $\call_\cals(\tilde F,\phi)$ must then satisfy the following non-trivial `integrability' condition. If ones uses $\call_\cals(\tilde F,\phi)$ to  compute $\tilde G_I$ as a function $\tilde F^I$ and  $\phi^i$,  and $\call(F,\phi)$ to compute $G_I$ as a function $F^I$ and  $\phi^i$, then the result must be compatible with  \eqref{gentrb}. Note that \eqref{Lvariation2} does not impose any restriction on the original Lagrangian. Rather, it identifies $\call_\cals(\tilde F,\phi)$ with the Lagrangian that one gets by making a  general symplectic {\em reparametrization} of the field strengths as in \eqref{gentrb} -- see for instance \cite{deWit:2001pz} and the following discussion  for an explicit verification of this statement.

Now, the key point is that the GZ condition \eqref{Lvariation} can be rewritten as 
\be\label{tildecall} 
\call( F,f_\cals(\phi))=\call_\cals( F,\phi)\quad,\quad \forall \cals\in \mathscr{G}\,.
\ee
 It is only the combination of \eqref{Lvariation2} and of the non-trivial condition  \eqref{tildecall} that implies the {\em invariance} of the  equations of motion of $\call$ under the combined transformations \eqref{gentr} -- see Appendix \ref{app:twoderiv} for a more explicit realization of these ideas in bosonic two-derivative models.\footnote{We thank Gianluca Inverso for discussions about these aspects.}  We can then interpret    \eqref{Lvariation2} and \eqref{tildecall}  as follows. Since in general $\call_\cals\neq \call$, \eqref{tildecall} tells us that, in general, $\call$ is not invariant under the transformation \eqref{gentra} of the neutral sector. However, because of \eqref{Lvariation2}, one can compensate for the corresponding variation of the equations of motion by performing a symplectic reshuffle \eqref{gentrb} of $F^I$ and $G_J$.  

For later convenience, we also observe that
\be\label{LSS'}
\left(\call_\cals\right)_{\cals'}=\call_{\cals'\cals}
\ee
for any for any pair  $\cals,\cals'\in{\rm Sp}(2n,\mathbb{R})$. This can be seen  by applying   \eqref{Lvariation2} to $\call_{\cals'\cals}$ and writing $\tilde{\mathbbm{F}}$ as $(\cals'\cals)\mathbbm{F}=\cals'(\cals\,\mathbbm{F})$.  Consider instead $\call_\cals(F,f_{\cals'}(\phi))$. By  \eqref{tildecall} and the composition rule 
\be\label{fSS'} 
f_\cals(f_{\cals'}(\phi))=f_{\cals\cals'}(\phi)\,,
\ee
this must be equal to $\call(F,f_\cals(f_{\cals'}(\phi)))=\call(F,f_{\cals\cals'}(\phi))=\call_{\cals\cals'}(F,\phi)$. Hence,
\be\label{fphiSS'} 
\call_\cals(F,f_{\cals'}(\phi))=\call_{\cals\cals'}(F,\phi)\,.
\ee
Notice  the different ordering in the product of $\cals$ and $\cals'$ on the r.h.s.\ of \eqref{LSS'} and \eqref{fphiSS'}.\footnote{Equations  \eqref{LSS'} and \eqref{fphiSS'} also imply that, if $\call$ satisfies the GZ condition \eqref{tildecall} then, for any given $\hat\cals\in\mathscr{G}$, $\call_{\hat\cals}$ satisfies the GZ condition \eqref{tildecall} with $f_\cals(\phi)$ replaced by the conjugated map $ \hat f_\cals(\phi)\equiv f_{\hat\cals^{-1}\cals\hat\cals}(\phi)$.}
 
It is also useful to consider the infinitesimal realization of \eqref{tildecall}:
 \be\label{delphiL} 
\delta^\phi_T\call=\frac1{4\pi} \mathbbm{F}^{\rm t}\wedge \Omega T\mathbbm{F}\,,
 \ee
where $\delta^\phi_T$ acts only on the neutral sector. The r.h.s.\ of this identity is obtained by expanding \eqref{Lvariation2}, and encodes in a precise way how the field strengths $F^I$ and  $G_I$ break the invariance of the action under the transformation of the neutral sector. This symmetry breaking mechanism can be represented in terms of a corresponding Noether three-form current 
\be\label{JTcurr} 
\calj_T=\xi_T^i(\phi)\frac{\del\call}{\del\d\phi^i}\,.
\ee 
In our conventions, it can be obtained by requiring that
\be\label{defJ} 
\delta_\alpha\call=\alpha\delta_T^\phi\call+\d\alpha\wedge \calj_T\,,
\ee 
under a point-dependent infinitesimal variation $\delta_\alpha\phi\equiv \alpha(x)\xi^i_T(\phi)$ of the neutral sector, and using \eqref{delTphi}. Taking \eqref{delphiL} into account, the standard Noether argument then implies the on-shell identity
\be\label{dJnon0a}
\d \calj_T=\frac1{4\pi} \mathbbm{F}^{\rm t}\wedge \Omega T\mathbbm{F}\,,
\ee
which quantifies the non-conservation of $\calj_T$. 
By using the decomposition \eqref{Tsplit}, \eqref{dJnon0a} coincides with \eqref{dJnon0}.

%%%%%%%%%%%%%%%%%%%%%%%%%%%%%

\subsection{GZ symmetry breaking by flux quantization}
\label{sec:GZ2}

 So far, the discussion has been completely classical, as in \cite{Gaillard:1981rj}. What can we say at the quantum level? Clearly,   the continuous transformations generated by \eqref{delTbF} are  incompatible with the standard 
flux quantization conditions associated with U(1) gauge fields,\footnote{\label{foot:torsion} The condition \eqref{qcond} for $\oint G_I$ is more precisely meant as a condition on its expectation value within the path integral, possibly including Wilson lines.  Additionally, in this work, we make the slightly simplifying assumption that the bulk cohomology group $H^2(X,\mathbb Z)$ has no torsion, so that the cohomology class of the field strengths $F^I$ and $G_I$ is completely determined by the flux quanta \eqref{qcond}. On the other hand, most of the results that will be discussed in this work do not depend on this assumption. }
\be \label{qcond}
\frac1{2\pi}\oint F^I\in \mathbb{Z}\quad,\quad \frac1{2\pi}\oint G_I\in \mathbb{Z}\,.
\ee
 The most obvious way out is to restrict ourselves to  {\em integral} symplectic transformations $\cals\in {\rm Sp}(2n,\mathbb{Z})$. As reviewed in Appendix \ref{app:Spdualities},  in this case  the corresponding Lagrangian $\call_\cals$ introduced above is physically equivalent to $\call$ also at the quantum level. 

More explicitly, one can always decompose a general symplectic matrix
\be\label{genS} 
\cals= \left(\begin{array}{cc} \mathsf{A} & \mathsf{B} \\
\mathsf{C} & \mathsf{D}\end{array}\right)\,,
\ee
as a product of  symplectic matrices of the restricted form
\be\label{Sfactor} 
\cals_{\mathsf{A}}=\left(\begin{array}{cc} \textsf{A} &  0\\
0 & \mathsf{A}^{-1{\rm t}}\end{array}\right)\quad,\quad \cals_{\mathsf{C}}=\left(\begin{array}{cc} \mathds{1} & 0 \\
\mathsf{C} & \mathds{1}\end{array}\right)\quad,\quad \cals_\Omega=\left(\begin{array}{rr} 0 &  \mathds{1} \\
-\mathds{1} & 0\end{array}\right)\,,
\ee
where $\sfA^I{}_J$ is invertible and $\mathsf{C}_{IJ}=\mathsf{C}_{JI}$. 
This decomposition is possible for symplectic matrices taking values in $\mathbb{R}$, $\mathbb{Q}$ and $\mathbb{Z}$ -- see for instance \cite{o1978symplectic,mumford2007tata}.  We can then, in turn, focus on the corresponding transformed Lagrangians,  denoting them   as $\call_\sfA\equiv \call_{\cals_{\mathsf{A}}}$, $\call_\sfC\equiv \call_{\cals_{\mathsf{C}}}$ and $\call_\Omega\equiv \call_{\cals_{\Omega}}$, respectively. 

Consider first $\call_\sfA$. The symplectic transformation   $\cals_{\mathsf{A}}$ in \eqref{Sfactor}   just corresponds to a field  redefinition $\tilde F^I=\mathsf{A}^I{}_JF^J$, so that we simply have 
\be\label{LSA} 
\call_{{\mathsf{A}}}(\tilde F,\phi)=\call(F,\phi)\,,
\ee 
compatibly with \eqref{Lvariation2}. However,  it is clear that $\tilde F^I=\mathsf{A}^I{}_JF^J$ preserves \eqref{qcond} if and only if $\mathsf{A}\in {\rm GL}(n,\mathbb{Z})$, that is, if $\mathsf{A}$ is unimodular. It is only in this case that $\call_{{\mathsf{A}}}$ is physically equivalent to $\call$. Otherwise $\call_{{\mathsf{A}}}(F,\phi)$, with properly quantized $F^I$, describes a different theory. 

Turning to $\call_\sfC$, we observe that $\cals_{\mathsf{C}}$ does not act on $F^I$ but transforms $G_I$ into $\tilde G_I=G_I+\sfC_{IJ}F^J$. Recalling \eqref{G_I}, the corresponding transformed Lagrangian should hence take the form
\be\label{LSC}
\call_{\sfC}( F,\phi)=\call(F,\phi)+\frac{1}{4\pi}\mathsf{C}_{IJ}F^I\wedge F^J\,,
\ee
again compatibly with \eqref{Lvariation2} (with $\tilde F^I=F^I$). In this paper we restrict to spin  manifolds. Hence, if $\mathsf{C}_{IJ}\in \mathbb{Z}$ then $\frac{1}{4\pi} \mathsf{C}_{IJ}\int F^I\wedge F^J\in 2\pi\mathbb{Z}$, and  the transformed Lagrangian is physically equivalent to the old one. On the other hand, if $\mathsf{C}_{IJ}\notin \mathbb{Z}$ this is no-longer true and $\call$ and $\call_\sfC$ are physically inequivalent.

Finally, $\cals_\Omega$ describes a simultaneous S-duality of all gauge-fields -- see Appendix \ref{app:Spdualities} -- which generates the physically equivalent Lagrangian 
\be 
\call_{\Omega}(\tilde F,\phi)=\big[\call(F,\phi)-\frac1{2\pi}\delta_{IJ}F^I\wedge \tilde F^J\big]|_{F=-\tilde G}\,,
\ee 
again in agreement with \eqref{Lvariation2}. It is important  to recall that the S-duality experiences a mixed 't Hooft anomaly with a curved background metric \cite{Witten:1995gf, Verlinde:1995mz,Seiberg:2018ntt} -- see also \cite{Meynet:2025zem} for a recent discussion in the context of non-invertible symmetries. In the rest of this paper, we will neglect this subtlety, implicitly ignoring the corresponding curvature contributions, which would encode the effect of the mixed 't Hooft anomaly.  A detailed study of these and other 't Hooft anomalies, like for example the ones studied in \cite{Hsieh:2019iba}, are left to the future.

Since any $\cals\in {\rm Sp}(2n,\mathbb{Z})$ can be decomposed as a product of transformations \eqref{Sfactor} with $\mathsf{A}\in {\rm GL}(n,\mathbb{Z})$ and $\sfC_{IJ}\in\mathbb{Z}$, the corresponding transformed Lagrangian $\call_\cals(F,\phi)$ is physically equivalent to $\call(F,\phi)$. The condition \eqref{tildecall} then implies that the transformation \eqref{gentra} with $\cals\in \mathscr{G}_{\mathbb{Z}}\equiv \mathscr{G}\cap{\rm Sp}(2n,\mathbb{Z})$ is an honest symmetry of the theory. In many string theory models,  $\mathscr{G}_{\mathbb{Z}}$ is in fact gauged, so that the actual field space is given by the orbifold $\calm/\mathscr{G}_{\mathbb{Z}}$.\footnote{As discussed in \cite{Witten:1995gf} in the case of a single gauge field, the quantum equivalence of $\call$ and $\tilde\call$ Lagrangians can involve `anomalous' effects. Of course, the gauging by $\mathscr{G}_{\mathbb{Z}}$ requires that these effects are harmless.} However, we will mostly consider  $\mathscr{G}_{\mathbb{Z}}$ as a standard group of (invertible) zero-form symmetries,  postponing its gauging to later comments.      

We can now more precisely phrase the question that will be addressed in this paper: does any other element $\cals\notin \mathscr{G}_{\mathbb{Z}}$ of the classical GZ symmetry group $\mathscr{G}$ survive at the quantum level? 
In the following sections we will show that  {\em any} $\cals$ belonging to {\em rational} GZ symmetries
\be\label{G_Qdef} 
\mathscr{G}_{\mathbb{Q}} \equiv \mathscr{G}\cap {\rm Sp}(2n,\mathbb{Q})\,,
\ee 
corresponds to a zero-form symmetry also at the quantum level, which is however non-invertible if $\cals$ does not belong to $\mathscr{G}_{\mathbb{Z}}$.   

%%%%%%%%%%%%%%%%%%%%%%%%%%%%%%%%%%%%%%%%%%%%%%%%%%%%%%%%%%

\section{GZ topological defects: general strategy}
\label{sec:GZdefects}

As discussed above, the GZ transformations \eqref{gentr} with $\cals\notin\mathscr{G}_\mathbb{Z}$ cannot be promoted to invertible zero-form symmetries. One of our main goals is to show that,  for any $\cals\in \mathscr{G}_{\mathbb{Q}}$ and any codimension-one surface $\Sigma$, we can construct  a topological defect $\cald_\cals(\Sigma)$ that  generates the GZ transformation, though generically in a non-invertible way. We will refer to these  as Gaillard-Zumino (GZ) defects.  In this section we illustrate the general strategy that we will follow for constructing GZ defects, postponing some  technical aspects to the following sections. 

\begin{figure}[h]
\centering
\begin{tikzpicture}[scale=1.2,>=Latex]

% --- Left: initial configuration ---
% surface defect
\filldraw[red!30!white,fill opacity=0.4,draw=red,thick]
  (-3.5,1.2) -- (-3.5,-1.2) -- (-2.7,-1.5) -- (-2.7,0.9) -- cycle;
\node[red,above] at (-3.1,1.2) {$\mathcal{U}_\cals$};

% --- Middle arrow ---
\draw[<->,thick] (-2.5,0) -- (-1.9,0);

% --- Right: after crossing ---
% surface defect (again)
\filldraw[blue!30!white,fill opacity=0.4,draw=blue,thick]
  (-1.7,1.2) -- (-1.7,-1.2) -- (-0.9,-1.5) -- (-0.9,0.9) -- cycle;
\node[blue,above] at (-1.3,1.2){$\mathcal{W}_\cals$};

\node[black] at (0,0){$=$};

\filldraw[purple!30!white,fill opacity=0.4,draw=purple,thick]
  (0.9,1.2) -- (0.9,-1.2) -- (1.7,-1.5) -- (1.7,0.9) -- cycle;
\node[purple,above] at (1.3,1.2){$\mathcal{D}_\cals$};

\end{tikzpicture}
\caption{Construction of a $\mathcal D_{\cals}$ defect by fusing  a $\calu_\cals$ defect and a  $\calw_\cals$ interface.}
\label{fig:UW=D}
\end{figure}
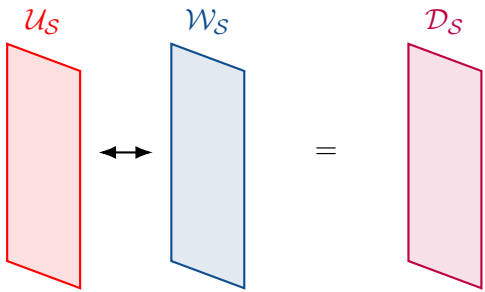

 We will construct the GZ defects by fusing two commuting defects: 
 \be\label{Doperator} 
\cald_\cals(\Sigma)=\calu_\cals(\Sigma)\times \calw_\cals(\Sigma)\,.
 \ee 
 Here $\calu_\cals(\Sigma)$  is an invertible defect implementing the transformation \eqref{gentra} of the neutral sector.  On the other hand, the defect  $\calw_\cals(\Sigma)$ should rather be considered as an interface  connecting the theories specified by $\call$ and $\call_\cals$, which are related as in \eqref{Lvariation2} -- see Figure \ref{fig:UW=D}. As discussed in Section \ref{sec:GZ2}, these theories are equivalent only if $\cals\in{\rm Sp}(2n,\mathbb{Z})$, and only in this case we expect the corresponding defect $\cald_\cals(\Sigma)$  to be invertible. We will explicitly check this expectation in Section \ref{sec:GZdefects2}.

 Let us start with $\calu_\cals(\Sigma)$. It acts only on the neutral sector and, therefore, does not affect the flux quantization condition and can be defined for any $\cals\in \mathscr{G}$. Since $\mathscr{G}$ is connected, we can decompose any $\cals\in \mathscr{G}$  as a product of elements of the form $\exp T$, with $T\in\mathfrak{g}$. Hence, without loss of generality,  we can focus on elements of the form \be\label{TSrel} 
\cals=\exp T\,.
\ee
The corresponding defect  
$\calu_\cals(\Sigma)$ is obtained  by  exponentiating the  charge $\oint_\Sigma\calj_T$, where $\calj_T$ is defined by  \eqref{defJ}. Note that, because of  \eqref{dJnon0a}, neither $\oint_\Sigma\calj_T$ nor  $\calu_\cals(\Sigma)$ are generically topological. This is of course expected, since generically \eqref{gentra} alone is not a symmetry of the theory. 

 The precise expression for $\calu_\cals(\Sigma)$ can be derived by following standard textbook arguments, which are however mostly applied to the case of classically unbroken symmetries. Hence, for the sake of clarity on how to  fix coefficients and signs, we find it useful to repeat the following well-known procedure. 

We will only consider orientable sufaces $\Sigma$. We  can focus on a tubular  neighborhood $\calt$ of $\Sigma$, and introduce on it a transverse coordinate $y$, such that $\Sigma=\{y=0\}$. The tubular  neighborhood splits in two parts $\calt=\calt_{\text{\tiny L}}\cup\calt_{\text{\tiny R}}$, corresponding to $y\geq 0$ and $y\leq 0$ respectively, with common boundary $\Sigma$. We fix the orientations in such a way that $ \del\calt_{\text{\tiny L}}$ contains $\Sigma$ while $\del\calt_{\text{\tiny R}}$ contains $-\Sigma$.\footnote{\label{foot:orient} Notice that a change of orientation of $\Sigma$ implies a change of orientation of the transverse $y$ coordinate, hence exchanging $\calt_{\text{\tiny L}}$ and $\calt_{\text{\tiny R}}$. }

 Let us start from a path integral description of the partition function defined by the Lagrangian $\call(F,\phi)$ and, using the notation introduced in \eqref{bF}, denote by $(\mathbbm{F}_{\text{\tiny L}},\phi_{\text{\tiny L}})$ and $(\mathbbm{F}_{\text{\tiny R}},\phi_{\text{\tiny R}})$ the fields supported on $\calt_{\text{\tiny L}}$ and $\calt_{\text{\tiny R}}$, respectively. The path integral splits in two parts, defined on $\calt_{\text{\tiny L}}$ and $\calt_{\text{\tiny R}}$, with gluing conditions $(\mathbbm{F}_{\text{\tiny L}},\phi_{\text{\tiny L}})|_\Sigma=(\mathbbm{F}_{\text{\tiny R}},\phi_{\text{\tiny R}})|_\Sigma$. Now let us make an infinitesimal transformation  \eqref{delTphi} of the right neutral sector, $\phi^i_{\text{\tiny R}}\rightarrow \phi^i_{\text{\tiny R}}+\varepsilon\xi^i_T(\phi_{\text{\tiny R}})$,  regarding it as a  change of path integration variable.  Assuming that the path integral measure is invariant under such transformation, we have only to take into account the transformation of the Lagrangian, which is provided by \eqref{defJ} with $\alpha(x)=\varepsilon \theta(y)$. Since $\d\alpha(x)=\delta(y)\d y$ the action splits into three contributions,
\be 
\int_{\calt_{\text{\tiny L}}}\call(F_{\text{\tiny L}},\phi_{\text{\tiny L}})+\varepsilon\oint_\Sigma\calj_T+\int_{\calt_{\text{\tiny R}}}\call\left(F_{\text{\tiny R}},\phi_{\text{\tiny R}}+\varepsilon\xi_T(\phi_{\text{\tiny R}})\right)\,,
\ee
where we now have the gluing condition $\phi^i_{\text{\tiny L}}|_\Sigma =[\phi^i_{\text{\tiny R}}+\varepsilon\xi^i_T(\phi_{\text{\tiny R}})]|_\Sigma$. This can be rewritten as $\phi^i_{\text{\tiny L}}|_\Sigma=f^i_{\cals_\varepsilon}(\phi_{\text{\tiny R}})|_\Sigma$, with $\cals_\varepsilon=\exp(\varepsilon T)$. 

We can now integrate the infinitesimal result to deduce that, for any finite transformation of the form \eqref{TSrel}, the corresponding defect 
\be\label{caluS} 
\calu_\cals(\Sigma)\equiv \exp \left(\ii\oint_\Sigma\calj_T\right)\, ,
\ee 
separates the theory defined by $\call(F_{\text{\tiny L}},\phi_{\text{\tiny L}})$ on $\calt_{\text{\tiny L}}$ to  the theory defined by  $ 
 \call(F_{\text{\tiny R}},f_\cals(\phi_{\text{\tiny R}}))$ on $\calt_{\text{\tiny R}}$, with
 $\phi^i_{\text{\tiny L}}|_\Sigma=f^i_\cals(\phi_{\text{\tiny R}})|_\Sigma$. The insertion of the $\calu_\cals(\Sigma)$ defect imposes the gluing condition on the neutral sector, whereas the transformation of $\phi^i_{\text{\tiny R}}$ on $\mathcal T_{\text{\tiny R}}$ makes the combination of the insertion and transformation a trivial operation. Infinitesimally -- or for nilpotent transformations like the axial symmetry in QED \cite{Choi:2022jqy,Cordova:2022ieu}  or the axion shift \cite{Choi:2022fgx}  -- this is a consequence of the non-conservation equation \eqref{dJnon0a}. The triviality of this operation ensures the fact that we can move the $\calu_\cals(\Sigma)$ defect provided that the right Lagrangian has been accordingly transformed. \footnote{In the case of a nilpotent transformation, like the axial symmetry in QED  \cite{Choi:2022jqy,Cordova:2022ieu} or the axion shift in axion-Maxwell \cite{Choi:2022fgx},   the transformation of $\phi^i_\tR$ produces a term which may be (improperly) rewritten as a boundary Chern-Simons term with non-properly quantized  level. Combined with $\calu_\cals$, this would give rise to a Page charge defect, which is topological but not gauge invariant.} 
 Furthermore the pull-back  of the field strengths $F^I$ and $G_J$ does not change across $\Sigma$, that is  
 \be\label{FGgluing} 
 F^I_{\text{\tiny L}}|_\Sigma=F^I_{\text{\tiny R}}|_\Sigma\quad,\quad G_{\tL\,I}|_\Sigma=G_{\tR\,I}|_\Sigma\,.
 \ee 
 This is probably more clearly seen in a Hamiltonian description, in which $y$ is identified with the time direction.\footnote{In order to get a direct operatorial interpretation of some of the following formulas, the standard time-ordering should correspond to {\em decreasing} $y$.} One can pick local coordinates $(y,\sigma^\alpha)$ such that on $\Sigma$ the metric takes the form $-\d y^2+h_{\alpha\beta}\d\sigma^\alpha\d\sigma^\beta$, and work in temporal gauge $A^I_y=0$.  Then the three components of the three-dimensional Hodge dual of $G_I|_\Sigma=\frac12 G_{I\, \alpha\beta}\d\sigma^\alpha\d\sigma^\beta$ can be identified with the momenta $\Pi^\alpha_I$ canonically conjugated to $A^I_\alpha$. On the other hand, recalling \eqref{JTcurr}, the operator  $\oint_\Sigma\calj_T$ appearing in \eqref{caluS} is the canonical generator of the  infinitesimal transformation $\delta_T\phi^i$ of the neutral fields, and leaves  $A^I_\alpha$ and $\Pi_I^\alpha$ invariant. This implies that  \eqref{caluS}, regarded as a unitary operator,  commutes with $A^I_\alpha$ and $\Pi_I^\alpha$ and, hence, the gluing conditions \eqref{FGgluing} hold.  
This conclusion extends to more general $\cals\in\mathscr{G}$, which cannot be written as \eqref{TSrel} but  rather as product of elements of this form. Of course, the corresponding $\calu_\cals(\Sigma)$ is a product of operators of the form \eqref{caluS}. 

It is now important to recall the GZ condition \eqref{tildecall}, which  implies that the relevant Lagrangian on $\calt_{\text{\tiny R}}$ can be actually identified with $\call_\cals(F_{\text{\tiny R}},\phi_{\text{\tiny R}})$. 
 We can then summarize  the complete set of gluing conditions induced by $\calu_\cals(\Sigma)$ as follows
 \be\label{caludef} 
     \begin{split}
    &\quad~~~~{\color{red}\cU_{\mathcal{S}}(
\Sigma
    )}\\
   \call(F_{\text{\tiny L}},\phi_{\text{\tiny L}}) \qquad &\qquad~{\color{red} \quad\Bigg|} \qquad~~~~~~~~~~ \call(F_{\text{\tiny R}},f_\cals(\phi_{\text{\tiny R}})) = \call_\cals(F_{\text{\tiny R}},\phi_{\text{\tiny R}})\, ,\\
   &\!\!\!\!\!{\color{red}\scaleto{(\mathbbm{F}_{\text{\tiny L}},\phi_{\text{\tiny L}})|_\Sigma=(\mathbbm{F}_{\text{\tiny R}},f_\cals(\phi_{\text{\tiny R}}))|_\Sigma}{2ex}}
\end{split}
\ee
 where, as above, in our notation $|_\Sigma$ acts as a pull-back on the field strengths that  constitute  $\mathbbm{F}$ as in \eqref{bF}. As discussed in Section \ref{sec:GZ},
in general $\call_\cals(F_{\text{\tiny R}},\phi_{\text{\tiny R}})$ is not physically equivalent to $\call(F_{\text{\tiny R}},\phi_{\text{\tiny R}})$. Therefore,   as already emphasized, generically the defect  $\calu_\cals(\Sigma)$ is  {\em not} topological.

Let us now turn to the interfaces  $\calw_\cals(\Sigma)$. By definition, an interface $\calw_\cals(\Sigma)$ should connect the theory defined by 
$\call_\cals(F_{\text{\tiny L}},\phi_{\text{\tiny L}})$ on $\calt_{\text{\tiny L}}$ with the theory defined by  $ 
 \call(F_{\text{\tiny R}},\phi_{\text{\tiny R}})$ on $\calt_{\text{\tiny R}}$,
with gluing conditions    $\mathbbm{F}_{\text{\tiny L}}|_\Sigma=\cals \mathbbm{F}_{\text{\tiny R}}|_\Sigma$ and $\phi_{\text{\tiny L}}|_\Sigma=\phi_{\text{\tiny R}}|_\Sigma$. 
These conditions  can be summarized as follows:
\be\label{WSpicture} 
     \begin{split}
    &\quad~~~~{\color{blue}\cW_{\mathcal{S}}(
\Sigma
    )}\\
   \call_{\mathcal{S}}(F_{\text{\tiny L}},\phi_{\text{\tiny L}}) \qquad &\quad~~~~{\color{blue} \quad\Bigg|} \qquad~~~~~~~ \call( F_{\text{\tiny R}},\phi_{\text{\tiny R}})\, .\\
&\!\!\!\!\!{\color{blue}\scaleto{(\mathbbm{F}_{\text{\tiny L}},\phi_{\text{\tiny L}})|_\Sigma=(\cals\mathbbm{F}_{\text{\tiny R}},\phi_{\text{\tiny R}})|_\Sigma}{2ex}}
\end{split}
\ee

As in Section \ref{sec:GZ2}, here the key difficulty comes from the flux quantization condition \eqref{qcond}. Indeed, for generic $\cals\in \mathscr{G}$, \eqref{WSpicture} implies that if $F^I_\tR$ and $G_{\tR\,J}$   satisfy \eqref{qcond} along the wall, then generically $F^I_\tL$ and $G_{\tL\,J}$ do not, and viceversa. On the other hand, if $\cals\in \mathscr{G}_{\mathbb{Q}}$ the problem appears alleviated, since the flux gluing conditions appearing in  \eqref{WSpicture} can be satisfied compatibly with \eqref{qcond} by at least a sublattice of possible electric and magnetic fluxes. The rest of this paper will be devoted to arguing that, indeed,  for any $\cals\in {\rm Sp}(2n,\mathbb{Q})$ -- and hence any $\cals\in \mathscr{G}_{\mathbb{Q}}$ -- one can construct a corresponding interface $\calw_\cals(\Sigma)$.
If $\cals\in {\rm Sp}(2n,\mathbb{Z})$, $\call$ and $\call_\cals$ are physically equivalent and the  interface  $\calw_\cals(\Sigma)$ is just the natural generalization of the duality walls discussed in \cite{Ganor:1996pe,Gaiotto:2008ak,Kapustin:2009av} -- see Appendix \ref{app:Spdualities}.  On the other hand, if $\cals\in {\rm Sp}(2n,\mathbb{Q})$ is {\em not} in ${\rm Sp}(2n,\mathbb{Z})$, the theory defined by $\call_\cals$ is not physically equivalent to $\call$, but an interface $\calw_\cals(\Sigma)$ realizing \eqref{WSpicture} can still   be constructed. In the rest of this section we will  proceed assuming the validity of  this claim, which will be justified in detail in  Section \ref{sec:GZdefects2}.

Combing these considerations, we conclude that if $\cals\in \mathscr{G}_{\mathbb{Q}}$ defined as in \eqref{G_Qdef}, we can construct a defect $\cald_\cals(\Sigma)$ by fusing $\calu_\cals(\Sigma)$ and $\calw_\cals(\Sigma)$ as in Figure \ref{fig:UW=D}. More explicitly, 
 let us  insert  $\calu_\cals(\Sigma)$ at $\Sigma=\{y=0\}$ and the interface $\calw_\cals(\Sigma')$ at $\Sigma'=\{y=\varepsilon\}$. 
The combined action of the interfaces can be represented as follows
\be\label{UWSpicture} 
 \begin{split}
    &\!{\color{red}\cU_{\mathcal{S}}(
\Sigma
    )}\\
   \call(F_{\text{\tiny L}},\phi_{\text{\tiny L}}) \quad~~~~~~ &{\color{red} \quad\Bigg|} \quad ~~~~~~~~~\call(F_{\text{\tiny I}},f_\cals(\phi_{\text{\tiny I}})) = \call_{\cals}(F_{\text{\tiny I}},\phi_{\text{\tiny I}})\quad~~~\\
   &\!\!\!\!\!\!\!\!\!\!\!\!\!\!\!\!\!\!\!{\color{red}\scaleto{(\mathbbm{F}_{\text{\tiny L}},\phi_{\text{\tiny L}})|_\Sigma=(\mathbbm{F}_{\text{\tiny I}},f_\cals(\phi_{\text{\tiny I}}))|_\Sigma}{2ex}}
   \end{split}
     \quad \begin{split}
    &\;{\color{blue}\cW_{\mathcal{S}}(
\Sigma'
    )}\\
    &{\color{blue} \quad\Bigg|} \quad~~~~~~~~ \call( F_{\text{\tiny R}},\phi_{\text{\tiny R}})\, ,
    \\
&\!\!\!\!\!\!\!\!\!\!\!\!\!\!\!\!\!\!\!{\color{blue}\scaleto{(\mathbbm{F}_{\text{\tiny I}},\phi_{\text{\tiny I}})|_{\Sigma'}=(\cals\mathbbm{F}_{\text{\tiny R}},\phi_{\text{\tiny R}})|_{\Sigma'}}{2ex}} 
\end{split}
\ee
where  in the intermediate region we have crucially used the GZ condition \eqref{tildecall}. By taking the limit $\varepsilon\rightarrow 0$ and combining the gluing conditions \eqref{caludef} and \eqref{WSpicture},  we get a GZ symmetry defect $\mathcal D_{\cals}(\Sigma)$ which realizes the following gluing conditions:
\be\label{UWcollapse} 
 \begin{split}
    &\quad~~~~~~ {\color{purple}\cD_{\mathcal{S}}(
\Sigma
    )}\\
   \call(F_{\text{\tiny L}},\phi_{\text{\tiny L}}) &\quad~~~~~~~ {\color{purple} \quad\Bigg|} \qquad~~~~~~~~ \call( F_{\text{\tiny R}},\phi_{\text{\tiny R}})\,. \\
   &\,{\color{purple}\scaleto{(\mathbbm{F}_{\text{\tiny L}},\phi_{\text{\tiny L}})|_\Sigma=(\cals \mathbbm{F}_{\text{\tiny R}},f_\cals(\phi_{\text{\tiny R}}))|_\Sigma}{2ex}}
   \end{split}
\ee
Hence the GZ defect $\cD_{\mathcal{S}}(
\Sigma
    )$ does not change the theory and realizes a finite GZ transformation \eqref{gentr} of the bulk fields, as required.

In order to prove that  $\cald_\cals(\Sigma)$ is topological, it is sufficient to prove that it commutes with the energy-momentum tensor $T_{\mu\nu}$. In other words, one must prove that, even if the fields jump as in \eqref{UWcollapse} across $\cald_\cals(\Sigma)$, the energy-momentum tensor  does not. This property follows from the invariance  of the energy-momentum tensor under combined infinitesimal GZ deformations \eqref{delTphi} and \eqref{delTbF},
\be 
\delta_TT_{\mu\nu}=0\,,
\ee
which was proved in full generality in \cite{Gaillard:1981rj}. Since any finite GZ transformation can be obtained as a sequence of infinitesimal ones, the energy-momentum does not change across $\cald_\cals(\Sigma)$. 

We emphasize that the gluing conditions in \eqref{caludef}, \eqref{WSpicture} and \eqref{UWcollapse} involve  the pull-back of the field strengths $F^I|_\Sigma$ and $G_I|_\Sigma$. As already mentioned, this is natural from the Hamiltonian point of view and, a priori, this does not imply that the orthogonal components transform in the same way. Let us denote the pull-back of the field-strengths by $F^I_\|$ and $G_{\|\,I}$, respectively.  In the adapted coordinates $(y,\sigma^\alpha)$  introduced above, the orthogonal components of the field strengths are instead given by $F^I_\perp=F^I_{y\alpha}\d y\wedge \d\sigma^\alpha$ and $G_{\perp\,I}=(G_I)_{y\alpha}\d y\wedge \d\sigma^\alpha$. In a Hamiltonian description, these must be considered as functions of $F^I_\|$, $G_{\|\,I}$ and of the neutral sector $\phi^i$. Hence the gluing conditions of $F^I_\perp$ and $G_{\perp\,I}$ across a defect are dictated by the transformations of $F^I_\|$, $G_{\|\,I}$ and  $\phi^i$ in a possibly complicated way. On the other hand, the GZ condition  \eqref{tildecall} guarantees that  the mutual relation between $F^I$ and $G_J$  is  compatible with \eqref{gentr}.  So, if we take $F^I_\|$ and  $G_{\|\,I}$ (and $\phi^i$) as independent fields and impose the gluing condition \eqref{UWcollapse}, then also  $F^I_\perp$ and $G_{\perp \,I}$  on the left and right of $\cald_\cals(\Sigma)$ turn out to be   related  by the same symplectic transformation, that is,  $\mathbbm{F}_{\text{\tiny L}\,\perp}=\cals\mathbbm{F}_{\text{\tiny R}\,\perp}$ on $\Sigma$. Hence, $\cald_\cals(\Sigma)$  realizes the full GZ transformation \eqref{gentr}, in which all the components of $F^I$ and $G_I$ transform as in \eqref{gentrb}.

In addition, we observe  that by swapping the positions of $\calw_\cals$ and $\calu_\cals$ in \eqref{UWSpicture} the result does not change. That is, $\calw_\cals(\Sigma)$ and $\calu_\cals(\Sigma)$ in \eqref{Doperator} commute. Indeed $\calu_\cals(\Sigma)$ can be equivalently regarded as interpolating between $\call(F_{\text{\tiny L}},f_\cals^{-1}(\phi_{\text{\tiny L}}))=\call_{\cals^{-1}}(F_{\text{\tiny L}},\phi_{\text{\tiny L}})$ and $\call(F_{\text{\tiny R}},\phi_{\text{\tiny R}})$, and  $\calw_\cals(\Sigma)$ as interpolating between $\call(F_{\text{\tiny L}},\phi_{\text{\tiny L}})$ and $\call_{\cals^{-1}}(F_{\text{\tiny R}},\phi_{\text{\tiny R}})$, with the very same gluing conditions. More generically, following the same procedure and using \eqref{LSS'} and \eqref{fphiSS'}, one can also verify that the commutativity of these defects holds more generically, namely:
\be \label{commutativity}
\calu_\cals(\Sigma)\times\calw_{\cals'}(\Sigma)=\calw_{\cals'}(\Sigma)\times\calu_\cals(\Sigma)\,,
\ee
for any pair $\cals,\cals'\in \mathscr{G}_{\mathbb{Q}}$.

 On the other hand, generically, the interfaces  $\calw_\cals(\Sigma)$ (and the defects $\calu_\cals(\Sigma)$) do not commute among themselves and so, if one considers multiple insertions, the order is important. 
As a warm-up, consider first the following insertion of two $\calu$ defects
\be\label{UU'Spicture} 
 \begin{split}
    &\!{\color{red}\cU_{\mathcal{S}}(
\Sigma
    )}\\
   \call(F_{\text{\tiny L}},\phi_{\text{\tiny L}}) \quad~~~~~~ &{\color{red} \quad\Bigg|} \quad ~~~~~~~~~\call_{\cals}(F_{\text{\tiny I}},\phi_{\text{\tiny I}})\quad~~~\\
   &\!\!\!\!\!\!\!\!\!\!\!\!\!\!\!\!\!\!\!{\color{red}\scaleto{(\mathbbm{F}_{\text{\tiny L}},\phi_{\text{\tiny L}})|_\Sigma=(\mathbbm{F}_{\text{\tiny I}},f_\cals(\phi_{\text{\tiny I}}))|_\Sigma}{2ex}}
   \end{split}
     \quad \begin{split}
    &\;{\color{blue}\cU_{\mathcal{S}'}(
\Sigma'
    )}\\
    &{\color{blue} \quad\Bigg|} \quad~~~~~~~~ \call_{\cals\cals'}( F_{\text{\tiny R}},\phi_{\text{\tiny R}})\, ,
    \\
&\!\!\!\!\!\!\!\!\!\!\!\!\!\!\!\!\!\!\!{\color{blue}\scaleto{(\mathbbm{F}_{\text{\tiny I}},\phi_{\text{\tiny I}})|_{\Sigma'}=(\mathbbm{F}_{\text{\tiny R}},f_{\cals'}(\phi_{\text{\tiny R}}))|_{\Sigma'}}{2ex}} 
\end{split}
\ee
at $\Sigma=\{y=0\}$ and $\Sigma'=\{y=\varepsilon\}$. In \eqref{UU'Spicture} 
we have used \eqref{caludef} and \eqref{fphiSS'}. By \eqref{fSS'} we see that in the $\varepsilon\rightarrow 0$ limit we get precisely the expected composition rule
\be 
\calu_\cals(\Sigma)\times\calu_{\cals'}(\Sigma)=\calu_{\cals\cals'}(\Sigma)\,.
\ee
Consider now the insertion of two $\calw$ interfaces,
\be\label{WW'Spicture} 
 \begin{split}
    &\!{\color{red}\cW_{\mathcal{S}}(
\Sigma
    )}\\
   \call_{\cals\cals'}(F_{\text{\tiny L}},\phi_{\text{\tiny L}}) \quad~~~~~~~ &{\color{red} \quad\Bigg|} \quad ~~~~~~~~\call_{\cals'}(F_{\text{\tiny I}},\phi_{\text{\tiny I}})\quad~~\\
   &\!\!\!\!\!\!\!\!\!\!\!\!\!\!\!\!\!\!\!{\color{red}\scaleto{(\mathbbm{F}_{\text{\tiny L}},\phi_{\text{\tiny L}})|_\Sigma=(\cals\mathbbm{F}_{\text{\tiny I}},\phi_{\text{\tiny I}})|_\Sigma}{2ex}} 
   \end{split}
     \quad \begin{split}
    &\;{\color{blue}\cW_{\mathcal{S}'}(
\Sigma'
    )}\\
    &{\color{blue} \quad\Bigg|} \quad~~~~~ \call( F_{\text{\tiny R}},\phi_{\text{\tiny R}})\, ,
    \\
&\!\!\!\!\!\!\!\!\!\!\!\!\!\!\!\!\!{\color{blue}\scaleto{(\mathbbm{F}_{\text{\tiny I}},\phi_{\text{\tiny I}})|_{\Sigma'}=(\cals'\mathbbm{F}_{\text{\tiny R}},\phi_{\text{\tiny R}})|_{\Sigma'}}{2ex}} 
\end{split}
\ee
where we have used \eqref{LSS'}. Hence,
in the $\varepsilon\rightarrow 0$ limit, we get a $\calw_{\cals\cals'}$ interface.  

Since, as discussed above, the $\calu$ and $\calw$ defects commute, we conclude that by fusing two  topological GZ defects $\cald_\cals$ and $\cald_{\cals'}$, we get a GZ defect  $\cald_{\cals\cals'}$. 
However we will see that a given classical GZ symmetry $\cals\in\mathscr{G}_{\mathbb{Q}}$  corresponds to  several possible interfaces $\calw_\cals$ and hence, as common in the context of non-invertible symmetries, several possible topological  defects $\cald_\cals$. Assuming a given set  $\{\cald^a_\cals,\ a=1,2,\ldots\}$ of simple (i.e.\ non-decomposable) defects associated to the same $\cals\in\mathscr{G}_{\mathbb{Q}}$, we therefore a priori expect 
a more generic non-trivial fusion rule of the form
\be\label{gencomp} 
\cald^a_\cals(\Sigma)\times\cald^b_{\cals'}(\Sigma)= \sum_c\calt^{ab}_c(\Sigma) \otimes\cald^c_{\cals\cals'}(\Sigma)\,,
\ee
where 
$\calt^{ab}_c(\Sigma)$ represents the contribution of world-volume TQFTs decoupled from the bulk. 
In the following sections, we will provide further support for these general statements, discussing the non-invertible nature of the GZ defects  if $\cals\in \mathscr{G}_\mathbb{Q}$ is not in $\mathscr{G}_\mathbb{Z}$ and considering illustrative concrete examples.

%%%%%%%%%%%%%%%%%%%%%%%%%%%%%%%%%%%%%%%%%%%%%%%%%%%%%%%%%%

\section{Construction of  $\calw_\cals$ interfaces}
\label{sec:GZdefects2}

The  procedure for obtaining  the GZ topological defects $\cald_\cals(\Sigma)$ discussed in Section \ref{sec:GZdefects} assumes the existence of the interfaces $\calw_\cals(\Sigma)$   for any $\cals\in {\rm Sp}(2n,\mathbb{Q})$. We  now explain how to more concretely construct them. An alternative derivation of these interfaces, based on the half-space gauging procedure  \cite{Choi:2021kmx,Choi:2022zal,Hayashi:2022fkw,Kaidi:2021xfk, Choi:2022jqy}, is  presented in  Appendix \ref{app:doublegauging} and Appendix \ref{app:Cinterfaces}.  In this section, our discussion will remain entirely general and apply to any GZ model, while the analysis of concrete examples is postponed to the following sections. In particular, Section \ref{sec:Maxaxdil} examines two simple $n=1$ models and may be read first.

A key point is that any $\cals\in {\rm Sp}(2n,\mathbb{Q})$ can be decomposed into products of elements of the form \eqref{Sfactor}, with now $\mathsf{A}^I{}_J\in {\rm GL}(n,\mathbb{Q})$ and $\mathsf{C}_{IJ}=\mathsf{C}_{JI}\in\mathbb{Q}$ -- see e.g.\  \cite{o1978symplectic}. (Without loss of generality, one could actually assume that $\sfA^I{}_J\in {\rm SL}(n,\mathbb{Q})$, i.e.\ $\det\sfA=1$,  but in the following  we will not need to impose this restriction.) Hence,  an interface $\calw_\cals(\Sigma)$ corresponding to a general $\cals\in {\rm Sp}(2n,\mathbb{Q})$ can be decomposed into some ordered superposition of interfaces $\calw_{\mathsf{A}}(\Sigma)$, $\calw_{\mathsf{C}}(\Sigma)$ and $\calw_{\Omega}(\Sigma)$, corresponding to $\cals_{\mathsf{A}}$, $\cals_{\mathsf{C}}$ and $\cals_\Omega$, respectively. We will then focus on the construction of these `generating' interfaces, referring to them as {\sf A}, {\sf C} and $\Omega$ interfaces. The embedding map 
  \eqref{symprep} ensures that, for any $\cals\in\mathscr{G}_\mathbb{Q}$, a given superposition of interfaces $\calw_{\mathsf{A}}(\Sigma)$, $\calw_{\mathsf{C}}(\Sigma)$ and $\calw_{\Omega}(\Sigma)$ that  gives $\calw_\cals(\Sigma)$, leads to a  GZ defect $\cald_\cals(\Sigma)$  when fused with $\calu_\cals(\Sigma)$ as  in \eqref{Doperator}.

Finally,
 since there is no canonical decomposition of a general $\cals\in{\rm Sp}(2n,\mathbb{Q})$ into the product of {\sf A}, {\sf C} and $\Omega$ elements, there is no canonical way to define the general GZ topological defect  $\cald_\cals(\Sigma)$ either.   This  broadens the range of possibilities in building  GZ defects.  Different $\cald_\cals$ may differ by the type of non-invertibility as well as by the  topological sectors supported on them. While it would be very interesting to systematically identify the various possible GZ defects and study their properties, in this work we will content ourselves with focusing on some specific examples. 

\subsection{${\rm Sp}(2n,\mathbb{Z})$ generating interfaces.} 

Let us first briefly focus on the case $\cals\in {\rm Sp}(2n,\mathbb{Z})$. The corresponding interface $\calw_\cals(\Sigma)$ is invertible and   can be decomposed as an ordered superposition of interfaces $\calw_{\Omega}(\Sigma)$, $\calw_{\mathsf{A}}(\Sigma)$ and $\calw_{\mathsf{C}}(\Sigma)$, with  $\cals_{\mathsf{A}}, \cals_{\mathsf{C}}\in {\rm Sp}(2n,\mathbb{Z})$ -- see e.g.\  \cite{mumford2007tata}. These interfaces are the obvious generalization of the duality walls discussed in  \cite{Ganor:1996pe,Gaiotto:2008ak,Kapustin:2009av} -- see Appendix \ref{app:Spdualities}.   More explicitly, by using the notation of Section \ref{sec:GZdefects}, we have 
\begin{subequations}\label{WSZ}
\begin{align}
\calw_{\mathsf{A}}(\Sigma)&=\int\cald b\exp\left[\frac{\ii}{2\pi}\oint_\Sigma b_I\wedge (F^I_{\text{\tiny L}}-\mathsf{A}^I{}_JF^J_{\text{\tiny R}})\right]\,,&\label{WSZb}\\
\calw_{\mathsf{C}}(\Sigma)&=\exp\left(-\frac{\ii}{4\pi}\mathsf{C}_{IJ}\oint_\Sigma A^I_{\text{\tiny R}}\wedge F^J_{\text{\tiny R}}\right) \int\cald b\,\exp\left[\frac{\ii}{2\pi}\oint_\Sigma b_I\wedge (F^I_{\text{\tiny L}}-F^I_{\text{\tiny R}})\right]\,,&\label{WSZc}\\
\calw_\Omega(\Sigma)&=\exp\left(\frac{\ii}{2\pi}\delta_{IJ}\oint_\Sigma A^I_{\text{\tiny L}}\wedge F^J_{\text{\tiny R}}\right)\,,&\label{WSZa}\,,
\end{align}
\end{subequations}
where $\mathsf{A}\in {\rm GL}(n,\mathbb{Z})$ and $\mathsf{C}_{IJ}=\mathsf{C}_{JI}\in\mathbb{Z}$, and   the path integration over the world-volume U(1) gauge fields $b_I$ enforces the appropriate gluing conditions for the field strengths $F^I$ and $G_I$. Here we can be brief about this point, since in order to get $\calw_\cals(\Sigma)$ for a more general $\cals\in {\rm Sp}(2n,\mathbb{Q})$, we need to find a generalization of \eqref{WSZb} and \eqref{WSZc} which works for any $\mathsf{A}\in{\rm GL}(n,\mathbb{Q})$ and $\mathsf{C}_{IJ}\in\mathbb{Q}$.   
In the following subsections  we will describe in some detail the construction of these more general {\sf A} and {\sf C} interfaces, which will then encompass \eqref{WSZc} and \eqref{WSZb} as particular subcases.

\subsection{{\sf A} interfaces} 
\label{sec:GZAdefects}

Consider first elements $\cals_{\mathsf{A}}$ of the form \eqref{Sfactor}, with $\mathsf{A}\in{\rm GL}(n,\mathbb{Q})$. One may first try by just using $\mathsf{A}\in{\rm GL}(n,\mathbb{Q})$ in \eqref{WSZb}, but this would generically break the large gauge transformations of $b_I$ or, integrating by parts, of the $A^I_{\text{\tiny R}}$ gauge fields. 
One can however unfold such sick defect into an healthy one by choosing a factorization 
\be\label{AMEdec} 
\mathsf{A}=\mathsf{E}^{-1}\mathsf{M}\quad,\quad  \mathsf{M},\mathsf{E}\in{\rm Mat}(n,\mathbb{Z})\quad,\quad \det\sfM\neq 0\neq \det\sfE\,.
\ee
Note that $\mathsf{M}$ and $\mathsf{E}$ are invertible in $\mathbb{Q}$ but, generically,  not in $\mathbb{Z}$. Of course a factorization of the form \eqref{AMEdec} always exists but it is quite non-unique and allows for several possibilities.\footnote{\label{foot:Afact}For instance, choose $\mathsf{E}=N\mathds{1}$, where $N$ is the minimal positive integer such that $N\mathsf{A}$ is integer valued.}  One can significantly  reduce such a degeneracy by imposing that $\sfM$ and $\sfE$ provide a {\em left-coprime}  factorization of $\mathsf{A}$. See Appendix 
\ref{app:coprime} for the definition of left- and right-coprime factorization and a summary of  useful properties. As explained therein, if two   factorizations  $\mathsf{A}=\mathsf{E}{}^{-1}\mathsf{M}=\mathsf{E}'{}^{-1}\mathsf{M}'$ are both left-coprime, then they are related  by an unimodular integral matrix $\mathsf{T}\in{\rm GL}(n,\mathbb{Z})$, such that $\mathsf{E}'=\mathsf{T}\mathsf{E}$ and $\mathsf{M}'=\mathsf{T}\mathsf{M}$. On the other hand, if only $\mathsf{A}=\mathsf{E}{}^{-1}\mathsf{M}$  is left-coprime, then they are related in the same way though by a more general integral matrix $\sfT\in{\rm Mat}(n,\mathbb{Z})$, with $\det\sfT\neq0$. In particular, $\mathsf{E}'$ and $\mathsf{M}'$ are {\em not} left-coprime if and only if $|\det\sfT|\geq 2$.

Given a (not necessarily left-coprime)  factorization  \eqref{AMEdec}, we have the following possible realization  of an $\sfA$ interface, 
\be\label{WA} 
\calw^{(\sfE,\sfM)}_{\mathsf{A}}(\Sigma)=\int D b\exp\left[\frac{\ii}{2\pi}\oint_\Sigma b_I\wedge \left(\mathsf{E}^I{}_JF^J_{\text{\tiny L}}-\mathsf{M}^I{}_JF^J_{\text{\tiny R}}\right)\right]\,,
\ee
where $b_I$ is a set of $n$ U(1) gauge fields supported on $\Sigma$.
Note that any  transformation $\mathsf{E}\rightarrow \mathsf{T}\mathsf{E}$ and $\mathsf{M}\rightarrow \mathsf{T}\mathsf{M}$, with $\mathsf{T}\in{\rm GL}(n,\mathbb{Z})$, can be reabsorbed by the ${\rm GL}(n,\mathbb{Z})$ redefinition $b_I\rightarrow (\sfT^{-1})^J{}_I b_J$ of the world-volume gauge fields.  Hence, if we restrict to left-coprime factorizations,  \eqref{WA} is unique, in the sense that it depends only on $\mathsf{A}$ and not on the choice of the left-coprime factorization \eqref{AMEdec}. We will refer to this choice of \eqref{WA} as the {\em minimal} one, and in this case one may just write $\calw^{(\sfE,\sfM)}_{\mathsf{A}}(\Sigma)=\calw_{\mathsf{A}}(\Sigma)$.\footnote{Notice that, if  $\sfA\in{\rm GL}(n,\mathbb{Z})$,  the simplest left-coprime factorization is just given by $\sfE=\mathds{1}$ and $\sfM=\sfA$, and \eqref{WA} reduces to \eqref{WSZb}.}  On the other hand, by the above discussion we could relate non-left-coprime  factorizations to a left-coprime one by a non-unimodular transformation matrix $\sfT\in{\rm Mat}(n,\mathsf{Z})$, $|\det\sfT|\geq 2$, and the corresponding interface $\calw^{(\sfE,\sfM)}_{\mathsf{A}}(\Sigma)$ depends only on the ${\rm GL}(n,\mathbb{Z})$ conjugacy class of $\sfT$. In addition, a non-unimodular transformation changes the periodicities of $b_I$. This implies that there are different possible $\sfA$ interfaces  realizing the appropriate gluing conditions.  Even if the minimal $\sfA$ interfaces always exist and may be considered as the canonical ones, it may be difficult to obtain their explicit form. For this reason in the following sections we will also consider non-minimal $\sfA$ interfaces.

It is clear that \eqref{WA} imposes the constraint \be\label{MEeq} 
\mathsf{E}^I{}_JF^J_{\text{\tiny L}}|_\Sigma=\mathsf{M}^I{}_JF^J_{\text{\tiny R}}|_\Sigma\,.
\ee 
\eqref{MEeq} and \eqref{AMEdec} imply that a non-vanishing $\calw^{(\sfE,\sfM)}_{\mathsf{A}}(\Sigma)$
induces the correct gluing condition $F^J_{\text{\tiny L}}|_\Sigma=\mathsf{A}^I{}_JF^J_{\text{\tiny R}}|_\Sigma$ -- see \eqref{WSpicture}. On the other hand, \eqref{MEeq} also implies that \eqref{WA} does not vanish only if this gluing condition is compatible with the flux quantization conditions.  In other words,  $\calw^{(\sfE,\sfM)}_{\mathsf{A}}(\Sigma)$  acts as a projector that annihilates the flux configurations incompatible with the gluing conditions. Notice  that the gluing conditions \eqref{MEeq} do not fix possible torsional contributions to  $F^J_{\text{\tiny L}}|_\Sigma$  
and $F^J_{\text{\tiny R}}|_\Sigma$ in $H^2(X,\mathbb{Z})$,  which belong to the kernel of $\mathsf{E}^I{}_J$ and $\mathsf{M}^I{}_J$, respectively. Assuming the absence of $H^2(X,\mathbb{Z})$  torsion -- see Footnote \ref{foot:torsion} -- these contributions are vanishing. In any case, this freedom would not affect the interpretation of $\calw^{(\sfE,\sfM)}_{\mathsf{A}}$ as interface between two physical effective field theories, since they are insensitive to such torsional contributions. 

The bulk equations of motion of $A^I_{\text{\tiny L}}$ and $A^I_{\text{\tiny R}}$ get the localized contribution  
\be \label{glcondWAonG}
G_{\text{\tiny L},I}|_\Sigma=-\mathsf{E}^J{}_I\d b_J\quad,\quad G_{\text{\tiny R},I}|_\Sigma=-\mathsf{M}^J{}_I\d b_J\,.
\ee
These imply that $G_{\text{\tiny L},I}|_\Sigma=(\mathsf{M}^{-1}\mathsf{E}){}^J{}_IG_{\text{\tiny L},J}|_\Sigma=(\mathsf{A}^{-1}){}^J{}_IG_{\text{\tiny L},J}|_\Sigma$, which are indeed the correct gluing conditions. The conditions  \eqref{glcondWAonG} also impose stronger quantization conditions on $G_{\text{\tiny L},J}|_\Sigma$ and $G_{\text{\tiny R},J}|_\Sigma$, which make the gluing conditions sensible.

The $\sf A$ interfaces has non-trivial fusion rules, which contribute  to the non-invertibility of the corresponding GZ topological defects $\cald_\sfA$ constructed as in Figure \ref{fig:UW=D}. The simplest possibility is provided by the fusion of the interfaces associated with  $\mathsf{A}=\mathsf{E}^{-1}$ and $\mathsf{A}=\mathsf{M}$, with $\mathsf{M}$ and $\mathsf{E}$ invertible integral matrices. Notice that these cases correspond to two trivial left-coprime factorizations \eqref{AMEdec}, in which either  $\sfE=\mathds{1}$ or $\sfM=\mathds{1}$, respectively. 
Consider 
the insertion of  $\calw^{(\sfE,\mathbbm1)}_{\mathsf{A}}(\Sigma)$ and $\calw^{(\mathbbm1,\mathsf{M})}_\sfA(\Sigma')$, at $\Sigma=\{y=0\}$ and $\Sigma'=\{y=\varepsilon\}$,
as in \eqref{WW'Spicture}. 
In the limit $\varepsilon\rightarrow 0$ limit the intermediate bulk gauge fields $A^I_{\text{\tiny I}}$ become world-volume gauge fields $a^I$:
\be\label{WMWE} 
\begin{aligned}
\calw^{(\sfE,\mathbbm1)}_{\mathsf{A}}\times\calw_\sfA^{(\mathbbm1,\mathsf{M})}&=\int D a D bDb' \exp\Big[\frac{\ii}{2\pi}\oint_\Sigma b_I\wedge \left(\mathsf{E}^I{}_JF^J_{\text{\tiny L}}-\d a^I\right)\\ &\quad~~~~~~~~~~~~~~~~~~~~~~~~~~~~~~~~~~~~+\frac{\ii}{2\pi}\oint_\Sigma b'_I\wedge \left(\d a^I-\mathsf{M}{}^I{}_JF^J_{\text{\tiny R}}\right)\Big]\,.
\end{aligned}
\ee
Integrating out $a^I$ imposes that $b_I'=b_I$ and so we get  
\be\label{WAfact} 
\calw_\sfA^{(\mathsf{E},\mathbbm1)}\times\calw_\sfA^{(\mathbbm1,\mathsf{M})}=\calw^{(\sfE,\sfM)}_{\mathsf{A}}(\Sigma)\,.
\ee
In Appendix \ref{app:doublegauging} we obtain $\calw^{(\mathsf{E},\mathbbm1)}$ and $\calw^{(\mathbbm1,\mathsf{M})}$ from  appropriate  half-space gaugings \cite{Choi:2022jqy, Cordova:2023ent} of finite discrete subgroups of the electric and magnetic U(1) one-form symmetries. Through  \eqref{WAfact}, this provides a conceptually clearer way of interpreting \eqref{WA} as an interface connecting different   theories.

As a more interesting but still tractable example, we can consider the fusion of $\calw_{\mathsf{A}}(\Sigma)$ with its orientation reversal $\overline\calw_{\mathsf{A}}(\Sigma)$. The orientation reversal of defect is obtained by changing the orientation of $\Sigma$. We recall that our identification  of the left/right regions $\calt_\tL$ and $\calt_\tR$, introduced in Section \ref{sec:GZdefects}, is fixed by the orientation of $\Sigma$. Hence the orientation reversal also exchanges $\calt_\tL$ and $\calt_\tR$ and -- see Footnote \ref{foot:orient} -- and, hence, $F^I_\tL$ and $F^I_\tR$. From \eqref{WA} it then follows that
\be 
\overline\calw^{(\sfE,\sfM)}_{\mathsf{A}}(\Sigma)=\calw^{(\sfM,\sfE)}_{\mathsf{A}^{-1}}(\Sigma)\,.
\ee
Concatenating $\calw^{(\sfE,\sfM)}_{\mathsf{A}}(\Sigma)$ and $\overline\calw^{(\sfE,\sfM)}_{\mathsf{A}}(\Sigma')$ 
as in \eqref{WW'Spicture},  and taking the superposition limit,  we get the corresponding fusion product:
\be\label{WAWA-1} 
\begin{aligned}
\calc^{(\sfE,\sfM)}_\sfA&\equiv \calw^{(\sfE,\sfM)}_{\mathsf{A}}\times\overline\calw^{(\sfE,\sfM)}_{\mathsf{A}}=\int D a D bDb'\exp\Big[\frac{\ii}{2\pi}\oint_\Sigma b_I\wedge \left(\mathsf{E}^I{}_JF^J_{\text{\tiny L}}-\mathsf{M}^I{}_J\, \d a^J\right)\\ &\quad~~~~~~~~~~~~~~~~~~~~~~~~~~~~~~~~~~~~+\frac{\ii}{2\pi}\oint_\Sigma b'_I\wedge \left(\mathsf{M}{}^I{}_J\,\d a^J-\mathsf{E}{}^I{}_JF^J_{\text{\tiny R}}\right)\Big]\,.
\end{aligned}
\ee
This can be slightly simplified by making the field redefinition $b'_I\rightarrow b'_I+b_I$, which allows the factorization of the  r.h.s.\ of \eqref{WAWA-1}:
\be\label{WAWA-12} 
\begin{aligned}
\calc^{(\sfE,\sfM)}_\sfA=&\,\int D b\exp\Big[\frac{\ii}{2\pi}\mathsf{E}^I{}_J\oint_\Sigma b_I\wedge \left(F^J_{\text{\tiny L}}-F^J_{\text{\tiny R}}\right)\Big] \\
&\times\int Da D b' \exp\Big[\frac{\ii}{2\pi}\oint_\Sigma b'_I\wedge \left(\mathsf{M}{}^I{}_J\,\d a^J-\mathsf{E}{}^I{}_JF^J_{\text{\tiny R}}\right)\Big]\,.
\end{aligned}
\ee
Integrating out $b_I$ from the first factor constrains  $A^I_\tL$ and $A^I_\tR$ to differ only  by a (flat) dynamical connection localized on the defect, 
\be\label{AR-AL} 
\rho^I\equiv  \frac1{2\pi} \sfE^I{}_J\left(A^J_\tR|_\Sigma-A^J_\tL|_\Sigma\right)\ \in H^1(\Sigma,\mathbb{Z})\,.
\ee 
In particular, we can set $F^J_\tL=F^J_\tR\equiv F^I$. Similarly, integrating out the world-volume gauge fields $a^I$ imposes that $b'_J$ is flat and such that 
\be\label{bprimeint} 
\eta_I\equiv \frac1{2\pi}\mathsf{M}^J{}_I b'_J\in  \ H^1(\Sigma,\mathbb{Z})\,.
\ee 
 Taking into account large gauge transformations, the non-equivalent choices of $\rho^I$ and $\eta_I$ correspond to cohomology classes  with coefficients in a finite group. Furthermore, by integrating out $b_I'$ one sees that the defect is non-vanishing only if 
 \be\label{FRquandcond} 
\sfE^I{}_JA^J_\tR|_\Sigma=\sfM^I{}_Ja^J\quad{\rm mod}\  2\pi H^1(\Sigma,\mathbb{Z})\,,
 \ee 
 for some world-volume U(1)  gauge  fields $a^I$.

 We can formalize these observations by reformulating the problem in the framework of singular cohomology. For later purposes, it is convenient to start from some more general definition. First define the lattice   $V_\mathbb{Z}\simeq \mathbb{Z}^n$ of  integers $p^I\in\mathbb{Z}$, and a dual lattice $V^*_\mathbb{Z}\simeq \mathbb{Z}^n$ of integers $ \tilde p_J\in\mathbb{Z}$. Given an $n \times n$ invertible integral matrix $\mathsf{R}^{I}{}_J\in\mathbb{Z}$, we can define sublattices $\mathsf{R} V_\mathbb{Z}\subset V_\mathbb{Z}$ and $\mathsf{R}^{\rm t} V_\mathbb{Z}^*\subset V^*_\mathbb{Z}$ of integers of the form $p^I=\mathsf{R}^I{}_{J} k^I$ and $\tilde p_I=\mathsf{R}^J{}_{I} \tilde k_J$ ($ k^I,\tilde k_J\in\mathbb{Z}$), respectively, and the corresponding finite discrete subgroups
\be
\begin{aligned}
\Gamma_\mathsf{R}\equiv V_\mathbb{Z}/\mathsf{R} V_\mathbb{Z}\quad,&\quad \Gamma^*_\mathsf{R}\equiv V^*_\mathbb{Z}/\mathsf{R}^{\rm t} V^*_\mathbb{Z}\,.
\end{aligned}
\ee
The order of $\Gamma_\sfR$ and $\Gamma^*_\sfR$ is given by $|\det \sfR|$.\footnote{Write $\sfR$ in Smith normal form, $\sfR=\sfT\sfD\sfW^{-1}$, with $\sfD=\text{diag}(d_1,\ldots,d_n)$ and $\sfT,\sfW\in {\rm GL}(n,\mathbb{Z})$. The image of $\sfR$ is equal to the image of $\sfR\sfW$, and then equal to the image of $\sfT\sfD$. Hence $\Gamma_\sfR=V_\mathbb{Z}/\sfR V_\mathbb{Z}=V_\mathbb{Z}/\sfR \sfW V_\mathbb{Z}=V_\mathbb{Z}/\sfT\sfD V_\mathbb{Z}$. But $ V_\mathbb{Z}=\sfT V_\mathbb{Z}$. Hence $\Gamma_\sfR$ is isomorphic to $V_\mathbb{Z}/\sfD V_\mathbb{Z}$. Since $\sfD$ is non-degenerate, the image of $\sfD$ is isomorphic to ${d_1}\mathbb{Z}\times \ldots\times d_n\mathbb{Z}$. We conclude that $\Gamma_\sfR$ is isomorphic to $\mathbb{Z}_{d_1}\times \cdots \times \mathbb{Z}_{d_n}$, which has order $d_1 \cdots d_n = |\det D| = |\det R|$. Hence, $\Gamma_\sfR$ has order $|\det R|$, and the same holds for $\Gamma^*_\sfR$.}  
These abelian groups can be defined through  short exact sequences, which then induce long exact sequences of the corresponding cohomology groups.  

In the case at hand, let us first choose $\mathsf{R}=\mathsf{E}$ and consider the short exact sequence
\be 
0\ \longrightarrow \  V_\mathbb{Z}\ \xrightarrow[]{\ \mathsf{E}\ } \  V_\mathbb{Z}\ \longrightarrow \ \Gamma_\mathsf{E}\ \longrightarrow\  0\,.
\ee
This induces the long exact sequence 
\be 
\ldots \ \longrightarrow \  H^1(\Sigma,V_\mathbb{Z})\ \xrightarrow[]{\ \mathsf{E}\ } \ H^1(\Sigma,V_\mathbb{Z})\ \longrightarrow \  H^1(\Sigma,\Gamma_\mathsf{E})\ \xrightarrow[]{\ \beta\ }\  H^2(\Sigma,V_\mathbb{Z})\ \longrightarrow \ \ldots
\ee
where the $\beta$ is the Bockstein map. Taking into account large gauge transformations, we can then more precisely identify the physically inequivalent choices of $\rho^I$ defined in \eqref{AR-AL} with the $H^1(\Sigma,V_\mathbb{Z})$ uplift of $\bm{\rho}\in   H^1(\Sigma,\Gamma_\mathsf{E})$. 
Similarly, the physically inequivalent choices of $\eta_I$ defined in \eqref{bprimeint} correspond to the $H^1(\Sigma,V^*_\mathbb{Z})$ uplift of $\bm{\eta}\in  H^1(\Sigma,\Gamma^*_\mathsf{M})$.

We can now discuss  how \eqref{WAWA-12} can be interpreted as a combination of condensation defects obtained by one-gauging  the bulk electric and magnetic one-form symmetries \cite{Roumpedakis:2022aik}.
Assume first that $\sfE$ is unimodular, so that we can set $\sfE=\mathbbm{1}$ with no loss of generality. In this case \eqref{AR-AL} reduces to $A^I_\tL=A^I_\tR\equiv A^I$ (up to large gauge transformations).  Taking into account \eqref{bprimeint} and the above observations, and denoting $F^I_\tR$ as $F^I$,  the second factor appearing in \eqref{WAWA-12} can be written as\footnote{The normalization factor \eqref{CCcondgen2} takes into account the residual  gauge transformations of the (flat) gauge fields $b'_I$ corresponding to zero-form parameters in $ H^0(\Sigma,\Gamma^*_\sfM)$ \cite{Gaiotto:2014kfa,Kaidi:2021xfk}.}  \be\label{Mcondensate} 
\begin{aligned}
\calc^{(\mathbbm1,\sfM)}_\sfA(\Sigma)\equiv \frac{1}{|H^0(\Sigma,\Gamma_{\mathsf{M}}^*)|}\sum_{{\bm\eta}\in  H^1(\Sigma,\Gamma_{\mathsf{M}}^*) }\exp\left[\ii(\sfM^{-1})^I{}_J\oint_\Sigma \eta_I\cup F^J\right]\,.
\end{aligned}
\ee
This defect can be interpreted  as implementing the one-gauging   along $\Sigma$ of the finite one-form symmetry group subgroup $\Gamma^{*(1)}_\sfM\subset [{\rm U}(1)^{(1)}_{\rm m}]^n$, where $[{\rm U}(1)^{(1)}_{\rm m}]^n$ represents the magnetic one-form symmetry group of the GZ model. This can also be understood from \eqref{FRquandcond}, which implies that the insertion of $\calc^{(\mathbbm1,\sfM)}_\sfA(\Sigma)$ gives a non-vanishing contribution only if, along $\Sigma$, the bulk fluxes satisfy the stronger quantization condition  $(\sfM^{-1})^I{}_J\oint F^J|_\Sigma\in2\pi\mathbb{Z}$.

We next assume that, up to an irrelevant unimodular transformation, we can set $\sfM=\mathbbm1$, so that $\sfA=\sfE^{-1}$. In this case, integrating out $b'_I$ in \eqref{WAWA-12} sets $a^I=\sfE^I{}_JA^J|_\Sigma$ and \eqref{WAWA-12}  reduces to its first factor.  Its effect can be understood  by straightforwardly extend the discussion of  \cite{Cordova:2023ent}, which encountered the same situation in the $n=1$ case, i.e.\  with just one gauge field. The main point is that, for fixed $\rho^I$, the gluing condition \eqref{AR-AL} induced by that term  precisely 
corresponds to the effect of the codimension-one  defect $\exp\left[\ii (\sfE^{-1})^I{}_J\oint_\Sigma \rho^J\cup G_I\right]$. Summing over the possible physically inequivalent   $\rho^I$, we conclude the effect of the first factor  in \eqref{WAWA-12} is equivalent to  inserting the defect  
\be \label{condAE0}
\calc_\sfA^{(\sfE,\mathbbm1)}(\Sigma)\equiv\frac{1}{|H^0(\Sigma,\Gamma_{\mathsf{E}})|}\sum_{{\bm\rho}\in H^1(\Sigma,\Gamma_\mathsf{E}) }\exp\left[\ii (\sfE^{-1})^I{}_J\oint_\Sigma \rho^J\cup G_I\right]\,.
\ee
This is just the electromagnetic dual of \eqref{Mcondensate} and has an analogous dual interpretation. Namely, $\calc_\sfA^{(\sfE,\mathbbm1)}(\Sigma)$ realizes the one-gauging   along $\Sigma$ of the finite subgroup $\Gamma^{(1)}_\sfE\subset [{\rm U}(1)^{(1)}_{\rm e}]^n$ of the  electric one-form symmetry group of the GZ model. Note for instance that, dually to what seen above for $\calc_\sfA^{(\mathbbm1,\sfM)}$, the insertion of \eqref{condAE0} is non-vanishing only if  $(\sfE^{-1})^J{}_I\oint G_J|_\Sigma\in 2\pi\mathbb{Z}$. This may be seen from \eqref{WAWA-12}, by imposing the vanishing of the contribution localized on $\Sigma$ to the saddle point equations for the bulk gauge fields.

Let us finally consider the  case of more general  $\sfE$ and $\sfM$ matrices. The first factor appearing in \eqref{WAWA-12} can still be identified with \eqref{condAE0}, as in the  $\sfM=\mathbbm1$ case.  
On the other hand, the second factor  produces a modified version of \eqref{Mcondensate}: 
\be \label{WAWA3} 
\widetilde\calc^{(\mathbbm1,\sfM)}_{\sfE}(\Sigma)\equiv \frac{1}{|H^0(\Sigma,\Gamma_{\mathsf{M}}^*)|}\sum_{{\bm\eta}\in  H^1(\Sigma,\Gamma_{\mathsf{M}}^*) } \exp\left[\ii(\sfM^{-1})^I{}_J\oint_\Sigma \eta_I\cup (\sfE^J{}_K F^K)\right]\,.
\ee
This again corresponds to a one-gauging of the finite one-form symmetry group $\Gamma^{(1)}_\sfM$, but now associated with a non-minimal  choice of `charges' $\sfE^I{}_J$. 

In conclusion, the general condensation defect \eqref{WAWA-1} can be written as  
\be\label{WAWA-fin} 
\begin{aligned}
\calc^{(\sfE,\sfM)}_{\mathsf{A}}=\calc^{(\sfE,\mathbbm{1})}_{\mathsf{A}}\times \widetilde\calc^{(\mathbbm1,\sfM)}_{\sfE}\,.
\end{aligned}
\ee
Finally, we observe that, if  $\sfE$ and $\sfM$ are left-coprime and can be simultaneously put in Smith normal form, \eqref{WAWA3} is in fact equal to \eqref{Mcondensate}, 
so that $\calc^{(\sfE,\sfM)}_{\mathsf{A}}=\calc^{(\sfE,\mathbbm{1})}_{\mathsf{A}}\times \calc^{(\mathbbm1,\sfM)}_{\sfA}$. This is in agreement with the analogous result of \cite{Cordova:2023ent} for the $n=1$ case.

\subsection{{\sf C} interfaces from TQFTs} 
\label{sec:GZCdefects}

Consider now the {\sf C} interfaces  $\calw_{\mathsf{C}}(\Sigma)$. As in \eqref{WSpicture}, $\calw_{\mathsf{C}}(\Sigma)$ should interpolate between
\be\label{callSC} 
\call_{\sfC}(F_{\text{\tiny L}},\phi_{\text{\tiny L}})\equiv\call(F_{\text{\tiny L}},\phi_{\text{\tiny L}})+\frac1{4\pi}\mathsf{C}_{IJ}F^I_{\text{\tiny L}}\wedge F^J_{\text{\tiny L}}\,,
\ee
and $\call(F_{\text{\tiny R}},\phi_{\text{\tiny R}})$. The corresponding gluing conditions in \eqref{UWcollapse} require that the fields $F^I,\phi^i$ are continuous across $\Sigma$, while 
\be\label{Cgluingcond} 
G_{\text{\tiny L}\,I}|_\Sigma=G_{\text{\tiny R}\,I}|_\Sigma+\mathsf{C}_{IJ}F^J_{\text{\tiny R}}|_\Sigma\,,
\ee 
in agreement with  \eqref{callSC}. 

A first naive candidate for $\calw_{\mathsf{C}}(\Sigma)$ could be provided by \eqref{WSZc} with $\mathsf{C}_{IJ}\in\mathbb{Q}$, as one can easily realize by integrating out the world-volume $b_I$ and computing the boundary contributions to the equations of motion of the bulk gauge fields. However, this possibility would  not be invariant under large gauge transformations. As in \cite{Choi:2022jqy,Cordova:2022ieu}, the way out is to  replace \eqref{WSZc} with an interface 
\be\label{WC2} 
\calw_{\mathsf{C}}(\Sigma)= \cali_{\mathsf{C},\Sigma}\left[F_\tR\right]\int\cald b\,\exp\left(\frac{\ii}{2\pi}\oint_\Sigma b_I\wedge (F^I_{\text{\tiny L}}-F^I_{\text{\tiny R}})\right)\,,
\ee
where $\cali_{\mathsf{C},\Sigma}\left[F_\tR\right]$ encodes the coupling of a TQFT supported on $\Sigma$ to the bulk field strengths $F_\tR$, assuming for simplicity that the manifolds are spin. This coupling  should be responsible for the generation of the gluing conditions \eqref{Cgluingcond}. This happens precisely if $\cali_{\mathsf{C},\Sigma}\left[F_\tR\right]$ is such that  
\be\label{deltaAC} 
\delta\log\cali_{\mathsf{C},\Sigma}\left[F_\tR\right]=-\frac\ii{2\pi}\mathsf{C}_{IJ}\oint_\Sigma \delta A^I_\tR\wedge F^J_\tR\,,
\ee
 under a general small variation $\delta A^I_\tR$. Indeed, taking  \eqref{deltaAC}  into account, the  
 boundary contributions to the equations of motion of  $A^I_\tL$ and $A^I_{\text{\tiny R}}$ 
are  $G_{\text{\tiny L}\,I}|_\Sigma+\d b_I=0$ and   $G_{\text{\tiny R}\,I}|_\Sigma+\sfC_{IJ}F^J_\tR|_\Sigma+\d b_I=0$, which imply \eqref{Cgluingcond}. 

In order to realize \eqref{deltaAC}, let us first introduce an integral  matrix factorization 
\be\label{Cfact} 
\mathsf{C}=\sfP\mathsf{N}^{-1}\,,
\ee
where $\sfP_{IJ}\in\mathbb{Z}$ and $\mathsf{N}^I{}_J\in\mathbb{Z}$, with $\det\mathsf{N}\neq 0$. 
As in the discussion of the {\sf A} interfaces, one has a large freedom in the choice of the factorization \eqref{Cfact},  which can be substantially  removed by requiring that \eqref{Cfact} defines a {\em right-coprime} factorization -- see Appendix \ref{app:coprime}. This condition  reduces the freedom in the choice of the right-coprime $\sfP$ and $\mathsf{N}$ to 
\be\label{PNfreedom} 
\sfP\rightarrow \sfP\mathsf{T}\quad,\quad  \mathsf{N}\rightarrow \mathsf{N}\mathsf{T}\,,
\ee
with $\mathsf{T}\in{\rm GL}(n,\mathbb{Z})$, while  a non-right-coprime factorization is related to a right-coprime one by \eqref{PNfreedom}  with $\sfT\in{\rm Mat}(n,\mathbb{Z})$ and $|\det\sfT|\geq 2$. 

One can then  rewrite  \eqref{deltaAC} as
\be\label{deltaAC2} 
\begin{aligned}
\delta\log\cali_{\mathsf{C},\Sigma}\left[F_\tR\right]=&\,-\frac\ii{2\pi}(\sfN^{\rm t}\sfP)_{IJ}\oint_\Sigma  (\mathsf{N}^{-1}\delta A_\tR)^I\wedge (\mathsf{N}^{-1} F_\tR)^J\,.
\end{aligned}
\ee
Since $\mathsf{C}_{IJ}$ is symmetric, $(\sfN^{\rm t}\sfP)_{IJ}\equiv \mathsf{N}^K{}_I\sfP_{KJ}=(\mathsf{N}^{\rm t}\mathsf{C}\mathsf{N})_{IJ}$ is symmetric too. Employing  the notation introduced after \eqref{AR-AL},  we can identify $\frac1{2\pi}F_\tR^I$ appearing in \eqref{deltaAC2} with the integral uplift of an element  of $ H^2(X,\Gamma_{\mathsf{N}})$. 
This suggests the following  physical realization of  \eqref{deltaAC2}. 

Consider on $\Sigma$ a TQFT with $\Gamma^{(1)}_{\mathsf{N}}$ one-form symmetry generated by line operators $L_{{\bf n}}(\gamma)$, where $\gamma\subset \Sigma$ is a closed curve and ${\bf n}=\{n^I\}\in V_{\mathbb{Z}}$,  with ${\bf n}\simeq {\bf n}+\mathsf{N}{\bf k}$ for any ${\bf k}=\{k^I\}\in V_{\mathbb{Z}}$,  represents an element of $\Gamma_{\mathsf{N}}\equiv V_{\mathbb{Z}}/\mathsf{N}V_{\mathbb{Z}}$. These lines must realize the conditions   $L_{{\bf n}_1}(\gamma)L_{{\bf n}_2}(\gamma)=L_{{\bf n}_1+{\bf n}_2}(\gamma)$ and $L_{\mathsf{N}{\bf k}}(\gamma)=\mathds{1}$. 
 We assume that the rational numbers $\sfC_{IJ}\in\mathbb{Q}$ mod $\mathbb{Z}$ determine the mutual braiding of these lines as follows  
\be\label{genbraid} 
L_{{\bf n}}(\gamma)L_{{\bf n}'}(\gamma')=\exp\left(-2\pi\ii \mathsf{C}_{IJ}n^In'^J\right)L_{{\bf n}'}(\gamma')\,,
\ee
where $\gamma$ loops $\gamma'$. 
This brainding respects the periodicities ${\bf n}\simeq {\bf n}+\mathsf{N}{\bf k}$ because of \eqref{Cfact},  and is consistent with the exchange of the two lines since $\mathsf{C}_{IJ}$ is symmetric.\footnote{The braiding rule \eqref{genbraid} also determines the spin of the line operators: $h[L_{{\bf n}}]=\frac12\mathsf{C}_{IJ}n^In^J$ mod $\frac12$.} 
More generically, the charged objects are line operators $W(\gamma)$ such that
\be\label{gammaNcharge} 
L_{{\bf n}}(\gamma)W(\gamma')=\exp\left[2\pi\ii (\mathsf{N}^{-1})^J{}_In^Iq_J(W)\right]W(\gamma')\,,
\ee 
where the integers $q_I(W)\in\mathbb{Z}$ represent the charges of the line. Note that we must make the identification  $q_I(W)\simeq q_I(W)+\mathsf{N}^J{}_Ik_J$, with $k_J\in\mathbb{Z}$, and hence the charges of $W(\gamma)$ are in one-to-one correspondence with the elements of $\Gamma^*_\sfN\equiv V^*_\mathbb{Z}/\mathsf{N}^{\rm t}V^*_\mathbb{Z}$. 
Comparing \eqref{gammaNcharge} and \eqref{genbraid}, we see that the topological lines  $L_{\bf n}$ are charged under their own one-form symmetry, with charges 
\be\label{chargeL} 
q_I(L_{\bf n})=-n^J\sfP_{JI}\quad~~~ \text{mod}\quad \mathsf{N}^{\rm t}V^*_\mathbb{Z}\,.
\ee
Therefore the charges of the topological lines are determined by the integers $\sfP_{IJ}$ appearing in the factorization \eqref{Cfact}, which  hence also determine the anomaly of the  $\Gamma^{(1)}_{\mathsf{N}}$ one-form symmetry of the three-dimensional TQFT. More precisely, 
these data are determined by the equivalence class of $\sfP_{IJ}$, or equivalently $\sfC_{IJ}$,  modulo  the identifications 
\be\label{PCeqclass} 
\sfP_{IJ}\simeq \sfP'_{IJ}=\sfP_{IJ}+\sfZ_{IK}\sfN^K{}_J\quad,\quad \sfC_{IJ}\simeq \sfC'_{IJ}=\sfC_{IJ}+\sfZ_{IJ}\,,
\ee 
for any $\sfZ_{IK}=\sfZ_{KI}\in\mathbb{Z}$.   

The anomaly can be detected by coupling the TQFT to classical $\Gamma_{\mathsf{N}}$ background connections  $\bm{\omega}\in H^2(X,\Gamma_{\mathsf{N}})$. Their integral uplift $\omega^I$ can be identified with $ \frac1{2\pi}\sfN^I{}_J\calb^J\in H^2(X,\mathbb{Z})$, where $\calb^I$ are flat background   two-form gauge fields, defined modulo $2\pi H^2(X,\mathbb{Z})$.
Following  \cite{Hsin:2018vcg},\footnote{Our discussion can be connected to  Appendix F of \cite{Hsin:2018vcg} by choosing $\mathsf{N}^I{}_J=\delta^I_J N_J$. With this specific choice, $\mathsf{C}_{IJ}=\sfP_{IJ}/N_J$, which  can be translated  to the notation of \cite{Hsin:2018vcg} by identifying  our $\sfP_{IJ}$ and $\mathsf{C}_{IJ}$ with  $m_{JI}$ and $P_{IJ}/N_{IJ}$ therein. One can always make such a choice  by combining  a transformation \eqref{PNfreedom} with $\sfT\in$GL$(n,\mathbb{Z})$ and a  GL$(n,\mathbb{Z})$ redefinition of the field strengths $F^I$.} 
 the anomaly 
is then determined by the four-dimensional term 
\be\label{Banomalypol} 
-\frac1{4\pi}(\sfN^{\rm t}\sfP)_{IJ}\int_X\calb^I\wedge  \calb^J\,,
\ee
where $Y$ is an auxiliary four-dimensional space such that $\del Y=\Sigma$, and we are assuming that the theory is spin. Under an infinitesimal  gauge transformation $\delta\calb^I=\d \delta\Lambda^I$, the TQFT partition function then transforms by the phase
\be\label{infanomaly} 
\exp\left(-\frac{\ii}{2\pi}(\sfN^{\rm t}\sfP)_{IJ}\int_\Sigma \delta\Lambda^I\wedge \calb^J\right)\,.
\ee
Comparing \eqref{infanomaly} and \eqref{deltaAC}, it is  clear that a good candidate for $\cali_{\sfC,\Sigma}[F_\tR]$ appearing in \eqref{WC2} is provided by the partition function of a TQFT with anomaly determined by the integers $\mathsf{P}_{IJ}$ and coupled to bulk two-form gauge fields 
\be\label{B-Frel} 
\calb^I=(\mathsf{N}^{-1})^I{}_JF^J_\tR\,.
\ee

Notice that, since the anomaly is determined by $\sfP_{IJ}$ only modulo \eqref{PCeqclass}, the partition function of a given TQFT with such an anomaly may satisfy \eqref{deltaAC} only up to an additional term of the form $\exp\left(\frac\ii{2\pi}\sfZ_{IJ}\oint_\Sigma\delta A^I_\tR\wedge F^J_\tR\right)$, for some  $\sfZ_{IJ}=\sfZ_{IJ}\in\mathbb{Z}$. 
However such a term can always be reabsorbed by a counterterm of the form  \be\label{Zcounterterm} 
\exp\left(-\frac{\ii}{4\pi}\sfZ_{IJ}\oint_\Sigma A^I_\tR\wedge F^J_\tR \right)\,,
\ee  
corresponding to an additional integral shift of the constants $\sfC_{IJ}$. In the following, we will always assume that this possible counterterm has been included in $\cali_{\sfC,\Sigma}[F_\tR]$, so that the the resulting interface $\calw^{(\sfN,\sfP)}_\sfC$ induces the correct gluing condition \eqref{Cgluingcond}  for any $\sfC_{IJ}\in\mathbb{Q}$.

As a simple example, we can use the $\calz^{(\sfN,\sfP)}_\Sigma$ TQFT defined by the  action 
\be\label{simpleTQFT}
-\frac1{4\pi}(\sfN^{\rm t}\sfP)_{IJ}\oint_\Sigma a^I\wedge f^J +\frac1{2\pi}\mathsf{N}^I{}_J  \oint_\Sigma c_I\wedge f^J\,,
\ee
where $a_I$ and $c_I$ are U(1) one-form potentials, and $f^I=\d a^I$.  
The topological lines generating the $\Gamma^{(1)}_{\mathsf{N}}$ one-form symmetry are explicitly given by
\be\label{simpleLn} 
L_{\bf n}(\gamma)=\exp\left(\ii n^I\oint_\gamma c_I\right)\,,
\ee
and the bulk two-form gauge fields $\calb^I$ can be coupled  by adding a term $\frac1{2\pi}N^I{}_J\oint_\Sigma c_I\wedge \calb^J$ to \eqref{simpleTQFT}.
According to our general prescription \eqref{B-Frel}, we then get the following choice of the first factor on the r.h.s.\ of \eqref{WC2}:
\be\label{Ictheory}
\begin{aligned}
\cali_{\mathsf{C},\Sigma}[F_\tR]&\equiv \calz^{(\sfN,\sfP)}_\Sigma[\sfN^{-1}F_\tR]\\
&=\int \cald a\cald c\exp\Big[-\frac{\ii }{4\pi}(\sfN^{\rm t}\sfP)_{IJ}\oint_\Sigma a^I\wedge f^J +\frac\ii{2\pi} \oint_\Sigma c_I\wedge (\mathsf{N}^I{}_J f^J + F^I_\tR)\Big]\,.
\end{aligned}
\ee

  We can now more explicitly check  our general arguments. First, the $a^I$ and $c_I$ equations of motion impose
\be\label{aceom} 
\sfP_{IJ}f^J=\d c_I\quad,\quad \mathsf{N}^I{}_Jf^J=-F^I_\tR|_\Sigma\quad\Rightarrow\quad\d c_I=-C_{IJ}F^J_\tR|_\Sigma\,.
\ee
On the other hand,  the boundary contribution to the bulk $A^I_{\text{\tiny L},\text{\tiny R}}$ equations of motion is
\be\label{ALReom}  
G_{\text{\tiny L}\, I}|_\Sigma+\d b_I=0\quad,\quad G_{\text{\tiny R}\, I}|_\Sigma-\d c_I+\d b_I=0\,.
\ee
Combining them with \eqref{aceom} and \eqref{Cfact}, we get the correct gluing conditions \eqref{Cgluingcond}, without any need to include a counterterm of the form \eqref{Zcounterterm}.

Notice that $\sfN^{\rm t}\sfP\rightarrow \mathsf{T}^{\rm t}(\sfN^{\rm t}\sfP)\mathsf{T}$ under \eqref{PNfreedom}. If  $\sfT\in{\rm GL}(n,\mathbb{Z})$,  such a transformation can be reabsorbed by a redefinition $a^I\rightarrow (\mathsf{T}^{-1})^I{}_Ja^J$. 
This means that all right-coprime factorizations \eqref{Cfact} give the same  \eqref{Ictheory}, which  is then unique and depends only on $\sfC$. We may then  write $\cali^{(\sfN,\sfP)}_{\mathsf{C},\Sigma}=\cali_{\mathsf{C},\Sigma}$.\footnote{As a further  simple check, suppose that $\sfC_{IJ}\in\mathbb{Z}$.  In this case, the easiest right-coprime factorization \eqref{Cfact} is obtained by picking $\sfN=\mathds{1}$ and $\sfP=\sfC$. One can then explicitly integrate out $c_I$ from \eqref{Ictheory}, which amounts to just setting $a^I=A^I_\tR|_\Sigma$. Using the resulting \eqref{Ictheory} in \eqref{WC2}, the latter reduces to \eqref{WSZc}.} For non-right-coprime  factorizations, \eqref{Ictheory} and the corresponding interface \eqref{WC2} are instead in one-to-one correspondence with the $\sfT\in{\rm GL}(n,\mathbb{Z})$ conjugacy classes of the integral matrices $\sfT\in{\rm Mat}(n,\mathbb{Z})$ with $|\det\sfT|\geq 2$.

\subsubsection{Minimal {\sf C} interfaces}
\label{sec:minimalC}

The choice \eqref{Ictheory} is concrete and simple, but  any other TQFT with anomaly determined by  \eqref{Banomalypol} could be used to get the appropriate gluing conditions. It would be natural to require some notion of `minimality' on the choice of such TQFT. Following \cite{Hsin:2018vcg}, a minimal TQFT can be identified if and only if the braiding \eqref{genbraid} is non-degenerate, namely, if and only if the only line that braids trivially with all the other lines is the trivial identity line.  One can more precisely formulate this condition as follows \cite{Hsin:2018vcg}. Consider  the relation \eqref{chargeL} as defining a linear map $M: \Gamma_\mathsf{N}\rightarrow \text{Hom}(\Gamma_\mathsf{N},\text{U}(1))$. The braiding \eqref{genbraid} is non-degenerate if and only if $\ker M$ is trivial. In such a case the symmetry lines $L_{\bf n}(\gamma)$ form a  modular three-dimensional TQFT, which one can identify with the {\em minimal} TQFT with $\Gamma^{(1)}_\mathsf{N}$ one-form symmetry. As we are now going to show,  $\ker M$ is trivial, and therefore there  exists a minimal TQFT, precisely when the factorization \eqref{Cfact} is  right-coprime!  Hence, given any $\sfC_{IJ}$, by picking a right-coprime factorization \eqref{Cfact}, one can costruct a corrisponding minimal TQFT formed just by the  $\Gamma^{(1)}_{\sfN}$ topological lines $L_{\bf n}$ with brading \eqref{genbraid}.

The proof goes as follows.  First notice that the  triviality of $\ker M$ is equivalent to requiring that any integral vector ${\bf n}=\{n^I\}\in V_\mathbb{Z}\simeq \mathbb{Z}^n$ is such that  $\mathsf{C}{\bf n}\in V^*_\mathbb{Z}$ only if  ${\bf n}\in \mathsf{N}V_\mathbb{Z}$.
But from \eqref{Cfact}, this is equivalent to require that $\sfP{\bm\alpha}\in V^*_\mathbb{Z}$, with ${\bm\alpha}\equiv\mathsf{N}^{-1}{\bf n}$ for some ${\bf n}\in V_\mathbb{Z}$, necessarily implies that ${\bm\alpha}\in V_\mathbb{Z}$. Hence  $\ker M$ is trivial if and only if any ${\bm\alpha}\in V_\mathbb{Q}$ such that $\sfP{\bm\alpha}\in V^*_\mathbb{Z}$ and $\mathsf{N}{\bm\alpha}\in V_\mathbb{Z}$ is necessarily integral,  ${\bm\alpha}\in V_\mathbb{Z}$. That is, $\ker M$ is trivial if and only if $\sfP$ and $\sfN$ define a right-coprime factorization -- see Appendix \ref{app:coprime} -- as we wanted to show.    

For any given  $\sfC_{IJ}$,  we can pick a right-coprime factorization \eqref{Cfact} and construct the corresponding minimal TQFT, which actually depends only on the equivalence class  under \eqref{PCeqclass}.\footnote{Note that $(\sfP,\sfN)$ define a right-coprime factorization of $\sfC$ if and only if $(\sfP',\sfN)$ define a right-coprime factorization of $\sfC'$, with $\sfP'$ and $\sfC'$ defined as in \eqref{PCeqclass}.} This minimal TQFT is just formed by the $\Gamma^{(1)}_N$ topological  lines $L_{\bf n}$ with braiding \eqref{genbraid}. Notice that a general change of right-coprime factorization, which is given by  \eqref{PNfreedom} with $\sfT\in{\rm GL}(n,\mathbb{Z})$, corresponds just to a relabeling of the lines $L_{\bf n}$. Hence, the minimal TQFT is in fact unique and depends only on $\sfC_{IJ}$ mod $\mathbb{Z}$. For clarity, we will nevertheless denote  it as $\cala^{(\sfN,\sfP)}_\Sigma$.

As a simple example of minimal TQFT, suppose that \eqref{Cfact} can be realized with $\sfP_{IJ}=\delta_{IJ}$, i.e.\ that there exists an integral matrix $\sfN^{IJ}\equiv (\sfN\sfP^{-1})^{IJ}= \mathsf N^I{}_K\delta^{KJ}$ which is just the inverse of  $\sfC_{IJ}$. It is easy to check that this is a trivial example of right-coprime factorization. Then the minimal TQFT can be relized as the following abelian Chern-Simons theory
\be\label{minimalTQFT} 
\frac{\ii}{4\pi}\sfN^{IJ}\oint_\Sigma a_I\wedge f_J\,.
\ee
The topological lines generating $\Gamma^{(1)}_{\mathsf{N}}$ are
\be 
L_{\bf n}(\gamma)=\exp\left(\ii n^I\oint_\gamma a_I\right)\,.
\ee
The bulk two-form gauge fields $\calb^I$ can be coupled  by adding $\frac1{2\pi}\sfN^I{}_J\oint_\Sigma a_I\wedge \calb^J$  to \eqref{minimalTQFT}. Hence, applying the prescription \eqref{B-Frel}, the corresponding $\sfC$ interface is given by \eqref{WC2} with
\be\label{minCtheory}
\begin{aligned}
\cali_{\mathsf{C},\Sigma}[F_{\text{\tiny L}}]&=\cala^{(\sfN,\mathbbm1)}_{\Sigma}[\sfN^{-1}F_\tR]\\
&=\int \cald a\exp\Big[\frac{\ii }{4\pi}\sfN^{IJ}\oint_\Sigma a_I\wedge f_J +\frac\ii{2\pi} \oint_\Sigma a_I\wedge  F^I_\tR\Big]\,.
\end{aligned}
\ee
One can easily verify that, with such a choice, \eqref{WC2} provides the correct gluing conditions. 

 More generically,  one can realize the minimal TQFT on the boundary of a  four-dimensional theory, along the lines of    \cite{Kapustin:2014gua,Gaiotto:2014kfa,Hsin:2018vcg}.  This construction  is  discussed in detail in Appendix \ref{app:Cinterfaces}. By applying the general prescription \eqref{B-Frel}, we then get the following minimal realization of the  $\sfC$ defect \eqref{WC2}.   Pick an auxiliary four-dimensional space $Y$ such that $\del Y=\Sigma$ and a right-coprime integral factorization \eqref{Cfact}. Then identify $\cali_{\sfC,\Sigma}[F_\tL]$ in \eqref{WC2} with the partition function $\cala^{(\sfN,\sfP)}_\Sigma[\sfN^{-1}F_\tR]$ of the topological theory \eqref{4dmin2} with $\calb^I=(\sfN^{-1})^I{}_JF^J_\tR$:
\be\label{minpartfunct} 
-\frac1{4\pi}(\sfN^{\rm t}\sfP)_{IJ}\int_Y B^I\wedge  B^J+\frac1{2\pi}\sfN^I{}_J\int_Y  B^J\wedge  \tilde F_I+\frac1{2\pi}\oint_\Sigma  \tilde A_I\wedge F^I_\tR\,,
 \ee
where $\tilde A_I$ and $B^I$ are dynamical one- and two-form potentials defined on $Y$. The integration over $\tilde A_I$  imposes that $B^I$ are flat and identify $\Gamma_\sfN$ Wilson lines, and also enforces the  modified Dirchlet boundary conditions for $B^I$:
\be\label{modDirich} 
\sfN^I{}_J B^J|_\Sigma=-F^I_\tR|_\Sigma\,.
\ee
Together with the $B^I$ equations of motion $(\sfN^{\rm t}\sfP)_{IJ}B^J=\sfN^J{}_I\tilde F_J$, we have that
\be\label{Bbulkeq}  
\quad \sfN^J{}_I\tilde F_J|_\Sigma=-\sfP_{JI}F^J_\tR|_\Sigma\,.
\ee
which follows from $\sfN^t\sfP \sfN^{-1}= \sfP^t $ and the fact that  $\sfC_{IJ}=\sfC_{JI}$. 
Moreover, the last term in 
\eqref{minpartfunct}, together with the last factor appearing in \eqref{WC2}, affects the boundary $A^I_\tR$ equations of motion as follows: 
\be 
G_{\tR\, I}|_\Sigma-\tilde F_I|_\Sigma+\d b_I=0\,.
\ee
Combining them  with \eqref{Bbulkeq} and the boundary $A^I_\tL$ equations of motion $G_{\tL\, I}|_\Sigma+\d b_I=0$,    we again get the correct gluing conditions \eqref{Cgluingcond}, without having to include a counterterm of the form \eqref{Zcounterterm}.\footnote{For $\sfC_{IJ}\in\mathbb{Z}$, one can again use the right-coprime factorization $\sfN=\mathds{1}$ and $\sfP=\sfC$ and check that,  using the TQFT defined by \eqref{minpartfunct},  \eqref{WC2} reduces to \eqref{WSZc}.}

Following \cite{Hsin:2018vcg}, we can also show that any other three-dimensional TQFT $\tilde\cala^{(\sfN,\sfP)}_\mathsf{C}$ with  $\Gamma^{(1)}_\mathsf{N}$ one-form symmetry satisfying \eqref{chargeL},  can be factorized into
\be\label{calafact} 
\tilde\cala^{(\sfN,\sfP)}_\Sigma=\cala^0_\Sigma\otimes \cala^{(\sfN,\sfP)}_\Sigma\,,
\ee
where  $\cala^0_\Sigma$ is a TQFT invariant under the $\Gamma^{(1)}_\mathsf{N}$ one-form symmetry.  
By assumption  $\tilde\cala^{(\sfN,\sfP)}_\Sigma$ includes the topological lines $L_{\bf n}$,  which by themselves form the minimal TQFT $\cala^{(\sfN,\sfP)}_\Sigma$. Consider next any other line $W$ of $\cala^{(\sfN,\sfP)}_\Sigma$, with charges given by $q_I(W)$ mod $\mathsf{N^{\rm t}}V^*_\mathbb{Z}$. By fusing $W$ and  with any $L_{\bf n}$, one gets another line $W'=WL_{\bf n}$ of charges $q_I(W')=q_I(W)-n^J\sfP_{JI}$ mod $\mathsf{N^{\rm t}}V^*_\mathbb{Z}$. We would like to show that one  can always pick $n^I$ so that the 
$\Gamma^{(1)}_\mathsf{N}$ charge of $W'$ is vanishing, namely such that 
\be\label{zeroqcond}  
q_I(W)-n^J\sfP_{JI}=k_J\mathsf{N}^J{}_I\,, 
\ee
for some $k_J\in\mathbb{Z}$. Again, this follows from the fact that $\sfP$ and $\mathsf{N}$ are right-coprime. Indeed, by definition -- see Appendix \ref{app:coprime} --  there exist two integral matrices $\mathsf{X}^{IJ}$ and $\mathsf{Y}^I{}_J$ such that
\be
(\mathsf{X}\sfP-\mathsf{Y}\mathsf{N})^I{}_J=\delta^I{}_J\,.
\ee
By using this identity it is easy to see that $n^I=q_J(W)X^{JI}$ satisfies \eqref{zeroqcond}, with $k_I=-q_J(W)\mathsf{Y}^J{}_I$. 
Hence, with such a choice, the line $W'=WL_{\bf n}$ is $\Gamma^{(1)}_\mathsf{N}$ neutral.
We can then identify the set of such neutral lines with $\cala^0_\Sigma$ in   \eqref{calafact}, since it is clear that any line $W$ is the product of a line of $\cala^0_\Sigma$  and a line of $\cala^{(\sfN,\sfP)}_\Sigma$.

Consider, for instance, the $\calz^{(\sfP,\sfN)}_\Sigma$ theory \eqref{simpleTQFT}. In addition to the $\Gamma^{(1)}_{\sfN}$ lines  \eqref{simpleLn}, it contains also the topological lines 
\be 
W_{\bf n}(\gamma)=\exp\left[\ii n^I\oint_\gamma\left(\sfP_{IJ}a^J-c_I\right)\right]\,.
\ee
One can check that the $L_{\bf n}$ and $W_{\bf m}$ lines are mutually transparent. Hence the $W_{\bf m}$ lines form the decoupled sector $\cala^0_\Sigma$. Note that  $W_{\bf m}$ have braiding of the form \eqref{genbraid} with $\sfC_{IJ}\rightarrow -\sfC_{IJ}$. Hence, we can identify $\cala^0_\Sigma$ with $\cala^{(-\sfP,\sfN)}_{\Sigma}$.

In conclusion, for any $\mathsf{C}_{IJ}\in\mathbb{Q}$ we can construct a minimal interface $\calw_{\mathsf{C}}(\Sigma)$, by choosing
\be\label{calicala}
\cali_{\sfC,\Sigma}[F_\tR]=\cala^{(\sfN,\sfP)}_\Sigma[\sfN^{-1}F_\tR]\,,
\ee
in \eqref{WC2}, i.e.\ the partition function of the minimal TQFT  coupled to bulk two-form gauge potentials $\calb^I$ of the form \eqref{B-Frel}, plus a possible counterterm \eqref{Zcounterterm}. Furthermore, any other $\widetilde\calw^{(\sfN,\sfP)}_{\mathsf{C}}(\Sigma)$ constructed by using  any other TQFT associated with a right-coprime factorization,  can be written as 
\be 
\widetilde\calw^{(\sfN,\sfP)}_{\mathsf{C}}(\Sigma)=\calt_0(\Sigma)\,\calw^{(\sfN,\sfP)}_{\mathsf{C}}(\Sigma)\,,
\ee
where $\calt_0(\Sigma)$ is the partition function of the neutral sector $\cala^0_\Sigma$ appearing in \eqref{calafact}.
More generically, one can also construct $\sfC$ interfaces by using TQFT with $\Gamma^{(1)}_N$ one-form symmetry associated with non-necessarily right-coprime factorizations, as for instance the TQFT defined by \eqref{Ictheory}, which works for any factorization. 

As in the case of $\sfA$ defect, in concrete examples, the minimal interfaces may often not be the simplest ones. In these cases, one may use non-minimal ones, possibly associated with non-coprime factorizations of $\sfC$. In particular, \eqref{Ictheory} provides a simple concrete choice, which works for any factorization.

%%%%%%%%%%%%%%%%%%%%%%%%%%%%%%%%%%%%%%%%%%%%%%%%%%%%%%%%%%%%%%%%%%%%

\subsubsection{Condensation defects} \label{sec:Ccondefects}

Similarly to what already seen for the $\sfA$ defects, the non-trivial TQFT supported by the $\sfC$ interface implies that, if $\sfC_{IJ}\notin\mathbb{Z}$,  the fusion $\calw^{(\sfN,\sfP)}_{\mathsf{C}}\times\overline\calw^{(\sfN,\sfP)}_{\mathsf{C}}(\Sigma)$ does not give the identity, but rather a condensation defect $\calc^{(\sfN,\sfP)}_\sfC(\Sigma)$ of the type discussed in \cite{Roumpedakis:2022aik}. 

In order to illustrate this point, let us restrict ourselves to the case $\sfP=\mathbbm1$, that is, $\sfC=\sfN^{-1}$. In this case the minimal TQFT contribution to \eqref{WC2} is given  by \eqref{minCtheory}, with $\sfN^{IJ}\equiv \sfN^I{}_K\delta^{KJ}$. By fusing $\calw^{(\sfN,\mathbbm1)}_{\mathsf{C}}$ and $\overline\calw^{(\sfN,\mathbbm1)}_{\mathsf{C}}$  we then get the condensation defect
\begin{equation}
\begin{aligned}
    \mathcal C^{(\sfN,\mathbbm1)}_\mathsf{C}  (\Sigma) =&\, \int Da \,D\tilde{a}\,Db \,D\tilde{b}\, D c \; {\rm exp}\bigg[ \frac{\ii }{4\pi}\sfN^{IJ}\oint_{\Sigma}  \left( a_I\wedge \d a_J  - \tilde a_I\wedge \d\tilde a_J \right)+ \frac{\ii}{2\pi}\oint_\Sigma b_I \wedge (\d c^I - F^I_\tL)\\
    &\quad+\frac{\ii}{2\pi} \oint_\Sigma\left(a_I \wedge \d c^I - \tilde a_I \wedge F^I_\tR\right)  - \frac{\ii}{2\pi}\oint_\Sigma\tilde b_I \wedge (F^I_\tR - \d c^I) \bigg]\,,
    \end{aligned}
\end{equation}
where $c^I$ represents the world-volume remnant of the bulk gauge field enclosed by  the two interfaces. Integrating  it out sets $\tilde b_I=-b_I-a$. Furthermore, integrating out $b_I$ imposes $F^I_\tR|_\Sigma=F^I_\tL|_\Sigma$, so that we can just set $F^I_\tL=F^I_\tR\equiv F^I$. Then, redefining  $a_I\rightarrow a_I+\tilde a_I$ one gets  
\begin{equation}\label{CCcondgen}
   \calc^{(\sfN,\mathbbm1)}_\sfC(\Sigma) =\, \int Da \,D\tilde{a}   \; {\rm exp}\left[\frac{\ii}{4\pi}\sfN^{IJ}\oint_{\Sigma}   a_I\wedge \d a_J +\frac{\ii}{2\pi}\sfN^{IJ}\oint_{\Sigma}   a_I\wedge \d \tilde a_J +  \frac{\ii}{2\pi}\oint_{\Sigma}   a_I\wedge F^I\right]\,.
\end{equation}
This is the natural generalization to $n\geq 1$ of the condensation defect found in \cite{Choi:2022jqy}, and can be interpreted as a 
`one-gauging' \cite{Roumpedakis:2022aik} of a discrete magnetic  one-form symmetry group $\Gamma^{*(1)}_\sfN\subset [U(1)^{(1)}_{\rm m}]^n$.

More explicitly, by
integrating out $\tilde a_I$ one gets the constraints $\d a_I=0$ and $\sfN^{J}{}_I a_J\in 2\pi H^1(X,\mathbb{Z})$. That is, the closed one-forms $\eta_I\equiv \frac1{2\pi}\sfN^J{}_Ia_J$ can be regarded as the integral uplift of  a $\Gamma^*_\sfN$ cohomology class  $\bm{\eta}\in H^1(X,\Gamma^*_\sfN)$. We can then rewrite the condensation defect as follows 
\be\label{CCcondgen2} 
\calc^{(\sfN,\mathbbm1)}_\sfC(\Sigma)=\frac{1}{\left|H^0(\Sigma,\Gamma^*_\sfN)\right|}\sum_{\bm\eta\in H^1(\Sigma,\Gamma^*_\sfN)}\exp\left[\ii(\sfN^{-1})^I{}_J\oint_\Sigma \eta_I\cup F^J\right]\exp\left[\pi\ii\delta^{IJ}\oint_\Sigma \eta_I\cup \beta_J(\bm\eta)\right]\,,
\ee
where $\beta_I(\bm\eta)=(\sfN^{-1})^J{}_I\delta\eta_J\in H^2(\Sigma,\mathbb{Z})$ gives the Bockstein image of $\bm\eta\in H^1(X,\Gamma^*_\sfN)$.  Notice that the last factor in \eqref{CCcondgen2} introduces a discrete  torsion phase  for this one-gauging.

%%%%%%%%%%%%%%%%%%%%%%%%%%%%%%%%%%%%%%%%%%%%%%%%%%%%%%%%%%%%%%%%%%%%

\section{Action on line defects \label{sec:AactWH}}

The action of the GZ  defects $\cald_\cals(\Sigma)$ on local operators can be immediately read from \eqref{UWcollapse}. For instance, suppose that $\cald_\cals(S^3)$ is inserted into a correlator, where $S^3$ surrounds a local gauge-invariant operator $\calo(x)$ constructed out of $F^I,G_J$ and $\phi^i$, and the orientation is fixed so that the interior of $S^3$ corresponds to the left side of $S^3$.
Collapsing $S^3$ to a point leaves the operator $\tilde\calo(x)$ obtained by applying the GZ transformations \eqref{gentra}. So, the action on local operators is the same as that expected from a standard invertible operator.

The action of GZ defects  on  Wilson, 't Hooft lines and more general dyonic line operators is more interesting, and will further highlight the non-invertible nature of these defects. It is convenient to directly consider a general dyonic line.  
\begin{equation} \label{defdyonline0}
    \begin{aligned}
L_{\mathbbm{L}}(\gamma)=\exp{\left(\ii\,\tilde\ell^I\oint_\gamma\tilde A_I-\ii\, \ell_I\oint_{\gamma} A^I\right)}\,,\\
   \end{aligned} 
\end{equation}
where $\tilde A_I$ are the S-dual gauge fields, such that $\d\tilde A_I=G_I$, and  we have introduced the $2n$-dimensional vector 
\be
\mathbbm{L}=\left(\begin{array}{c}
\tilde{\ell}^I \\ {\ell}_J
\end{array}\right)\,,
\ee
which is formed by the magnetic and electric charges of the dyonic line. These charges must be integral for \eqref{defdyonline0} to be invariant under large gauge transformations of $A^I$ and $\tilde A_I$. The signs in \eqref{defdyonline0} are chosen for later convenience. As particular subcases, by setting $\tilde\ell^I=0$ or $\ell_I=0$, $L_{\mathbbm{L}}$ reduces to a  Wilson  or 't Hooft lines,
\be\label{Wilsonthooft} 
W_{\bm\ell}(\gamma)\equiv \exp{\left(-\ii\, \ell_I\oint_{\gamma} A^I\right)}\quad,\quad H_{\tilde{\bm\ell}}(\gamma)\equiv \exp{\left(\ii\, \tilde\ell^I\oint_{\gamma} \tilde A_I\right)}\,,
\ee
where $\bm\ell$ and $\tilde{\bm\ell}$ denote the charge vectors of components  $\ell_I$ and $\tilde\ell^I$, respectively. 

The action of the GZ defects on these line operators is cleaner if we restrict ourselves to homologically trivial curves $\gamma$, which can hence be written as $\gamma=\del M_2$ for some two-dimensional surface $M_2$. In this case \eqref{defdyonline0} can be rewritten as 
\be\label{dyongen} 
 \widehat L_{\mathbbm{L}}(\gamma, M_2)=\exp\left(\ii\,\tilde\ell^I\int_{M_2} G_I-\ii\, \ell_I\int_{M_2} F^I\right)\equiv \exp\left(\ii\,\int_{M_2} \mathbb{L}^{\rm t}\Omega \mathbbm{F}\right)\,,
\ee
and we can correspondingly denote the purely electric and magnetic subcases as 
\be\label{Wilsonthooft1} 
\widehat W_{\bm\ell}(\gamma,M_2)\equiv \exp{\left(-\ii\, \ell_I\int_{M_2} F^I\right)}\quad,\quad \widehat H_{\tilde{\bm\ell}}(\gamma,M_2)\equiv \exp{\left(\ii\, \tilde\ell^I\int_{M_2} G_I\right)}\,.
\ee
 For these operators we use a different symbol since they actually make sense for any, not necessarily integral, value of the charges $\ell_I$ and $\tilde\ell^J$. Only for integral charges do \eqref{dyongen}  correspond to the genuine line operators \eqref{defdyonline0}, and are completely independent of the choice of $M_2$:
\be\label{genWLine}
\widehat L_{\mathbbm{L}}(\gamma, M_2)\equiv L_{\mathbbm{L}}(\gamma)\quad~~~\Leftrightarrow\quad~~~ \tilde\ell^I,\ell_J\in\mathbb{Z}\,.
\ee
Otherwise, \eqref{dyongen}  must be interpreted as open surface operators, which depend on both $\gamma$ and the relative  homology class of $M_2$.  

The action of a topological GZ defect $\cald_\cals(\Sigma)$ on the surface operators  \eqref{dyongen} immediately follows from the gluing conditions described by \eqref{UWcollapse}. Suppose that we first put $\widehat L_{\mathbbm{L}}(\gamma,M_2)$ on the right of $\Sigma$. Hence, if we sweep $\cald_\cals(\Sigma)$ past  $\widehat L_{\mathbbm{L}}(\gamma,M_2)$, the latter will undergo a transformation  
\be\label{hatLLtr} 
\widehat L_{\mathbbm{L}}(\gamma,M_2)= \exp\left(\ii\,\int_{M_2} \mathbb{L}^{\rm t}\Omega \mathbbm{F}_\tR\right)\quad\longrightarrow\quad \widehat L_{\cals\mathbbm{L}}(\gamma,M_2)= \exp\left(\ii\,\int_{M_2} (\cals\mathbb{L})^{\rm t}\Omega \mathbbm{F}_\tL\right)\,,
\ee
where we have used the fact that $\mathbbm{F}_\tR\rightarrow \cals^{-1}\mathbbm{F}_\tL$ across the defect, and the defining identity \eqref{calScond} rewritten as $\Omega\cals^{-1}=\cals^{\rm t}\Omega$.  In other words,  sweeping $\cald_\cals(\Sigma)$ past a  surface operator \eqref{dyongen}, from the left to the right,  turns it into a new  surface operator  with charge vector $\mathbbm{L}'=\cals\mathbbm{L} $. More explicitly, using the decomposition \eqref{genS}, the new charges are 
\be\label{trasell} 
\tilde\ell'^I=\sfA^I{}_J\tilde\ell^J+\sfB^{IJ}\ell_J\quad,\quad \ell'_I=\sfC_{IJ}\tilde\ell^J+\sfD_I{}^J\ell_J\,.
\ee

\begin{figure}[h]
\centering
\begin{tikzpicture}[scale=1.2,>=Latex]

\tikzset{
  surface/.style={thick,purple},
  lineop/.style={thick,green!60!black},
  fluxop/.style={thick,orange!80!black,fill=orange!30},
  ->/.style={-{Latex[length=3mm,width=2mm]}}
}

% --- Left: initial configuration ---
% surface defect
\filldraw[purple!30!white,fill opacity=0.4,draw=purple,thick]
  (-3,1.2) -- (-3,-1.2) -- (-2.2,-1.5) -- (-2.2,0.9) -- cycle;
\node[purple,above] at (-2.6,1.2) {$\mathcal{D}_\cals(\Sigma)$};

% line operator L
\draw[lineop] (-1.1,0) ellipse (0.18 and 0.35);
\node[green!40!black,below right=-2pt] at (-1.5,-0.52) {$L_{\mathbb L}(\gamma)$};

% --- Middle arrow ---
\draw[->,thick] (-0.6,0) -- (0.6,0);

% --- Right: after crossing ---
% surface defect (again)
\filldraw[purple!30!white,fill opacity=0.4,draw=purple,thick]
  (2.2,1.2) -- (2.2,-1.2) -- (3,-1.5) -- (3,0.9) -- cycle;
\node[purple,above] at (2.6,1.2){$\mathcal{D}_\cals(\Sigma')$};

% Parameters
\def\L{1.5}  % cylinder length
\def\R{0.35}  % cylinder radius
\def\colorc{orange!40}

% Front ellipse
\draw[orange!80!black,thick,fill=\colorc,opacity=0.6]
  (1.1,0) ellipse ({0.18} and {\R});

% Label
\node[orange!80!black,below right=2pt] at (0.2,-\R) {$\hat L_{\cals \mathbb{L} }(\gamma,M_2)$};

\end{tikzpicture}
\caption{Action of the  defect $\mathcal D_{\cals}$ on the line operator $L_{\mathbb{L}}$.}
\label{fig:line-surface-flux-filled}
\end{figure}
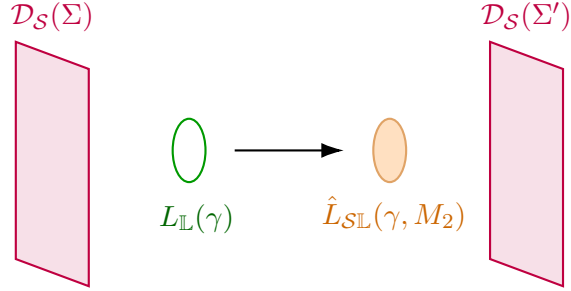

It is now clear that a GZ defect corresponding to a {\em non}-integral symplectic matrix $\cals\in{\rm Sp}(2n,\mathbb{Q})$ generically turns a genuine dyonic line \eqref{defdyonline0} into a non-genuine one, 
\be\label{actiononlines} 
L_{\mathbbm{L}}(\gamma) \quad~~~\longrightarrow\quad~~~ \widehat L_{\cals\mathbbm{L}}(\gamma,M_2)\,,
\ee
see Figure \ref{fig:line-surface-flux-filled}.
Of course, this is not always the case as, for any choice of rational matrices appearing in \eqref{genS}, one can always find charges $\tilde\ell^I,\ell_J\in\mathbb{Z}$ such that the new charges \eqref{trasell} are integral too. 

Suppose for instance that a GZ model admits an element $\cals_\sfC\in\mathscr{G}_\mathbb{Q}$, with $\cals_\sfC$ as in \eqref{Sfactor}, for some  $\sfC_{IJ}\in\mathbb{Q}$. Then \eqref{actiononlines} implies that a Wilson line $W_{\bm\ell}(\gamma)$ -- see \eqref{Wilsonthooft} -- is transparent to $\cald_\sfC(\Sigma)$. On the other hand,  by sweeping $\cald_\sfC(\Sigma)$ past a 't Hooft line $H_{\tilde{\bm\ell}}(\gamma)$ from the left to the right, one gets the open surface operators  
\be\label{actiononlinesC} 
  H_{\tilde{\bm\ell}}(\gamma)\widehat W_{\sfC\tilde{\bm\ell}}(\gamma,M_2)=H_{\tilde{\bm\ell}}(\gamma)\exp\left(-\ii \sfC_{IJ}\tilde\ell^J\int_{M_2}F^I\right)\,.
\ee
Analogously, if a GZ model admits element an $\cals_\sfA\in\mathscr{G}_\mathbb{Q}$ of the form \eqref{Sfactor}, then sweeping the corresponding topological defect $\cald_\sfA(\Sigma)$ past  $W_{\bm\ell}(\gamma)$ and $H_{\tilde{\bm\ell}}(\gamma)$   turns them into the open surface operators  
\be 
 \widehat   W_{\sfA^{-1}\bm\ell}(\gamma,M_2)=\exp\left[-\ii (\sfA^{-1})^I{}_{J}\ell_I\int_{M_2}F^J\right]  \quad,\quad \widehat H_{\sfA\tilde{\bm\ell}}(\gamma,M_2)=\exp\left(\ii \sfA^I{}_{J}\tilde\ell^J\int_{M_2}G_I\right)\,,
\ee
respectively.

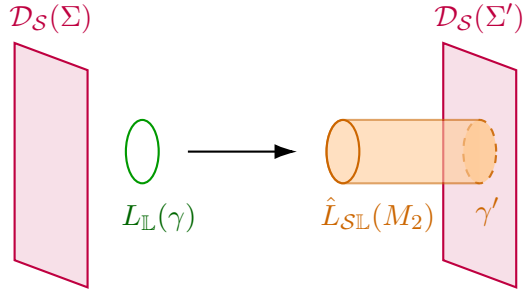
\begin{figure}[h]
\centering
\begin{tikzpicture}[scale=1.2,>=Latex]

\tikzset{
  surface/.style={thick,purple},
  lineop/.style={thick,green!60!black},
  fluxop/.style={thick,orange!80!black,fill=orange!30},
  ->/.style={-{Latex[length=3mm,width=2mm]}}
}

% --- Left: initial configuration ---
% surface defect
\filldraw[purple!30!white,fill opacity=0.4,draw=purple,thick]
  (-2.5,1.2) -- (-2.5,-1.2) -- (-1.7,-1.5) -- (-1.7,0.9) -- cycle;
\node[purple,above] at (-2.1,1.2) {$\mathcal{D}_\cals(\Sigma)$};

% line operator L
\draw[lineop] (-1.1,0) ellipse (0.18 and 0.35);
\node[green!40!black,below right=-2pt] at (-1.4,-0.5) {$L_{\mathbb L}(\gamma)$};

% --- Middle arrow ---
\draw[->,thick] (-0.6,0) -- (0.6,0);

% --- Right: after crossing ---
% surface defect (again)
\filldraw[purple!30!white,fill opacity=0.4,draw=purple,thick]
  (2.2,1.2) -- (2.2,-1.2) -- (3,-1.5) -- (3,0.9) -- cycle;
\node[purple,above] at (2.6,1.2){$\mathcal{D}_\cals(\Sigma')$};

% Parameters
\def\L{1.5}  % cylinder length
\def\R{0.35}  % cylinder radius
\def\colorc{orange!40}

% Back ellipse (dashed)
\draw[dashed,orange!80!black,thick,fill=\colorc,opacity=0.9] (1.1+\L,0) ellipse ({0.18} and {\R});

% Body (filled rectangle with curved ends)
\path[fill=\colorc,draw=orange!80!black,thick,opacity=0.6]
  (1.1+\L,\R) -- (1.1,\R)
  arc (90:270:{0.18} and {\R})
  -- (1.1+\L,-\R);
%  arc (270:90:{0.18} and {\R})
%  -- cycle;

% Front ellipse
\draw[orange!80!black,thick,opacity=1]
  (1.1,0) ellipse ({0.18} and {\R});

% Label
\node[orange!80!black,below right=2pt] at (0.7,-\R) {$\hat L_{\cals \mathbb{L} }(M_2)$};
\node[orange!80!black,below right=2pt] at (2.4,-\R) {$\gamma'$};

\end{tikzpicture}
\caption{Action of the  defect $\mathcal D_{\cals}$ on the line operator $L_{\mathbb{L}}$.}
\label{fig:line-surface-flux}
\end{figure}

Something even more interesting happens if the homology class of $\gamma$ is non-trivial. In this case the line operators \eqref{defdyonline0}  are still well defined, while the surface operators \eqref{dyongen} are not, so that the transition \eqref{actiononlines} appears obstructed. However, one can  allow $M_2$ to end on $\Sigma$ -- see Figure \ref{fig:line-surface-flux}. Indeed, if $\gamma$ is contained in the right region of a tubular neighborhood of $\Sigma$, we can assume it to be homologous to a curve $\gamma'\subset \Sigma$, and we can take $M_2$ such that $\del M_2=\gamma-\gamma'$. In this case \eqref{actiononlines} can  still be realized and would describe a transition from a genuine line operator to a  non-genuine `disorder' line operator, which is linked to the topological defect by the surface $M_2$. 

These types of effect are one of the hallmarks of non-invertible symmetries.
We will further expand upon them in Section \ref{sec:Maxaxdil}, which focuses on a simple $n=1$ GZ model and whose main  conclusions can be immediately extended to more complicated GZ models -- see also   \cite{Choi:2022jqy,Cordova:2022ieu,Roumpedakis:2022aik, Choi:2022fgx,Hasan:2024aow} for  similar results that are more directly connected with our $n=1$ examples.

%%%%%%%%%%%%%%%%%%%%%%%%%%%%%%%%%%%%%%%%%%%%%%%%%%%%%%%%%%

\section{Examples with one gauge field}
\label{sec:Maxaxdil}

So far the discussion has been rather general.  In order to better illustrate it,  in this section we apply our general results  to two simple concrete models with $n=1$, i.e.\ involving only a single U(1) gauge field $A$. The first one is the axion-Maxwell theory, whose generalized symmetry structure has already been thoroughly studied in the literature \cite{Choi:2022jqy,Cordova:2022ieu,Choi:2022fgx,Yokokura:2022alv,DelZotto:2024ngj,Pace:2023mdo}. It will provide a simple warm-up example, which will allow us to show  how our framework reproduces some known results \cite{Choi:2022jqy,Cordova:2022ieu}, without dwelling much on them. The second model is an immediate generalization of the first one, which includes a dilaton and has GZ symmetry group $\mathscr{G}={\rm SL}(2,\mathbb{R})$. The neutral sector is enlarged to include a dilaton, and can be represented by a complex field $\tau$  parametrizing the coset SL$(2, \mathbb{R})/{\rm U}(1)$. We will therefore refer to this model as the $\tau$-Maxwell model.  We will discuss the construction of  GZ topological defects $\cald_\cals$ with $\cals\in\mathscr{G}_\mathbb{Q}={\rm SL}(2,\mathbb{Q})$, and study some of their properties in details. 

\subsection{Warm-up: the axion-Maxwell model}
\label{sec:warmup}

The axion-Maxwell theory is defined by the  Lagrangian
\be\label{axL} 
\call=-\frac1{2g^2}F\wedge *F-\frac{1}{2}\,\mathsf{f}^2\d\vartheta\wedge *\d\vartheta-\frac{\vartheta}{8\pi^2}F\wedge F\,,
\ee
where $\mathsf{f}$ is some mass scale.
This is probably  the simplest realization of the GZ general setting reviewed in Section \ref{sec:GZ}, with one gauge field ($n=1$) and neutral sector provided by $\vartheta$. 
In this case the action \eqref{gentra} of the GZ  group $\mathscr{G}$ reduces to a constant  axion shift 
\be\label{axionshift} 
\vartheta\rightarrow f_{\mathsf{c}}(\vartheta)=\vartheta-2\pi\mathsf{c}\,,
\ee
with $\mathsf{c}\in\mathbb{R}$.
The dual field strength can be easily computed by applying  \eqref{G_I}:
\be\label{axionMaxG} 
G=-\frac{2\pi}{g^2}*F-\frac{\vartheta}{2\pi}F\,.
\ee
It is clear that under \eqref{axionshift}, the doublet \eqref{bF} formed by $F$ and $G$ transforms as in  \eqref{gentrb}, with 
\be\label{axionS} 
\cals_{\sfC}=\left(\begin{array}{rr} 1 &  0 \\
\mathsf{c} & 1\end{array}\right)\in{\rm Sp}(2,\mathbb{R})\,.
\ee
Furthermore, we get a simple realization of the GZ condition \eqref{tildecall}, with 
\be\label{tauMLc} 
\call_{\cals_{\mathsf{C}}}\equiv\call_{{\mathsf{C}}}=\call+\frac{\mathsf{c}}{4\pi}F\wedge F\,,
\ee 
which is just a simple subcase of \eqref{LSC}. Here and in the following, we use the lowercase letter $\sfc$, rather than $\sfC$, to emphasize that it is a number rather than a matrix. Clearly $\call$ and $\call_{\sfC}$ are physically equivalent only if $\mathsf{c}\in\mathbb{Z}$ (on spin manifolds), and only in this case does the GZ transformation correspond to a standard  zero-form symmetry of the theory. In our general notation, these integral shifts identify the integral subgroup $\mathscr{G}_{\mathbb{Z}}\simeq \mathbb{Z}$. We may quotient (or gauge) this group, i.e.\ by assuming that the axion is periodic,  $\vartheta\simeq\vartheta+2\pi$, but from a purely QFT viewpoint this is not compulsory. 

Let us now restrict to the rational GZ subgroup $\mathscr{G}_{\mathbb{Q}}$, corresponding to  shifts \eqref{axionshift} with $\mathsf{c}\in \mathbb{Q}$. We can then set
\be\label{ratiosmallc}  
\mathsf{c}=\frac{\mathsf{p}}{\mathsf{n}}\,. 
\ee 
This is the $n=1$ form of the  factorization \eqref{Cfact}, and again we use lower case letters since we are dealing with numbers rather than matrices. 
Following Section \ref{sec:GZdefects},  the GZ topological defects $\cald^{(\sfn,\sfp)}_{\mathsf{C}}$ for rational values \eqref{ratiosmallc} of $\mathsf{c}$ can be obtained by fusing the corresponding $\calu^{(\sfn,\sfp)}_\sfC$ defects and $\calw^{(\sfn,\sfp)}_\sfC$ interfaces as in Figure \ref{fig:UW=D}. Let us then revisit the construction of these objects for this specific setting. We emphasize that  the rest of this section is just a mild rephrasing, following our more general framework, of  known results \cite{Choi:2022jqy,Cordova:2022ieu}.

Let us start with the $\calu_\sfC$ defect, which can actually be constructed for any real $\mathsf{c}$. We can write  \eqref{axionS} as in \eqref{TSrel} by taking  
\be\label{axionT}
T=\left(\begin{array}{rr} 0 &  0 \\
\mathsf{c} & 0\end{array}\right)\in{\rm sp}(2,\mathbb{R})\,.
\ee 
This corresponds to the infinitesimal transformation $\delta_T\vartheta=-2\pi{\mathsf{c}}$. These equations provide simple examples of  \eqref{Tsplit} and \eqref{delTphi}, respectively. The corresponding current 
 \be \label{JTam}
\calj_T=2\pi\mathsf{c}\,\mathsf{f}^2*\d\vartheta\,,
\ee
can be computed by applying \eqref{JTcurr}. The axion equation of motion can be written as non-conservation equation 
\be\label{axdJ} 
\d \calj_T=\frac{{\mathsf{c}}}{4\pi }F\wedge F\,,
\ee
which matches the general formula \eqref{dJnon0a}. 
 With the choice \eqref{ratiosmallc}, the defect \eqref{caluS} takes the form
\begin{equation}\label{UaxioM}
\calu^{(\sfn,\sfp)}_\sfC(\Sigma)= {\rm exp}\left(\frac{ 2\pi \ii \sfp}{\sfn}\mathsf{f}^2\oint_\Sigma \, \,*\d\vartheta \right)
\end{equation}
According to \eqref{caludef}, it  acts as follows,
\be\label{axionU} 
     \begin{split}
    &\quad~~~~{\color{red}\cU^{(\sfn,\sfp)}_\sfC(
\Sigma
    )}\\
   \call(F_{\text{\tiny L}},\phi_{\text{\tiny L}})  &\qquad~{\color{red} \quad\Bigg|} \qquad~~~~ \call\left(F_{\text{\tiny R}},\vartheta_{\text{\tiny R}}-\frac{2\pi\sfp}{\sfn}\right)= \call^{(\sfn,\sfp)}_\sfC(F_{\text{\tiny R}},\vartheta_{\text{\tiny R}})= \call(F_{\text{\tiny R}},\vartheta_{\text{\tiny R}}) + \frac{\sfp}{4\pi \sfn}F\wedge F \, .\\
   &\!\!\!\!\!\!\!\!\!\!\!\!\!{\color{red}\scaleto{(F_\tL,G_\tL,\vartheta_\tL)|_\Sigma = (F_\tR,G_\tR,\vartheta_\tR-2\pi \sfp/\sfn)|_\Sigma}{2ex}}
\end{split}
\ee
More directly, the  $\cU^{(\sfn,\sfp)}_\sfC(
\Sigma
    )$ defect in \eqref{axionU} can also be obtained  by integrating the trivial operator  $-\d\calj +\frac{\sfp}{4\pi\sf n}F\wedge F=0$ on the right of $\Sigma$ \cite{Choi:2022jqy}.

Consider now the corresponding $\calw$ interface, focusing on the minimal ones, as in Section \ref{sec:GZCdefects}. They can be defined by picking a factorization \eqref{Lvariation} that is {\em coprime}, i.e.\ such that $\gcd(\sfp,\sfn)=1$. This fixes $\sfp$ and $\sfn$ up to an overall sign. According to the general discussion,  this sign does not affect the definition of minimal interface, and we can then remove this small degeneracy by imposing $\sfn>0$. Following  \cite{Hsin:2018vcg}, one can then define a minimal three-dimensional TQFT,  realizing a $\mathbb{Z}^{(1)}_\sfn$ one-form symmetry with anomaly fixed by $\sfp$ mod $\sfn$, which can be detected by coupling it to a background $\mathbb{Z}^{(1)}_\sfn$ gauge potential $\calb$. Since coprime $\sfn>0$ and $\sfp$ are  uniquely fixed by $\mathsf{c}$, the corresponding partition function $\cala^{(\sfn,\sfp)}_{\Sigma}[\calb]$ depends just on $\sfc$. Our general formula \eqref{WC2} with the choice \eqref{calicala} becomes 
\begin{equation}\label{WcaxioM}
    \calw^{(\sfn,\sfp)}_\mathsf{C}(\Sigma)= \cali^{(\sfn,\sfp)}_{\mathsf{C},\Sigma}[F_{\text{\tiny R}}]  \int Db\; \exp\left[\frac{\ii}{2\pi} \oint_{\Sigma}b\wedge (F_\tR- F_\tL)\right]\,,
\end{equation}
with\footnote{Since the anomaly of the TQFT is fixed by $\sfp$ mod $\sfn$, in order to ensure the gluing condition $G_\tL|_\Sigma=G_\tR|_\Sigma+\frac{\sfp}{\sfn}F_\tR|_\Sigma$, the partition function $\mathcal \cali^{(\sfn,\sfp)}_{\mathsf{C},\Sigma}[F_{\text{\tiny R}}]$ generically includes also a possible counterterm of the form $\exp\left(-\frac{\ii\mathsf{z}}{4\pi}\oint_\Sigma A_\tR\wedge F_\tR\right)$, with $\mathsf{z}\in\mathbb{Z}$  -- cf.\ the discussion around \eqref{Zcounterterm}.} 
\be 
\mathcal \cali^{(\sfn,\sfp)}_{\mathsf{C},\Sigma}[F_{\text{\tiny R}}]=\cala^{(\sfn,\sfp)}_\Sigma\left[ F_\tR/\sfn\right]\,.
\ee
As a  simple subcase, suppose that $\sfp=1$. Then $\mathcal A^{(\sfn,1)}_\Sigma\left[ F_\tR/\sfn\right]$  reduces to   the $U(1)_\sfn$ Chern-Simons theory coupled to $F_\tR$, so that
\begin{equation}\label{SCn=1}
    \calw^{(\sfn,1)}_\sfC(\Sigma)= \int Da \, Db\; {\rm exp}\left(\frac\ii{2\pi}\oint_{\Sigma} \left[\frac{1}{2} \,\sfn\,a\wedge \d a +  a\wedge F_\tR + b \wedge  (F_\tR - F_\tL) \right] \right)\, .
\end{equation}
More generically, $\mathcal A^{(\sfn,\sfp)}_\Sigma\left[ F_\tR/\sfn\right]$ can be defined in terms of  the four-dimensional description \eqref{minpartfunct} of the minimal three-dimensional TQFT. In the present case this reduces to
\be 
-\frac{\sfn\sfp}{4\pi}\int_Y B\wedge B+\frac\sfn{2\pi}\int_Y B\wedge \tilde F+\frac1{2\pi}\oint_\Sigma \tilde A\wedge F_\tR\,,
\ee
with $Y$ such that $\del Y=\Sigma$. As in \eqref{WSpicture}, the interface $\calw^{(\sfn,\sfp)}_{\mathsf{C}}$ separates two 
inequivalent theories, rather than implementing a transformation within a given theory:
\be\label{WSpictureaxion} 
     \begin{split}
    &\quad~~~~{\color{blue}\cW^{(\sfn,\sfp)}_\sfC(
\Sigma
    )}\\
   \call^{(\sfn,\sfp)}_\sfC(F_{\text{\tiny R}},\vartheta_{\text{\tiny R}})= \call(F_{\text{\tiny R}},\vartheta_{\text{\tiny R}}) + \frac{\sfp}{4\pi \sfn}F\wedge F  &\qquad~{\color{blue} \quad\Bigg|} \qquad~~~~ \call\left(F_{\text{\tiny R}},\vartheta_{\text{\tiny R}}\right) \, .\\
   &\!\!\!\!\!\!\!\!\!\!\!\!\!{\color{blue}\scaleto{(F_\tL,G_\tL,\vartheta_\tL)|_\Sigma = (F_\tR,G_\tR+(\sfp/\sfn) F_\tR,\vartheta_\tR)|_\Sigma}{2ex}}
\end{split}
\ee

We can now fuse \eqref{UaxioM} and \eqref{WcaxioM}  as in \eqref{Doperator}, obtaining the GZ topological defect \be
\begin{aligned}
\cald^{(\sfn,\sfp)}_\sfC(\Sigma)&=\calu^{(\sfn,\sfp)}_\sfC(\Sigma)\times \calw^{(\sfn,\sfp)}_\sfC(\Sigma)\\
&=\,\mathcal \cali^{(\sfn,\sfp)}_{\mathsf{C},\Sigma}[F_{\text{\tiny R}}]\int Db\; \exp\left[\frac{\ii}{2\pi} \oint_{\Sigma}b\wedge (F_\tR- F_\tL)\right]
\times {\rm exp}\left(\frac{ 2\pi \ii \sfp}{\sfn}\mathsf{f}^2\oint_\Sigma \, \,*\d\vartheta \right)\,,
\end{aligned}
\ee 
which acts as follows
\be\label{axionDaction} 
 \begin{split}
    &\quad~~~~~~ {\color{purple}\cD^{(\sfn,\sfp)}_{\sfC}(
\Sigma
    )}\\
   \call(F_{\text{\tiny L}},\vartheta_{\text{\tiny L}}) &\quad~~~~~~~ {\color{purple} \quad\Bigg|} \qquad~~~~~~~~ \call( F_{\text{\tiny R}},\vartheta_{\text{\tiny R}})\quad~~~~~~~~~. \\
   &\,\!\!\!\!\!\!\!\!{\color{purple}\scaleto{(F_\tL,G_\tL,\vartheta_\tL)|_\Sigma = (F_\tR,G_\tR+(\sfp/\sfn) F_\tR,\vartheta_\tR+ 2 \pi\sfp/\sfn)|_\Sigma}{2ex}}
   \end{split}
\ee
Hence $\cald^{(\sfn,\sfp)}_\sfC(\Sigma)$ realizes the complete GZ transformations defined by \eqref{axionshift} and \eqref{axionS}, with $\sfc$ as in \eqref{ratiosmallc}. As anticipated above, the GZ defect  $\cald^{(\sfn,\sfp)}_\sfC(\Sigma)$ coincides with the non-invertible topological defect  already found in \cite{Choi:2022jqy,Cordova:2022ieu}. This topological defect can also be obtained through a half-space gauging procedure \cite{Choi:2022jqy}, which can be regarded as a particular subcase of the half-space gauging discussed in Appendix \ref{app:Chalfgaug}.  

\subsection{$\tau$-Maxwell model}
\label{sec:tauMaxwell}

 Let us now  turn to the $\tau$-Maxwell model, which provides a simple but non-trivial generalization of the axion-Maxwell model discussed above. We still have just one gauge field $A$, while the neutral sector is provided  by a dynamical complexified coupling
\be\label{taudec} 
\tau=\frac{\vartheta}{2\pi}+\ii e^{-\phi}\,,
\ee
parametrizing the complex upper-half plane $\mathbb{H}=\{\Im\tau >0\}$.  The Lagrangian is\footnote{The Lagrangian \eqref{adMlagr} is a particular case of \eqref{gen2derL}, with $n=1$ and $\caln=-\tau$.}     
\be\label{adMlagr} 
\begin{aligned}
\call&=-\frac1{4\pi}\Im\tau\,F\wedge *F -\frac{1}{4\pi}\Re\tau\, F\wedge F-\mathsf{f}^2\,\frac{\d\tau\wedge *\d\bar\tau}{(\Im\tau)^2
}\,,
\end{aligned}
\ee
where  $\mathsf{f}$ is a mass scale. Furthermore,  for the moment we are not considering any discrete identification in $\mathbb{H}$.

The kinetic term of $\tau$ corresponds to a hyperbolic metric on $\mathbb{H}$, which can be also written as the metric on the coset ${\rm SL}(2,
\mathbb{R})/{\rm U}(1)$. Its isometry group ${\rm SL}(2,
\mathbb{R})\simeq {\rm Sp}(2,\mathbb{R})$ can be identified with the GZ symmetry group $\mathscr{G}$ of the model, which therefore coincides with the entire symplectic group:
\be \label{Stransf}
\cals=\left(\begin{array}{rr} \sfa & \sfb \\ \sfc & \sfd\end{array}\right)\in \mathscr{G}={\rm SL}(2,\mathbb{R})\quad \Leftrightarrow\quad \sfa\sfd-\sfb\sfc=1\,.
\ee 
This representation of $\cals$ is the analogue of \eqref{genS}, and  as in Section \ref{sec:warmup} we use lowercase letters since they are numbers rather than matrices.  
The classical GZ transformation associated with any $\cals$ of the form \eqref{Stransf} acts 
on $\tau$ and on the doublet  
\be\label{tauMdoublet} 
\mathbbm{F}=\left(\begin{array}{c} F \\
 G\end{array}\right)
\ee
 as in \eqref{gentr}, with: 
\be\label{adtautr} 
f_\cals(\tau)=\frac{\sfd\tau-\sfc}{\sfa-\sfb\tau}\quad,\quad 
\cals\mathbbm{F}= \left(\begin{array}{rr} \sfa & \sfb \\ \sfc & \sfd\end{array}\right)\left(\begin{array}{c} F \\
 G\end{array}\right)\,.
\ee
The explicit form of dual field strength $G$ can be obtained from \eqref{G_I}:
\be\label{adMaxG} 
G=-\Im\tau\, *F-\Re\tau\, F\,.
\ee

The generic infinitesimal GZ transformations \eqref{delTphi} and \eqref{delTbF} now take the form  
\be \label{Tgeneral}
T=\left(\begin{array}{rr} \sfu & \sfv \\ \sfw & -\sfu\end{array}\right)\in{\rm sp}(2,\mathbb{R})
\quad,\quad \delta_T\tau=\xi_T(\tau)=-
\sfw-2\sfu\tau+\sfv\tau^2\,,
\ee
and the associated GZ current \eqref{JTcurr}  becomes
\be\label{adJ_T} 
\calj_T=-\sff^2\,\frac{\xi_T(\tau)*\d\bar\tau+\bar\xi_T(\bar\tau)*\d\tau}{(\Im\tau)^2}\,,
\ee
while its non-conservation equation \eqref{dJnon0a} reads
\be\label{adMaxJeq} 
\begin{aligned}
\d\calj_T&=\frac{1}{4\pi}\left(\sfw F\wedge F-\sfv G\wedge G-2\sfu F\wedge G\right)\,.
\end{aligned}
\ee

Since the GZ assumptions are satisfied, according to our general discussion we should be able to construct a topological defect $\cald_\cals$ implementing any GZ transformation $\cals\in \mathscr{G}_{\mathbb Q}= {\rm Sp}(2,\mathbb{Q})={\rm SL}(2,\mathbb{Q})$. The general  claim exploits the fact that any element of ${\rm Sp}(2n,\mathbb{Q})$ can be decomposed as the product of elements of the form \eqref{Sfactor}. Furthermore, we can use the integral factorizations \eqref{AMEdec} and \eqref{Cfact}. In the present $n=1$ setting, the generators \eqref{Sfactor} of ${\rm Sp}(2,\mathbb{Q})={\rm SL}(2,\mathbb{Q})$ can then be written as\footnote{In fact, any $\cals_\sfA$ may be decomposed into a product of elements of the type $\cals_\sfC$ and $\cals_\Omega$, and therefore we may focus on latter. It is nevertheless interesting to  discuss a direct construction of  $\sfA$ defects.}
\be\label{Sfactor2} 
\cals_\sfA=\left(\begin{array}{cc} \frac{\sfm}{\sfe} &  0\\
0 & \frac\sfe\sfm \end{array}\right)\quad,\quad \cals_\sfC=\left(\begin{array}{cc} 1 & 0 \\
\frac\sfp\sfn & 1\end{array}\right)\quad,\quad \cals_\Omega=\left(\begin{array}{rr} 0 & 1 \\
-1 & 0\end{array}\right)\,,
\ee
with no loss of generality, we can assume $\sfe,\sfn,\sfm\in\mathbb{Z}_{>0}$ and $\sfp\in\mathbb{Z}$.\footnote{We can assume $\sfm/\sfe>0$ since a negative value can be obtained by taking the product of  $\cals_{\sfA}$   and $\cals^2_\Omega=-\mathbbm{1}$.} 
As discussed in Sections \ref{sec:GZAdefects} and \ref{sec:GZCdefects},   minimal defects correspond to coprime factorizations, which in the present  model  means    ${\rm gcd}(\sfe ,\sfm)=1$ and  ${\rm gcd}(\sfn,\sfp)=1$. 

As already emphasized, in the $\tau$-Maxwell model the rational GZ group $\mathscr{G}_{\mathbb{Q}}$ coincides with the entire rational symplectic group ${\rm Sp}(2,\mathbb{Q})$. Therefore, there should exist a GZ topological defect $\cald_\cals$ for any of the three transformations  \eqref{Sfactor2}. We can then separately discuss each of them.

\subsubsection{$\sfA$ defect}

We first consider $\sfA$ GZ topological defects, associated with $\cals_\sfA$ in \eqref{Sfactor2}. 
 Following the general procedure of Section \ref{sec:GZdefects}, an $\sfA$ topological defect can be written in the form \eqref{Doperator}, that is
\be\label{taumaxDAdec} \cald^{(\sfe,\sfm)}_\sfA(\Sigma)=\calu^{(\sfe,\sfm)}_\sfA(\Sigma)\times \calw^{(\sfe,\sfm)}_\sfA(\Sigma)\,.
\ee 
The defect $\calu^{(\sfe,\sfm)}_\sfA(\Sigma)$ implements the transformation $\tau\rightarrow \frac{\sfe^2}{\sfm^2}\tau$ and  can be written in the form \eqref{caludef}. In order to do that, we need to find $T$ such that $\exp T=\cals_\sfA$. It is easy to see that we must take $T$ as in \eqref{Tgeneral}, with $\sfu=\log(\sfm/\sfe)$ and $\sfv=\sfw=0$. By using the corresponding current \eqref{adJ_T} in  \eqref{caluS} we get 
\begin{equation}\label{UAaxioM}
\calu^{(\sfe,\sfm)}_\sfA(\Sigma)= {\rm exp}\left[4\ii\mathsf{f}^2 \log\left(\frac{\sfm}{\sfe}\right)\oint_\Sigma \, \frac{\Re\left(\bar\tau*\d\tau\right)}{(\Im\tau)^2}\right]\,.
\end{equation}

The interface $\calw^{(\sfe,\sfm)}_\sfA$  can be obtained as $n=1$ subcase of the more general interface \eqref{WA}. Namely, 
\be\label{WAem} 
\calw^{(\sfe,\sfm)}_\sfA(\Sigma)=\int D b\,\exp\left[\frac\ii{2\pi}\oint_\Sigma b\wedge \left(\sfe F_\tL-\sfm F_\tR\right)\right]\,,
\ee
where $b$ is a world-volume U(1) gauge field defined on $\Sigma$. 
This interface was first introduced  in \cite{Cordova:2023ent} for studying non-invertible symmetries of pure electromagnetism with rational fine-structure constant. According to the general property \eqref{WSpicture}, it separates a theory defined by
\be 
\call^{(\sfe,\sfm)}_\sfA(F_\tL,\tau_\tL)\equiv \call\left(\frac{\sfe}{\sfm}F_\tL,\tau_\tL\right)\,,
\ee
and $\call(F_\tR,\tau_\tR)$, with gluing conditions $\sfe F_\tL|_\Sigma=\sfm F_\tR|_\Sigma$ and $\sfm G_\tL|_\Sigma=\sfe G_\tR|_\Sigma$. These can also been obtained from \eqref{WAem}, by integrating out $b$ and from the boundary contributions to the saddle point conditions for $A_\tL$ and $A_\tR$.

    In the present model, the general GZ condition \eqref{tildecall} boils down to  $\call(F,\frac{\sfe^2}{\sfm^2}\tau)=\call^{(\sfe,\sfm)}_\sfA(F,\tau)$, which can be easily verified. The action of $\cald^{(\sfe,\sfm)}_\sfA$ defined as in \eqref{taumaxDAdec} can then be obtained by concatenating $\calu^{(\sfe,\sfm)}_\sfA$ and $\calw^{(\sfe,\sfm)}_\sfA$ as in \eqref{UWSpicture}: 
\be\label{UWSpicturetauA} 
 \begin{split}
    &\!{\color{red}\calu^{(\sfe,\sfm)}_\sfA(\Sigma)}\\
   \call(F_{\text{\tiny L}},\tau_{\text{\tiny L}}) \quad~~~~ &{\color{red} \quad\Bigg|} \quad ~~~~~~~\call\left(F_{\text{\tiny I}},\frac{\sfe^2}{\sfm^2}\tau_{\text{\tiny I}}\right) = \call^{(\sfe,\sfm)}_{\sfA}(F_{\text{\tiny I}},\tau_{\text{\tiny I}})\quad~\\
   &\!\!\!\!\!\!\!\!\!\!\!\!\!\!\!\!\!\!\!\!\!\!{\color{red}\scriptstyle{(F_\tL,G_\tL,\tau_\tL)|_\Sigma=(F_{\text{\tiny I}},G_{\text{\tiny I}},(\sfe/\sfm)^2\tau_{\text{\tiny I}})|_\Sigma}}
   \end{split}
     \quad \begin{split}
    &\;\!\!\!\!\!{\color{blue}\calw^{(\sfe,\sfm)}_\sfA(
\Sigma'
    )}\\
    &{\color{blue} \quad\Bigg|} \quad~~~~~~ \call( F_{\text{\tiny R}},\tau_{\text{\tiny R}})\, ,
    \\
&\!\!\!\!\!\!\!\!\!\!\!\!\!\!\!\!\!\!\!\!\!{\color{blue}\scriptstyle{(F_{\text{\tiny I}},G_{\text{\tiny I}},\tau_{\text{\tiny I}})|_{\Sigma'}=((\sfm/\sfe) F_{\text{\tiny R}},(\sfe/\sfm)F_\tR,\tau_{\text{\tiny R}})|_{\Sigma'}}} 
\end{split}
\ee
By fusing  $\calu^{(\sfe,\sfm)}_\sfA$ and $\calw^{(\sfe,\sfm)}_\sfA$ we conclude that the GZ topological defect 
\be\label{tauMDA}
\cD^{(\sfe,\sfm)}_{\sfA}(
\Sigma
    )={\rm exp}\left[4\ii\mathsf{f}^2 \log\left(\frac{\sfm}{\sfe}\right)\oint_\Sigma \, \frac{\Re\left(\bar\tau*\d\tau\right)}{(\Im\tau)^2}\right]\times \int D b\,\exp\left[\frac\ii{2\pi}\oint_\Sigma b\wedge \left(\sfe F_\tL-\sfm F_\tR\right)\right]
\ee
induces the complete GZ  transformation \eqref{adtautr} associated with $\cals_\sfA$ in \eqref{Sfactor2}:
\be\label{tauDAaction} 
 \begin{split}
    &\quad~~~~~~ {\color{purple}\cD^{(\sfe,\sfm)}_{\sfA}(
\Sigma
    )}\\
   \call(F_{\text{\tiny L}},\tau_{\text{\tiny L}}) &\quad~~~~~~~ {\color{purple} \quad\Bigg|} \qquad~~~~~~~~ \call( F_{\text{\tiny R}},\tau_{\text{\tiny R}})\quad~~~~~~~~~. \\
   &\,\!\!\!{\color{purple}\scriptstyle{(F_\tL,G_\tL,\tau_\tL)|_\Sigma = \big((\sfm/\sfe)F_\tR,(\sfe/\sfm)G_\tR,(\sfe/\sfm)^2\tau_\tR\big)|_\Sigma}}
   \end{split}
\ee
Note that for $\sfe\neq 1$ or $\sfm\neq 1$ this topological  defect is non-invertible. We will come back to this point in Section \ref{sec:tauMfusions}. We also observe that  $\cald^{(\sfe,1)}_\sfA\times \cald^{(1,\sfm)}_\sfA=\cald^{(\sfe,\sfm)}_\sfA$ -- cf.\ the more general \eqref{WAfact} -- and that   $\cald^{(\sfe,1)}_\sfA$ and  $\cald^{(1,\sfm)}_\sfA$ can be obtained from the half-space gauging procedures described in Appendix \ref{app:Mhalfgaug} and \ref{app:Ehalfgaug}, respectively. 

\subsubsection{$\sfC$ defect}

We now turn to the $\sf C$ defects which, recalling again \eqref{Doperator}, can be written as   
\be\label{taumaxDdec} \cald^{(\sfn,\sfp)}_\sfC(\Sigma)=\calu^{(\sfn,\sfp)}_\sfC(\Sigma)\times \calw^{(\sfn,\sfp)}_\sfC(\Sigma)\,.
\ee They are almost  identical to the $\sfC$ defects constructed for the axion-Maxwell model in Section \ref{sec:warmup}. The only minor difference is the form of the current $\calj_T$ that one has to use  in \eqref{caluS}. We can still take $T$ as in \eqref{axionT}, with $\sfc=\sfp/\sfn$. This means that  in \eqref{Tgeneral}  we must take $\sfw=\sfp/\sfn$ and $\sfu=\sfv=0$. So, from \eqref{adJ_T} it follows that we now have 
\begin{equation}\label{UaxioM2}
\calu^{(\sfn,\sfp)}_\sfC(\Sigma)= {\rm exp}\left[\frac{2\ii \sfp}{\sfn}\mathsf{f}^2\oint_\Sigma \, \frac{*\d\Re\tau}{(\Im\tau)^2}\right]\,.
\end{equation}

The interface $\calw^{(\sfn,\sfp)}_\sfC$ can be chosen as in \eqref{WcaxioM}. Hence, fusing it with  \eqref{UaxioM2} as in \eqref{taumaxDdec}, we get the corresponding GZ topological defect
\be\label{tauMDC}
\begin{aligned}
\cald^{(\sfn,\sfp)}_\sfC(\Sigma)=&\,\mathcal \cali^{(\sfn,\sfp)}_{\mathsf{C},\Sigma}[F_{\text{\tiny R}}] \int Db\; \exp\left[\frac{\ii}{2\pi} \oint_{\Sigma}b\wedge (F_\tR- F_\tL)\right]\\
&
\times {\rm exp}\left[\frac{2\ii \sfp}{\sfn}\mathsf{f}^2\oint_\Sigma \, \frac{*\d\Re\tau}{(\Im\tau)^2}\right]\,.
\end{aligned}
\ee

By concatenating $\calu^{(\sfn,\sfp)}_\sfC$ and $\calw^{(\sfn,\sfp)}_\sfC$, it is easy to see that $\cald^{(\sfn,\sfp)}_\sfC$ acts as follows on the bulk theory
\be\label{tauDactionC} 
 \begin{split}
    &\quad~~~~~~ {\color{purple}\cD^{(\sfn,\sfp)}_{\sfC}(
\Sigma
    )}\\
   \call(F_{\text{\tiny L}},\tau_{\text{\tiny L}}) &\quad~~~~~~~ {\color{purple} \quad\Bigg|} \qquad~~~~~~~~ \call( F_{\text{\tiny R}},\tau_{\text{\tiny R}})\,, \\
   &\,\!\!\!\!\!\!\!{\color{purple}\scaleto{(F_\tL,G_\tL,\tau_\tL)|_\Sigma = (F_\tR,G_\tR+(\sfp/\sfn) F_\tR,\tau_\tR+ \sfp/\sfn)|_\Sigma}{2ex}}
   \end{split}
\ee
and hence realizes the complete  GZ transformation \eqref{adtautr} associated to $\cals_\sfC$ appearing in \eqref{Sfactor2}. 

As in axion-Maxwell,  the interface $\calw_\sfC^{(\sfn,\sfp)}$ -- and hence the GZ topological defect $\cald_\sfC^{(\sfn,\sfp)}$ -- can be obtained through the half-space gauging procedure discussed in \cite{Choi:2022jqy}, which  corresponds to  the $n=1$ subcase of the more general half-space gauging discussed  in Appendix \ref{app:Chalfgaug}. Furthermore, if $\sfn\neq 1$,  $\cald_\sfC^{(\sfn,\sfp)}$ is non-invertible. This will be made more explicit in Section \ref{sec:tauMfusions}.

\subsubsection{$\Omega$ defect}

We finally consider the $\Omega$ defect $\cald_\Omega$. Using again the decomposition \eqref{Doperator}, this can be written in the form 
\be\label{caldOmega} 
\cald_\Omega(\Sigma)= \calu_\Omega(\Sigma)\times \calw_\Omega(\Sigma)\,.
\ee
 By picking $T$ in \eqref{Tgeneral} such that $\Omega=\exp T$, and using the corresponding current \eqref{adJ_T}, we can construct $\calu_\Omega$ as in   \eqref{caluS}. More explicitly, in \eqref{Tgeneral} we must take $\sfu=0$ and $\sfv=-\sfw=\frac\pi2$. Using these values in  \eqref{adJ_T} we  get 
\be 
\calu_\Omega(\Sigma)=\exp\left[-\ii\pi\sff^2\,\Re\oint_\Sigma\frac{\left(1+\tau^2\right)*\d\bar\tau}{(\Im\tau)^2}\right]\,.
\ee
On the other hand, the $\calw_\Omega$ interface  \cite{Ganor:1996pe,Gaiotto:2008ak,Kapustin:2009av} can be obtained by taking the case $n=1$ of \eqref{WSZa}, \be 
\calw_\Omega(\Sigma)=\exp\left(\frac\ii{2\pi}\oint_\Sigma A_\tL\wedge F_\tR\right)\,, 
\ee
which separates the S-dual Lagrangian $\call_\Omega$ and $\call$ -- see Appendix \ref{app:Omegaduality}. The action of $\cald_\Omega$ can then be obtained by concatenating $\calu_\Omega$ and $\cald_\Omega$ as in \eqref{UWSpicture}: 
\be\label{UWSpicturetauOmega} 
 \begin{split}
    &\!{\color{red}\cU_{\Omega}(
\Sigma
    )}\\
   \call(F_{\text{\tiny L}},\tau_{\text{\tiny L}}) \quad~~~~~~ &{\color{red} \quad\Bigg|} \quad ~~~~~~~~~\call(F_{\text{\tiny I}},-1/\tau_{\text{\tiny I}}) = \call_{\Omega}(F_{\text{\tiny I}},\phi_{\text{\tiny I}})\quad~~~\\
   &\!\!\!\!\!\!\!\!\!\!\!\!\!\!\!\!\!\!\!\!\!\!\!\!\!\!{\color{red}\scaleto{(F_\tL,G_\tL,\tau_\tL)|_\Sigma=(F_{\text{\tiny I}},G_{\text{\tiny I}},-1/\tau_{\text{\tiny I}})|_\Sigma}{2ex}}
   \end{split}
     \quad \begin{split}
    &\;{\color{blue}\cW_{\Omega}(
\Sigma'
    )}\\
    &{\color{blue} \quad\Bigg|} \quad~~~~~~~~ \call( F_{\text{\tiny R}},\phi_{\text{\tiny R}})\, ,
    \\
&\!\!\!\!\!\!\!\!\!\!\!\!\!\!\!\!\!\!\!\!\!\!\!\!\!{\color{blue}\scaleto{(F_{\text{\tiny I}},G_{\text{\tiny I}},\tau_{\text{\tiny I}})|_{\Sigma'}=(G_{\text{\tiny R}},-F_\tR,\tau_{\text{\tiny R}})|_{\Sigma'}}{2ex}} 
\end{split}
\ee

The topological defect \eqref{caldOmega} is then given 
\be\label{tauMDOmega} 
\cald_\Omega(\Sigma)=\exp\left(-\ii\pi\sff^2\,\Re\oint_\Sigma\frac{\left(1+\tau^2\right)*\d\bar\tau}{(\Im\tau)^2}+\frac\ii{2\pi}\oint_\Sigma A_\tL\wedge F_\tR\right)\,, 
\ee
which acts on the bulk theory by implementing the complete GZ transformation \eqref{adtautr} defined by $\cals_\Omega$:
\be\label{tauDactionOmega} 
 \begin{split}
    &\quad~~~~~~ {\color{purple}\cD_{\Omega}(
\Sigma
    )}\\
   \call(F_{\text{\tiny L}},\tau_{\text{\tiny L}}) &\quad~~~~~~~ {\color{purple} \quad\Bigg|} \qquad~~~~~~~~ \call( F_{\text{\tiny R}},\tau_{\text{\tiny R}})\,. \\
   &\,\!\!\!\!\!\!\!\!{\color{purple}\scaleto{(F_\tL,G_\tL,\tau_\tL)|_\Sigma = (G_\tR,-F_\tR,-1/\tau_\tR)|_\Sigma}{2ex}}
   \end{split}
\ee
We emphasize that, differently from  $\sfA$ and $\sfC$ defects discussed  above,  $\cD_{\Omega}(
\Sigma
    )$ is invertible.

    So far we have constructed the symmetry defects without explicitly discussing  their 't Hooft anomalies. For instance there should be a non-trivial 't Hooft anomaly for some of the elements of the ${\rm SL}(2,\mathbb Z)$ group of invertible  symmetries, along the lines of  \cite{Hsieh:2019iba}.  The anomaly shows up in the presence of a non-trivial background for the ${\rm SL}(2,\mathbb Z)$ zero-form symmetry. 
    In addition, as mentioned in Section \ref{sec:GZ}, in curved spacetimes one should also take into account  the mixed anomaly between the $\Omega$ transformation and gravity \cite{Witten:1995gf}. For our present purposes, it is not essential to explicitly consider this effect and therefore, for simplicity, we will restrict to (e.g. flat) spacetimes in which this anomaly is absent. Of course, in order to promote the  gravitational background to a dynamical one, one should  explictly take this effect into account.

\subsubsection{Fusion of GZ topological  defects}
\label{sec:tauMfusions}

Let us now discuss some of the basic fusion rules of the GZ topological defects $\cald^{(\sfe,\sfm)}_\sfA$, $\cald^{(\sfn,\sfp)}_\sfC$ and $\cald_\Omega$ -- see \eqref{tauMDA}, \eqref{tauMDC} and \eqref{tauMDOmega}, respectively -- leaving a more systematic investigation for future work. We observe that any of the fusion below relies on the general commutativity  \eqref{commutativity}.

 As already mentioned, $\cald_\mathsf{\Omega}$ is an invertible defect, in the sense that  its fusion with $\cald_{\Omega^{-1}}=\overline{\cald}_\Omega$ gives the identity. We also observe that, since $\Omega^{-1}=-\Omega$, the fusion rule of $\cald_\mathsf{\Omega}$ with itself produces the charge conjugation defect $\cald_\mathsf{-\mathbbm 1}$,
\begin{equation}
   \cald_\mathsf{\Omega} \times \cald_\mathsf{\Omega} = \cald_\mathsf{-\mathbbm 1}\,.
\end{equation}
Another simple example is provided by:
\begin{equation}
\cald^{(\sfe,\sfm)}_\mathsf{A}\times \cald^{(\sfe',\sfm')}_\mathsf{A} = \cald^{(\sfe\sfe',\sfm\sfm')}_{\mathsf{A}}\quad~~~~~~\text{if}\quad~  \gcd(\sfm,\sfe')=\gcd(\sfm',\sfe)=1\,.
\end{equation}
The fusion of more general $\cald_\sfA$ defects  will be discussed below. A less trivial example  is provided by  the self-fusion of a defect of the type \eqref{tauMDC} with $\sfp=1$ and odd $\sfn$, which can be extracted from  \cite{Choi:2022fgx, Choi:2022zal}:
\begin{equation}
    \cald^{(\sfn,1)}_\mathsf{C} (\Sigma)\times \cald^{(\sfn,1)}_{\mathsf{C}}(\Sigma) = \mathcal{A}^{(\sfn, 2)}_\Sigma[F_\tR/\sfn] \,\cald^{(\sfn,2)}_\mathsf{C}(\Sigma)\,.
\end{equation}

More general fusions can be used to construct  topological defects corresponding to more general GZ transformations. As simple examples, consider the following three types of $\cals\in{\rm Sp}(2,\mathbb Q)$ matrices:
\be\label{D123}
\cals_1=\left(\begin{array}{cc} \frac{\sfm}{\sfe} &  0\\
\frac{\sfp}{\sfn} & \frac\sfe\sfm \end{array}\right)\quad,\quad \cals_2=\left(\begin{array}{cc} 0 &  \frac\sfm\sfe\\
-\frac\sfe\sfm & 0 \end{array}\right)\quad,\quad \cals_3=\left(\begin{array}{cc} 0 &  1\\
-1 & \frac\sfp\sfn \end{array}\right)\,,
\ee
with $\sfe,\sfm,\sfn\in\mathbb{Z}_{>0}$ and $\sfp\in\mathbb{Z}$. Corresponding GZ topological defects $\cald_1$, $\cald_2$ and $\cald_3$, implementing the respective GZ transformations \eqref{adtautr}, can be constructed as follows. (As already emphasized, a GZ transformation does not correspond to a unique GZ topological defect.)

Consider first $\cals_1$. It can be factorized as 
\be 
\cals_1=\left(\begin{array}{cc} \frac{\sfm}{\sfe} &  0\\
0 & \frac\sfe\sfm \end{array}\right)\left(\begin{array}{cc} 1 &  0\\
\frac{\sfm\sfp}{\sfe\sfn} & 1\end{array}\right)\,.
\ee
Hence, an example of associated GZ defect is obtained by fusing the following $\sfA$ and a $\sfC$ defects: 
\be\label{cald_1dec} 
\cald_1\equiv\cald^{(\sfe,\sfm)}_\sfA\times \cald^{(\sfe\sfn,\sfm\sfp)}_\sfC=\left(\calu^{(\sfe,\sfm)}_\sfA\times \calu^{(\sfe\sfn,\sfm\sfp)}_\sfC\right)\times \left(\calw^{(\sfe,\sfm)}_\sfA\times \calw^{(\tilde\sfn,\tilde\sfp)}_\sfC\right)\,,
\ee
with $\tilde\sfp\in\mathbb{Z}$ and $\tilde\sfn\in\mathbb{Z}_{>0}$ such that $\tilde\sfp/\tilde\sfn=\sfm\sfp/(\sfe\sfn)$ and $\gcd(\tilde\sfn,\tilde\sfp)=1$. 
In particular, the contribution of the resulting $\calw$ interface can be written more explicitly a
\be
\begin{aligned}
\calw^{(\sfe,\sfm)}_\sfA\times \calw^{(\tilde\sfn,\tilde\sfp)}_\sfC&= \cali^{(\tilde\sfn,\tilde\sfp)}_{\mathsf{C},\Sigma}[F_{\text{\tiny R}}] \int Db\; {\rm exp}\left[\frac{\ii}{2 \pi} \oint_{\Sigma} b \wedge \left(\sfe F_\tL - \sfm F_\tR\right) \right]\,.
    \end{aligned}
\end{equation}
We can then set  $\cali^{(\tilde\sfn,\tilde\sfp)}_{\mathsf{C},\Sigma}[F_{\text{\tiny R}}]=\cala_\Sigma^{(\tilde\sfn,\tilde\sfp)}[F_{\text{\tiny R}}/\tilde\sfn]$,  or use a non-minimal TQFT,  as the one defined in \eqref{Ictheory} for a general GZ model.

Next, consider the symplectic matrix $\cals_2$, which can be written as 
\be 
\cals_2=\left(\begin{array}{cc} \frac{\sfm}{\sfe} &  0\\
0 & \frac\sfe\sfm \end{array}\right)\left(\begin{array}{cc} 0 &  1\\
-1 & 0\end{array}\right)\,.
\ee
Then, a corresponding GZ defect $\cald_2$ is given by 
\be 
\cald_2\equiv\cald^{(\sfe,\sfm)}_\sfA\times \cald_\Omega=\left(\calu^{(\sfe,\sfm)}_\sfA\times \calu_\Omega\right)\times \left(\calw^{(\sfe,\sfm)}_\sfA\times \calw_\Omega\right)\,,
\ee
with
\begin{equation}
    \calw^{(\sfe,\sfm)}_\sfA\times \calw_\Omega= \int Da\, Db\,\; {\rm exp}\left[\frac{\ii}{2 \pi} \oint_{\Sigma} b\wedge(\sfe F_\tL -\sfm \d a) + \frac{\ii}{2 \pi} \oint_{\Sigma}a \wedge F_\tR \right]\,.
\end{equation}

Finally, $\cals_3$ can be factorized as
\be 
\cals_3=\left(\begin{array}{cc} 1 &  0\\
\frac\sfp\sfn & 1 \end{array}\right)\left(\begin{array}{cc} 0 &  1\\
-1 & 0\end{array}\right)\,.
\ee
and then a corresponding GZ defect $\cald_3$ can be defined as 
\be 
\cald_3\equiv\cald^{(\sfn,\sfp)}_\sfC\times \cald_\Omega=\left(\calu^{(\sfn,\sfp)}_\sfC\times \calu_\Omega\right)\times \left(\calw^{(\sfn,\sfp)}_\sfC\times \calw_\Omega\right)\,.
\ee
Assuming $\gcd(\sfp,\sfn)=1$, the $\calw$ component can be written as 
\begin{equation}
\begin{aligned}
\calw^{(\sfn,\sfp)}_\sfC\times \calw_\Omega&=\mathcal{A}_\Sigma^{(\sfn,\sfp)}[F_\tL/\sfn] \exp \left(\frac{\ii}{2 \pi}\oint_{\Sigma} A_\tL \wedge F_\tR\,. \right)
    \end{aligned}
\end{equation}

Other GZ topological defects, corresponding to more general $\cals\in{\rm Sp}(2,\mathbb{Q})$, can be obtained in a similar way. Namely, one can decompose $\cals$ into a product of elements of the form \eqref{Sfactor2} and compute the ordered fusion of the associated  GZ defects. Of course, the factorization  is not unique and the resulting GZ defect $\cald_\cals$ can a priori depend on the chosen factorization, as well as on the TQFT used to define the possible $\sfC$ defects involved in the fusion. Hence, we again see how a given GZ tranformation does not correspond to a unique GZ defect. It would be extremely interesting to more systematically classify the possible GZ defects and, in particular, identify the possible ``minimal" ones.

\subsubsection{Condensation Defects}

As already seen in Section \ref{sec:GZdefects2}, if $\cald_\cals$ is a non-invertible GZ defect, a fusion of the form  $\cald_\cals\times \overline\cald_\cals$ produces a non-trivial condensation defect.   Let us discuss a few simple examples of such condensation defects and how they can also appear in more general fusions.  

Take first the operator \eqref{taumaxDdec} with $\sfp=1$ and $\sfn\geq 2$, and $\calw^{(\sfn,1)}_\sfC$ as in \eqref{SCn=1}. The corresponding condensation defect is
\begin{equation}
  \calc^{(\sfn,1)}_{\mathsf{C}} \equiv \cald^{(\sfn,1)}_{\mathsf{C}}  \times \overline\cald^{(\sfn,1)}_{\mathsf{C}} =\calw^{(\sfn,1)}_{\mathsf{C}}  \times \overline\calw^{(\sfn,1)}_{\mathsf{C}} \,.
\end{equation}
This fusion can be straightforwardly computed, as done in Section \ref{sec:Ccondefects} for the more general $\sfC$ interface, and gives
the $n=1$ subcase of the condensation defect \eqref{CCcondgen}-\eqref{CCcondgen2}:
\begin{equation}\label{CCtaucond}
\begin{aligned}
     \mathcal C^{(\sfn,1)}_\mathsf{C}  (\Sigma) &=\int Da \,D\tilde{a}   \; {\rm exp}\left[\frac{\ii\sfn}{4\pi}\oint_{\Sigma}   a\wedge \d a +\frac{\ii\sfn}{2\pi}\oint_{\Sigma}   a\wedge \d \tilde a +  \frac{\ii}{2\pi}\oint_{\Sigma}   a\wedge F\right]\\ 
     &= \frac{1}{|H^0(\Sigma,\mathbb Z_\sfn)|}\sum_{\eta\in H^1(\Sigma,\mathbb{Z}_\sfn)}\exp\left(\ii\oint_\Sigma \eta\cup F\right)\exp\left[\pi\ii\sfn\oint_\Sigma \eta\cup \beta(\eta)\right]\,.
     \end{aligned}
\end{equation}
This coincides with the condensation defect found in \cite{Choi:2022jqy} -- see also  \cite{Choi:2022fgx, Choi:2022rfe, Choi:2022zal, Roumpedakis:2022aik} -- 
and can be interpreted as coming from a one-gauging of the finite magnetic one-form symmetry group $\mathbb Z^{(1)}_\sfn \subset U(1)^{(1)}_{\rm m}$. Notice that the discrete torsion phase in \eqref{CCtaucond} gives a non-trivial effect only if $\sfn$ is even \cite{Choi:2022jqy}.

Let us now consider the condensation defect
associated with the GZ defect \eqref{tauMDA}:
\be\label{tauAconddef}
\begin{aligned} 
\calc^{(\sfe,\sfm)}_\sfA&\equiv \cald^{(\sfe,\sfm)}_{\mathsf{A}}  \times \overline\cald^{(\sfe,\sfm)}_{\mathsf{A}} =\calw^{(\sfe,\sfm)}_{\mathsf{A}}  \times \calw^{(\sfm,\sfe)}_{\mathsf{A}^{-1}}
\end{aligned}
\ee
We can then proceed as explained for the general case above \eqref{WAWA-1}, namely, concatenating  $\calw^{(\sfe,\sfm)}_{\mathsf{C}}$ and $\calw^{(\sfm,\sfe)}_{\mathsf{C}}$, rewriting the enclosed bulk gauge field as a world-volume gauge field $a$, and making a simple field redefinition. As a result, we can write \eqref{tauAconddef} as 
\be\label{tauMCA}
\calc^{(\sfe,\sfm)}_\sfA=  \int Db \exp\left[\frac{\ii\sfe}{2\pi}\oint_\Sigma b\wedge (F_\tL-F_\tR)\right]\int Da D\tilde b \exp\left[\frac{\ii}{2\pi}\oint_\Sigma\left( \sfm\,\tilde b\wedge \d a -\sfe\, \tilde b\wedge F_\tR\right)\right] \,.
\ee
Assuming that $\gcd(\sfe,\sfm)=1$ and adapting the general results   of Section \ref{sec:GZAdefects}, one concludes that $\calc^{(\sfe,\sfm)}_\sfA$ can be rewritten as 
\be\label{Cem} 
\calc^{(\sfe,\sfm)}_\sfA=\calc^{(\sfe,1)}_\sfA\times\calc^{(1,\sfm)}_\sfA\,, 
\ee
with 
\be
\begin{aligned} \label{Condpqstand}
\calc^{(\sfe,1)}_\sfA&=\frac{1}{|H^0(\Sigma,\mathbb Z_\sfe)|} \sum_{C \in H_2(\Sigma,\mathbb Z_\sfe)} {\rm exp}\left (\frac{\ii}{\sfe} \oint_C G_\tR  \right)\,,\\
\calc^{(1,\sfm)}_\sfA&=\frac{1}{|H^0(\Sigma,\mathbb Z_\sfm)|} \sum_{\tilde C \in H_2(\Sigma,\mathbb Z_\sfm)} {\rm exp}\left ( \frac{\ii}{\sfm} \oint_{\tilde C} F_\tR \right)\,.
\end{aligned}
\ee
$\calc^{(\sfe,\sfm)}_\sfA$  coincides with the condensation defect  found in \cite{Cordova:2023ent} -- see also \cite{Choi:2022zal,Hasan:2024aow}.  

Finally, consider the more general fusion 
\begin{equation}
\label{AAfusion}
\begin{aligned}
\cald^{(\sfe,\sfm)}_\mathsf{A} \times \cald^{(\sfe',\sfm')}_{\mathsf{A}}&= \calu^{(\sfe \sfe',\sfm\sfm')}_\sfA\times\int Db \, D\tilde b\, Da\;  {\rm exp}\Big[ \frac\ii{2\pi} \oint_{\Sigma} b \wedge (\sfe F_\tL - \sfm \d a)\\
&\quad~~~~~~~~~~~~~~~~~~~~~~~~~~~~ + \frac\ii{2\pi} \oint_{\Sigma}\tilde b \wedge (\sfe'\d a -\sfm' F_\tR ) \Big] \,.
 \end{aligned}
 \end{equation}
with $\gcd(\sfe,\sfm)=\gcd(\sfe',\sfm')=1$. 
In order to better understand the effect of the resulting topological defect, let us first introduce the integers $ \sfn\equiv\gcd(\sfe,\sfm')$ and $\sfk\equiv \gcd(\sfm,\sfe')$, so that we can set 
\be\label{emred}
\sfe=\sfn\,\tilde\sfe\quad,\quad \sfm=\sfk\,\tilde\sfm\quad,\quad \sfe'=\sfk\,\tilde\sfe'\quad,\quad \sfm'=\sfn\,\tilde\sfm'\,, 
\ee
with 
\be\label{gcdcond}
\begin{aligned}
&\gcd(\sfn,\sfk)=\gcd(\sfn,\tilde \sfm)=\gcd(\sfk,\tilde\sfe)=\gcd(\sfn,\tilde \sfe')=\gcd(\sfk,\tilde\sfm')=1\,,\\
&\gcd(\tilde\sfe,\tilde\sfm)=\gcd(\tilde\sfe',\tilde\sfm')=\gcd(\tilde\sfe,\tilde\sfm')=\gcd(\tilde\sfm,\tilde\sfe')=1\,.
\end{aligned}
\ee 
Notice that
\be
\cals_\sfA\cals'_\sfA=\left(\begin{array}{cc} \frac{\tilde\sfm\tilde\sfm'}{\tilde\sfe\tilde\sfe'} &  0\\
0 & \frac{\tilde\sfe\tilde\sfe'}{\tilde\sfm\tilde\sfm'} \end{array}\right)\,,
\ee
where $\cals_\sfA$ and $\cals'_\sfA$ are the ${\rm Sp}(2,\mathbb{Q})$ matrices associated to the pairs $(\sfe,\sfm)$ and $(\sfe',\sfm')$ as in \eqref{Sfactor2}. Hence, the ${\rm Sp}(2,\mathbb{Q})$ transformation $\cals_\sfA\cals'_\sfA$ corresponds to the minimal defect $\cald^{(\tilde\sfe\tilde\sfe',\tilde\sfm\tilde\sfm')}_\mathsf{A}$ which, indeed, realizes the same gluing conditions of  \eqref{AAfusion}, and hence corresponds to the same GZ transformation $\cals_\sfA\cals'_\sfA$. In particular, $\cald^{(\tilde\sfe\tilde\sfe',\tilde\sfm\tilde\sfm')}_\mathsf{A}$ requires that
\be\label{FFgluing} 
\tilde\sfe\tilde\sfe' F_\tL|_\Sigma=\tilde\sfm\tilde\sfm' F_\tR|_\Sigma\quad,\quad \tilde\sfm\tilde\sfm' G_\tL|_\Sigma= \tilde\sfe\tilde\sfe'G_\tR|_\Sigma\,.
\ee
On the other hand, \eqref{AAfusion} imposes stronger constraints on the pull-back of the bulk field strengths along $\Sigma$.  Indeed,  integrating out  $b$ and $\tilde b$ in \eqref{AAfusion}, we get the conditions:
\be\begin{aligned} 
\sfn\tilde\sfe F_\tL|_\Sigma=\sfk\tilde\sfm \d a\quad,\quad \sfn\tilde\sfm' F_\tR|_\Sigma=\sfk\tilde\sfe' \d a \,. 
\end{aligned}
\ee
Given \eqref{gcdcond}, these conditions admit a solution only if we can set $\d a=\sfn\tilde\sfe\tilde\sfm'\d\hat a $, for some U(1) gauge field $\hat a$ on $\Sigma$, and therefore 
\be \label{additionalquantcondF}
F_\tL|_\Sigma=\sfk\tilde\sfm\tilde\sfm'\d\hat a\quad,\quad F_\tR|_\Sigma=\sfk\tilde\sfe\tilde\sfe'\d\hat a\,.
\ee
Not only do these conditions imply the first of \eqref{FFgluing}, but they also imply stronger flux quantization conditions along $\Sigma$ that, if violated, make the defect vanish. 

A similar conclusion holds for the second of \eqref{FFgluing}. Integrating out $a$  and extremizing with respect to the bulk fields $A_\tL$ and $A_\tR$, one gets the conditions: 
\be 
\sfn\tilde\sfm \d b=\sfn\tilde\sfe' \d\tilde b \quad,\quad G_\tL|_\Sigma=-\sfn\tilde\sfe \d b\quad,\quad G_\tR|_\Sigma=-\sfn\tilde\sfm' \d b\,.
\ee
The first condition has a solution only if we can set $\sfn\d b=\sfn \tilde\sfe'\d\hat b$ and $\sfn\d \tilde b=\sfn \tilde\sfm\d\hat b$, for some U(1) gauge field $\hat b$. Then, the other two conditions become 
\be \label{additionalquantcondG}
G_\tL|_\Sigma=-\sfn\tilde\sfe\tilde\sfe' \d \hat b\quad,\quad G_\tR|_\Sigma=-\sfn\tilde\sfm\tilde\sfm' \d \hat b\,.
\ee
These conditions imply the second condition in \eqref{FFgluing}, but impose stronger quantization conditions  along $\Sigma$ for $G_\tL$ and $G_\tR$.  

The additional flux quantization conditions \eqref{additionalquantcondF} and \eqref{additionalquantcondG} imposed by the fusion defect \eqref{AAfusion}  can be associated with the combined action of a condensation defect. More precisely, by matching the corresponding gluing conditions, one can check that
the two defects are related by 
\be 
\cald^{(\sfe,\sfm)}_\mathsf{A} \times \cald^{(\sfe',\sfm')}_{\mathsf{A}}=\cald^{(\tilde\sfe\tilde\sfe',\tilde\sfm\tilde\sfm')}_\mathsf{A}\times \calc^{(\sfn,\sfk)}_\sfA\,,
\ee
where $\calc^{(\sfn,\sfk)}_\sfA$ is as in \eqref{tauMCA}, or \eqref{Cem}, and $\hat{a}$  in \eqref{additionalquantcondF} can be identified with  the bulk gauge field living between the two defects on the r.h.s\ of the equation, before fusing them. For simplicity we have also assumed that $\gcd(\sfk,\tilde\sfe')=\gcd(\sfn,\tilde\sfm')=1$. 
Similar relations hold for more general gcd's.

%%%%%%%%%%%%%%%%%%%%%%%%%%%%%%%%%%%%%%%%%%%%%%%%%%%%%%

\subsection{Action of $\tau$-Maxwell non-invertible defects on line operators}
\label{sec:tauMlines}

In order to read off the action of the GZ defects on line operators we now specialize the discussion in \ref{sec:AactWH} to the $n=1$ case. The dyonic line operators  can be written as follows,
\begin{equation}\label{tauMlines}
    \begin{aligned}
    L_{\mathbb L}(\gamma)&=\exp{\left(\ii \tilde{\ell} \oint_{\gamma} \widetilde A-\ii \ell\oint_{\gamma} A \right)}\, ,
   \end{aligned} 
\end{equation}
where the integral dyonic charges $\ell,\tilde\ell$ are identified by the vector $\mathbb L=   (\tilde \ell, \ell)^{\rm t}$. If $\gamma=\partial M_2$, we can also define the following surface operators,
\begin{equation}
   \widehat L_{\mathbb L}(\gamma,M_2) = \exp{\left(\ii\tilde \ell \oint_{M_2} G-\ii \ell\oint_{M_2} F\right)}\, .
\end{equation}
These are well defined also for non-integral charges, and reduce to the genuine line operators \eqref{tauMlines} only for integral ones, that is $ \widehat L_{\mathbb L}(\gamma,M_2)= L_{\mathbb L}(\gamma)$ if and only if  ${\mathbb L}\in\mathbb{Z}^2$.

The gluing conditions \eqref{UWcollapse} allow us to read off when the line are projected out from the spectrum of genuine line operators or they survive as such. Suppose that $L_{\mathbb L}(\gamma)$ is on the right of $\cald_\cals(\Sigma)$ and $\gamma=\del M_2$. Sweeping $\cald_{\cals}(\Sigma)$  across $L_{\mathbb L}(\gamma)$ produces the transformation 
\begin{equation} \label{noninvactn1}
    \cald_{\cals} : \qquad  L_{\mathbbm{L}}(\gamma) \quad~~~\longrightarrow\quad~~~ \widehat L_{\cals\mathbbm{L}}(M_2,\gamma) \,,
\end{equation}
where $\mathbbm{L}'\equiv\cals\mathbbm{L}$ corresponds to the new charges 
\begin{equation}
\tilde\ell'=\sfa\,\tilde\ell+\sfb\,\ell\quad,\quad \ell'=\sfc\,\tilde \ell +\sfd\, \ell\, .
\end{equation}
 The new line operators are genuine, $\widehat L_{\mathbbm{L}'}= L_{\mathbbm{L}'}$, if and only if $\tilde\ell',\ell'\in\mathbb{Z}$. If $\cald_\cals$ is non-invertible, and hence $\cals$ is non-integral, there will always be  genuine lines that are mapped to non-genuine ones by $\cald_\cals$ as in \eqref{noninvactn1}. On the other hand, an invertible  $\cald_\cals$ acts  by a matrix $\cals \in {\rm SL}(2,\mathbb{Z})$ on electric and magnetic charges and therefore  maps genuine lines into genuine lines.

Let us now focus on some specific example. Consider first $\cald^{(\sfe,\sfm)}_{\mathsf{A}}$. It maps $L_{\mathbbm{L}}(\gamma)$ on the right to $\widehat L_{\mathbbm{L}'}$ on the left, with
\begin{equation} 
   \tilde \ell' = \frac{\sf m}{\sf e} \tilde \ell\quad,\quad  \ell' = \frac{\sf e}{\sf m}\ell\,.
\end{equation}
Therefore, when 
\begin{equation}
    \frac{\ell\mathsf{e}}{\mathsf{m}} \notin \mathbb{Z}\quad, \quad \frac{\tilde \ell\mathsf{m}}{\mathsf{e}} \notin \mathbb{Z}\,,
\end{equation}
the defect $ \widehat L_{\mathbbm{L}'}$ cannot be considered a genuine line. 

The action of $\cald^{(\sfn,\sfp)}_{\mathsf{C}}$ on line operators  is completely analogous to the action  already discussed in \cite{Choi:2022fgx, Choi:2022jqy}. In particular $\cald^{(\sfn,\sfp)}_{\mathsf{C}}$ acts non-invertibly on 't Hooft lines, $ H_{\tilde \ell}(\gamma)\equiv L_{\mathbbm{L}}(\gamma)$ with ${\mathbbm{L}=(\tilde \ell,0)^{\rm t}}$, when $\tilde \ell \,\mathsf{p} \neq 0$ mod $\mathsf{n}$, and the action reads 
\begin{equation}
   \cald^{(\sfn,\sfp)}_{\mathsf{C}} : \quad   H_{\tilde \ell} (\gamma)\quad~~~\longrightarrow\quad~~~    \exp{\left(-\ii \frac{\tilde \ell\, \mathsf{p}}{\mathsf{n}}\oint_{M_2} F\right)}H_{\tilde \ell} (\gamma)\, .
\end{equation}
If $\tilde \ell \,\mathsf{p}= 0$ mod $\mathsf{n}$, the line remains genuine. 

Finally, let us discuss the action of a condensation defect $\calc_{\mathsf{A}}^{(\sfe,\sfm)}$ -- see \eqref{tauMCA} or \eqref{Cem} -- on line operators. Notice that insertion of a line operator \eqref{tauMlines} can be seen as a source for $F$ and $G$:  
\be 
\d   F=2\pi\tilde\ell\,\delta_3(\gamma)\quad,\quad \d G=2\pi\ell\,\delta_3(\gamma)\quad\Leftrightarrow\quad \d\mathbbm{F}=2\pi\mathbbm{L}\,\delta_3(\gamma)\,,
\ee
where $\delta_3(\gamma)$ is a delta-like three-form localized on $\gamma$. Therefore, by using the description \eqref{Cem}-\eqref{Condpqstand} of the condensation defects, we get the identity
\begin{equation}
    \label{CLaction}
    \calc^{(\sfe,\sfm)}_{\mathsf{A}} (\Sigma)  \; L_{\mathbbm{L}}(\gamma) = \caln_{\sfe,\sfm}\sum_{C\in H_2(\Sigma,\mathbb{Z}_{\sfe})}\sum_{\tilde C\in H_2(\Sigma,\mathbb{Z}_{\sfm})}e^{\frac{2\pi\ii}{\sfe}\ell\, {\rm Link}(C,\gamma)}e^{\frac{2\pi\ii}{\sfm}\tilde\ell\, {\rm Link}(\tilde C,\gamma)}L_{\mathbbm{L}}(\gamma)
\end{equation}
with $\caln^{-1}_{\sfe,\sfm}={|H^0(\Sigma,\mathbb{Z}_{\sfe})||H^0(\Sigma,\mathbb{Z}_{\sfm})|}$. In particular, the sum of the phases can give zero, and therefore make the condensation defect act as a projector, as already pointed out in  \cite{Roumpedakis:2022aik, Hasan:2024aow}. For instance, suppose that  $H_2(\Sigma,\mathbb{Z}_{\sfe})$ is generated by a single cycle $C_0$, with  ${\rm Link}(C_0,\gamma)=1$.  Then the sum runs over $C=kC_0$ with $k=0,\ldots,\sfe-1$ and  the right-hand side of \eqref{CLaction} vanishes for $\ell\neq 0$ mod $\sfe$. A similar conclusion holds if $H_2(\Sigma,\mathbb{Z}_{\sfm})$ is generated by a single cycle, and also for  other values of the linking numbers and/or for other settings.

%%%%%%%%%%%%%%%%%%%%%%%%%%%%%%%%%%%%%%%%%%%%%%%%%%%%%%%%%%

\section{Perturbative Calabi-Yau models}
\label{sec:CYdefects}

In this section, we apply the general procedure outlined in Section \ref{sec:GZdefects2} to a class of models with $n$ vector fields and  GZ symmetry group
\be\label{CYG} 
\mathscr{G}\simeq \mathbb{R}^{n-1}\,.
\ee
In particular, we will construct concrete examples of  GZ topological defects $\cald_{\bm\alpha}(\Sigma)$ for any  $(n-1)$-dimensional rational vector ${\bm\alpha}=\{\alpha^i\}\in \mathscr{G}_\mathbb{Q}\simeq \mathbb{Q}^{n-1}$. We will later comment on the  gauging of $\mathscr{G}_\mathbb{Z}\simeq \mathbb{Z}^{n-1}\subset \mathbb{R}^{n-1}$, which would reduce the global GZ symmetry group  to ${\rm U}(1)^{n-1}$.  

These models correspond to the bosonic sector of $\caln=2$ supergravities with $n-1$ vector multiplets and holomorphic homogeneous  prepotential 
\be\label{CYprepot} 
\calf(X)=-\frac{1}{3!}\frac{\kappa_{ijk}X^i X^j X^k}{X^0}+\frac12 \mathsf{y}_{IJ}X^{I}X^J+\frac\ii2 \rho (X^0)^2\,.
\ee
Here $X^I=(X^0,X^i)$ ($I=0,\ldots,n-1$) are homogeneous complex projective coordinates over the field space parametrized by the $n-1$ complex scalar fields
\be\label{N=2phi^i} 
\phi^i=\frac{X^i}{X^0}\,.
\ee
In the superconformal formulation, $X^I$ include a non-physical conformal compensator, which may be identified with $X^0$ -- see for instance \cite{Freedman:2012zz}. The constants $\rho$, $\kappa_{ijk}$ and $\mathsf{y}_{IJ}$ appearing in \eqref{CYprepot} are real, and  $\kappa_{ijk}$ and $\mathsf{y}_{IJ}$ are totally symmetric in their indices. For the moment, we just require these constants to define a classically sensible supergravity. 
In fact, $\kappa_{ijk}$ and $\mathsf{y}_{IJ}$ have to satisfy some more stringent constraints, which will be specified presently. 

The prepotential  \eqref{CYprepot} encompasses a large class of models, closely related to the effective theories that one gets in  type IIA string compactifications on Calabi-Yau three-folds, neglecting non-perturbative corrections. For convenience, we will loosely speaking refer to these models as perturbative Calabi-Yau models, although our arguments will be independent of this possible realization. In fact, as in the rest of the paper, we will continue to adopt an agnostic approach regarding the UV completion.     

Ignoring the gravitational sector, the bosonic sector of these models can be identified as a subclass of the models of Appendix \ref{app:twoderiv}, with $n$ U(1) vectors $A^I$, and the $n-1$ complex scalar fields $\phi^i$, which we split in real and imaginary parts as follows
\be 
\phi^i\equiv a^i+\ii s^i\,.
\ee 
As will become clear shortly, we can refer to $a^i$ and $s^i$ and axions and saxions, respectively. Their kinetic terms take the form
\be\label{NLSM}
-\sff^2K_{i\bar\jmath}\,\d\phi^i\wedge *\d\bar\phi^{\bar\jmath}\,,\quad~~~~~\text{with}\quad K_{i\bar\jmath}=\frac{\del^2 K}{\del\phi^i\del\bar\phi^{\bar\jmath}}\ ,
\ee
where  $\sff$ identifies a reference mass-scale   (which in supergravity coincides with the Planck mass) and the  metric is associated with the K\"ahler  potential
\be\label{CYK} 
K(\phi,\bar\phi)=-\log\left[P(s)-\frac14\rho\right]\,,\quad~~~~~\text{with}\quad 
P(s)\equiv \frac1{3!}\kappa_{ijk}s^is^js^k\,.
\ee 
For us, the important point is that the K\"ahler potential, and hence the kinetic terms \eqref{NLSM}, are invariant under arbitrary axionic shifts $a^i\rightarrow a^i+\alpha^i$, $\alpha^i\in\mathbb{R}$, namely  
\be\label{axionshifts} 
\phi^i\rightarrow \phi^i+\alpha^i\,.
\ee
 For the moment we treat $a^i$ as non-periodic, postponing the periodicity requirements to a later step. Hence, adopting the terminology introduced in Section \ref{sec:GZ}, we can identify the symmetry group of the neutral sector with \eqref{CYG}.

We now turn to the gauge sector. The kinetic matrix of the gauge fields takes the form \eqref{gen2derL} with 
\be\label{N=2caln} 
\caln_{IJ}\equiv\overline \calf_{IJ}+2\ii\, \frac{(N X)_I (N X)_J}{X^TNX}\,.
\ee
where $\calf_{IJ}\equiv \del^2\calf(X)/\del X^I\del X^J$ and $N_{IJ}\equiv \Im \calf_{IJ}$. By computing the explicit form of $\caln_{IJ}$ associated with \eqref{CYprepot} one finds that it has a non-trivial, up to cubic, dependence on the axions $a^i$, which breaks the invariance of the Lagrangian under the axionic shifts \eqref{axionshifts}.\footnote{ See \cite{Grimm:2022xmj} for a detailed discussion of the breaking of invertible symmetries induced by these generalized theta-terms, and \cite{grieco2022noninvertible} for an earlier proposal to resurrect some of the axionic shift symmetries \eqref{axionshifts} as non-invertible ones.} 
We will not need the detailed form of such dependence, but rather the transformation that it induces on $\caln_{IJ}$  under  \eqref{axionshifts}. This takes the form 
\be\label{calntrCY} 
\caln_{IJ}(\phi+\alpha)=\big(\big[\mathsf{C}(\alpha)+\mathsf{D}(\alpha)\caln(\phi)\big]\big[\mathsf{A}(\alpha)+\mathsf{B}(\alpha)\caln(\phi)\big]^{-1}\big)_{IJ}\,,
\ee
where the $n\times n$ matrices $\mathsf{A,B,C,D}$, which identify a symplectic matrix $\cals$ according to  \eqref{genS}, are specified  by the following  block decomposition,
\begin{subequations}
\label{pertsympl0}
\begin{align}
\mathsf{A}^I{}_J(\alpha)&=\left(\begin{array}{cc}\mathsf{A}^0{}_0 & \mathsf{A}^0{}_j\\
\mathsf{A}^i{}_0 & \mathsf{A}^i{}_j
\end{array}\right)=\left(\begin{array}{cc}1 & 0\\
\alpha^i & \delta^i{}_j
\end{array}\right)\,,\label{pertsympl0a}\\
\mathsf{B}^{IJ}(\alpha)&=0\,,\label{pertsympl0b}\\
\mathsf{C}_{IJ}(\alpha)&=\left(\begin{array}{cc}\mathsf{C}_{00} & \mathsf{C}_{0j}\\
\mathsf{C}_{i0} & \mathsf{C}_{ij}
\end{array}\right)=\left(\begin{array}{cc}P(\alpha)+2\mathsf{y}_{0k}\alpha^k & P_j(\alpha)+\mathsf{y}_{jk}\alpha^k\\
\mathsf{y}_{ik}\alpha^k-P_i(\alpha) & -P_{ij}(\alpha)
\end{array}\right)\,,\label{pertsympl0c}\\
\mathsf{D}_I{}^J(\alpha)&=(\sfA^{-1})^J{}_I=\left(\begin{array}{rr}1 & -\alpha^j\\
0 & \delta_i{}^j
\end{array}\right)\,,\label{pertsympl0d}
\end{align}
\end{subequations}
with $P_i(\alpha)\equiv \frac12\kappa_{ijk}\alpha^j\alpha^k$ and $P_{ij}(\alpha)\equiv \kappa_{ijk}\alpha^k$.
Through \eqref{genS},  the matrices \eqref{pertsympl0} define a symplectic representation 
\be 
\mathscr{G}\ni \{\alpha^i\}\ \mapsto
 \ \cals(\alpha)\in {\rm Sp}(2n,\mathbb{R})
\ee
of  the GZ symmetry group \eqref{CYG}. As in the rest of the paper, we will simply identify $\mathscr{G}$ with its image in ${\rm Sp}(2n,\mathbb{R})$ under this map. Note that \eqref{calntrCY} takes the form \eqref{GZcond2der}, and therefore provides the realization of the GZ condition \eqref{tildecall}. 

\subsection{GZ topological defects}

The parameters $\alpha^i$ identify a basis of generators $T_i\in{\rm sp}(2n,\mathbb{R})$  of the symplectic representation of the  abelian group   \eqref{CYG}, so that \eqref{TSrel} is realized with $T=\alpha^iT_i$, namely,   $\cals(\alpha)=\exp\left(\alpha^iT_i\right)$.   We can then write the defect $ \calu_{{\alpha}}(\Sigma)$ implementing the axionic shift \eqref{axionshifts} as in \eqref{caluS}, by choosing  $\calj_T=\alpha^i\calj_i$ with
\be 
\calj_{i}=-\sff^2K_{ij}(s)*\d a^j\,. 
\ee
As in our general discussion, the currents  $\calj_i$ are not conserved, and hence $\calu_{\alpha}(\Sigma)$ is not topological, since
\be\label{CYaxioncurrent} 
\d\calj_{i}=-\frac1{4\pi}\kappa_{ijk}F^j\wedge F^k-\frac1{2\pi}F^0\wedge G_i+\frac1{2\pi}\mathsf{y}_{0i}F^0\wedge F^0+\frac1{2\pi}\mathsf{y}_{ij}F^0\wedge F^j\,.
\ee
This  codifies the breaking of the axionic shift symmetries \eqref{axionshifts} due to the non-invariance \eqref{calntrCY} of  the kinetric matrices \eqref{N=2caln}  of the gauge fields.

In order to construct GZ topological defects,  we must 
impose some further conditions on the constants $\kappa_{ijk}$ and $\mathsf{y}_{ij}$. More precisely, we require that the integral shifts $\alpha^i=m^i\in\mathbb{Z}$ identify the integral subgroup $\mathscr{G}_\mathbb{Z}\simeq \mathbb{Z}^{n-1}$. By consistency with our general construction, the matrices \eqref{pertsympl0} must take integral values and hence define elements $\cals({m})$ of $  {\rm Sp}(2n,\mathbb{Z})$. (Again, we will for simplicity use the same symbol for  $\mathscr{G}_\mathbb{Z}$ as well as for its representation in $ {\rm Sp}(2n,\mathbb{Z})$). It is easy to see that this is possible if and only if 
\begin{subequations}
\label{yCcond}
\begin{align}
& \kappa_{ijk}\in\mathbb{Z}\,,\label{yCconda}\\
&\frac12\kappa_{ijk}m^jm^k-\mathsf{y}_{ij}m^j\in\mathbb{Z}\,,\label{yCcondb}\\
&\frac16 \kappa_{ijk}m^im^jm^k+2\mathsf{y}_{0i}m^j\in\mathbb{Z}\,,\label{yCcondc}
\end{align}
\end{subequations}
for any choice of integers $m^i\in\mathbb{Z}$. In particular, these imply that 
\be\label{ycond} 
\mathsf{y}_{ij}\in\frac12\mathbb{Z}\quad,\quad \mathsf{y}_{0i}\in\frac1{12}\mathbb{Z}\,.
\ee The conditions \eqref{yCcond} are indeed  realized in type IIA Calabi-Yau compactifications. In that case, the condition \eqref{yCconda} is satisfied since the constants $\kappa_{ijk}$ can be identified with triple intersection numbers of the Calabi-Yau space. \footnote{Furthermore, the conditions \eqref{yCcondb} and \eqref{yCcondc} are a consequence of the {\em stronger} conditions derived in \cite{Alexandrov:2010ca}. More precisely, \eqref{yCcondb} is equivalent to the first  of \cite[Eq.(4.12)]{Alexandrov:2010ca}, while  \eqref{yCcondc} is implied by the second of \cite[Eq.(4.12)]{Alexandrov:2010ca}, since $P(q)+\frac1{12}c_{2,i}q^i\in\mathbb{Z}$.  Note also that, in our agnostic approach, the real constant $\rho$ appearing in \eqref{CYprepot} is arbitrary, while in type IIA  realizations it takes  the fixed value $\rho=\chi\, \zeta(3)/(2\pi)^3$, where $\chi$ is the Euler number of the Calabi-Yau three-fold \cite{Candelas:1990rm}.}  

Now, according to the general discussion of Sections \ref{sec:GZdefects} and \ref{sec:GZdefects2}, we can construct GZ topological defects for elements of $\mathscr{G}_\mathbb{Q}=\mathscr{G}\cap {\rm Sp}(2n,\mathbb{Q})$. In our setting, $\cals({\alpha})\in \mathscr{G}_\mathbb{Q}$ if 
\be\label{alphaQ} 
\alpha^i=\alpha^i_{{\bf q},{\bf p}}\equiv\frac{q^i}{p^i}\in \mathbb{Q}\,,
\ee
with $q^i,p^i\in\mathbb{Z}$. We may also impose $\gcd(q^i,p^i)=1$ for any $i=1,\ldots, n-1$, but most of the following results hold also for more general choices. 
The corresponding GZ topological defect can  be factorized as in \eqref{Doperator}, with $\calu$ defect
\be\label{caluCY} 
\calu_{{\bf q},{\bf p}}(\Sigma)\equiv\exp\left(\ii\sum_i\frac{q^i}{p^i}\oint_\Sigma\calj_i\right)\,.
\ee 
It hence remains to identify a realization of the interfaces $\calw_{{\bf q},{\bf p}}(\Sigma)\equiv \calw_{\cals(\alpha_{{\bf q},{\bf p}})}(\Sigma)$. 

As a first step, we need to factorize the symplectic matrix $\cals(\alpha)$ defined by \eqref{pertsympl0}  as a  product of matrices  \eqref{Sfactor}. As a simple possibility, we can write it as
\be\label{Salphadec} 
\cals({\alpha})=\cals_{\tilde{\mathsf{C}}({\alpha})}\cals_{\mathsf{A}({\alpha})}\,,
\ee
with  ${\sf A}(\alpha)$ is as in \eqref{pertsympl0a} and 
\be\label{tildeCCY}
\begin{aligned}
\tilde{\mathsf{C}}(\alpha)&\equiv(\mathsf{C}\mathsf{A}^{-1})(\alpha)=\left(\begin{array}{cc}\tilde{\mathsf{C}}_{00} & \tilde{\mathsf{C}}_{0j}\\
\tilde{\mathsf{C}}_{i0} & \tilde{\mathsf{C}}_{ij}
\end{array}\right)=\left(\begin{array}{cc}-2P(\alpha)+2\mathsf{y}_{0k}\alpha^k -\mathsf{y}_{kl}\alpha^k\alpha^l& P_j(\alpha)+\mathsf{y}_{jk}\alpha^k\\
P_i(\alpha)+\mathsf{y}_{ik}\alpha^k & -P_{ij}(\alpha)
\end{array}\right)\,,
\end{aligned}
\ee 
where we have used  \eqref{pertsympl0c}. We can then get a realization of $\calw_{\cals(\alpha)}(\Sigma)$ by concatenating  $\mathsf{A}$ and $\mathsf{C}$ interfaces according to  the decomposition \eqref{Salphadec}.

We first focus on the {\sf A} defect $\calw^{{\bf q},{\bf p}}_{\mathsf{A}}(\Sigma)\equiv \calw_{\mathsf{A}(\alpha_{{\bf q},{\bf p}})}(\Sigma)$. Following the construction outlined in Section \ref{sec:GZAdefects}  we need to identify a factorization of the form \eqref{AMEdec}. For instance, a simple choice is 
\be\label{CYME}
\begin{aligned} 
(\mathsf{E}_{{\bf q},{\bf p}})^I{}_J&=\left(\begin{array}{cc}\mathsf{E}^0{}_0 & \mathsf{E}^0{}_j\\
\mathsf{E}^i{}_0 & \mathsf{E}^i{}_j
\end{array}\right)=\left(\begin{array}{cc}1 & 0\\
0 & p^i\delta^i{}_j
\end{array}\right)\,,\\
(\mathsf{M}_{{\bf q},{\bf p}})^I{}_J&=\left(\begin{array}{cc}\mathsf{M}^0{}_0 & \mathsf{M}^0{}_j\\
\mathsf{M}^i{}_0 & \mathsf{M}^i{}_j
\end{array}\right)=\left(\begin{array}{cc}1 & 0\\
q^i & p^i\delta^i{}_j
\end{array}\right)
\,.
\end{aligned}
\ee
Using this or any other decomposition in \eqref{WA}, we get 
\be\label{WACY} 
\calw^{{\bf q},{\bf p}}_{\mathsf{A}}(\Sigma)=\int\cald b\exp\left(\frac\ii{2\pi}\oint_\Sigma b_I\wedge \left[(\sfE_{{\bf q},{\bf p}})^I{}_JF^J_{\text{\tiny L}}-(\sfM_{{\bf q},{\bf p}})^I{}_JF^J_{\text{\tiny R}}\right]\right)\,.
\ee
In general, the factorization defined by \eqref{CYME} is not left-coprime, and then  \eqref{WACY}  is not minimal in the sense discussed in Section \ref{sec:GZAdefects}. However, its simplicity allows us to  easily combine it with other interfaces and get a readable explicit general formula. 

Let us now turn to the $\cals_{\tilde\sfC(\alpha_{{\bf q},{\bf p}})}$ element appearing in \eqref{Salphadec}. Taking the conditions \eqref{yCcond} and \eqref{ycond} into account, from \eqref{tildeCCY} it is clear that $\tilde\sfC^{{\bf q},{\bf p}}_{IJ}\equiv \tilde\sfC_{IJ}(\alpha_{{\bf q},{\bf p}})\in\mathbb{Q}$. Again, we must pick an integral factorization of the form \eqref{Cfact} and, as discussed in Section \ref{sec:GZCdefects}, the canonical choice would correspond to a right-coprime factorization. This would allow us to use a  minimal TQFT and to 
construct the corresponding minimal $\sfC$ interface. However, as in the construction of \eqref{WACY},  for our purposes  it is  convenient to consider a more general factorization and use it in \eqref{Ictheory}. 

  So, in order to realize \eqref{Cfact} for any choice of $q^i$ and $p^i$,  we must find an invertible matrix $(\sfN_{{\bf q},{\bf p}})^I{}_J$ such that
\be\label{tildeCN} 
(\sfP^{{\bf q},{\bf p}})_{IJ}\equiv\tilde\sfC^{{\bf q},{\bf p}}_{IK}\,(\sfN_{{\bf q},{\bf p}})^K{}_J\in\mathbb{Z}\,,\quad \forall I,J=0,\ldots,n-1.
\ee
The simplest possibility is obtained by  picking a positive integer $ N_{{\bf q},{\bf p}}$ such that $N_{{\bf q},{\bf p}}\tilde\sfC^{{\bf q},{\bf p}}_{KL}\in\mathbb{Z}$ for any  $K,L=0,\ldots,n-1$, and setting  
\be\label{simpleNqp} 
(\sfN_{{\bf q},{\bf p}})^I{}_J= N_{{\bf q},{\bf p}}\delta^I{}_J\,.
\ee 

We can now use \eqref{Ictheory} in \eqref{WC2} to construct a (non-minimal) $\tilde\sfC$ interface $\calw^{{\bf q},{\bf p}}_{\tilde\sfC}(\Sigma)$. 
By concatenating it with \eqref{WACY} according to \eqref{Salphadec}, we get the complete $\calw$ interface:
\be\label{Wqp} 
\calw_{{\bf q},{\bf p}}=\calw_{\tilde\sfC}^{{\bf q},{\bf p}}\times \calw_\sfA^{{\bf q},{\bf p}}\,.
\ee
Combining it with \eqref{caluCY} as in \eqref{Doperator}, we then arrive  at  the  GZ topological defects 
\be\label{caldCY}
\begin{aligned}
\cald_{{\bf q},{\bf p}}(\Sigma)=&\,\exp\left(\ii\,\alpha^i_{{\bf q},{\bf p
}}\oint_\Sigma\calj_i\right)\int\cald b\exp\left(\frac\ii{2\pi}\oint_\Sigma b_I\wedge \left[(\sfM_{{\bf q},{\bf p}})^I{}_JF^J_{\text{\tiny L}}-(\sfE_{{\bf q},{\bf p}})^I{}_JF^J_{\text{\tiny R}}\right]\right) \\
&\int\cald a\cald c\,\exp\left(-\frac{\ii}{4\pi}\sfK_{IJ}^{{\bf q},{\bf p}}\oint_\Sigma  a^I\wedge f^J+\frac\ii{2\pi}\oint_\Sigma c_I\wedge \left[(\tilde\sfN_{{\bf q},{\bf p}})^I{}_J f^J-F^I_{\text{\tiny L}}\right]\right)\,,
\end{aligned}
\ee
where $\sfK_{IJ}^{{\bf q},{\bf p}}\equiv (\sfN_{{\bf q},{\bf p}})^K{}_I(\sfN_{{\bf q},{\bf p}})^L{}_J\tilde\sfC^{{\bf q},{\bf p}}_{KL}$. To make \eqref{caldCY}  more concrete, we can  pick $\sfM_{{\bf q},{\bf p}}$ and $\sfE_{{\bf q},{\bf p}}$ as in \eqref{CYME} and $\sfN_{{\bf q},{\bf p}}$ as in \eqref{simpleNqp}, so that $\sfK_{IJ}^{{\bf q},{\bf p}}=N^2_{{\bf q},{\bf p}}\tilde\sfC^{{\bf q},{\bf p}}_{IJ}$. 
This provides just one possible realization of GZ  defects  implementing rational  axionic shifts. Many other examples could of course be constructed and, 
in particular, it would be interesting to identify and study the minimal TQFT dressing these interfaces.

 Note that for integral axionic shifts, $\alpha^i=m^i\in\mathbb{Z}$, we can just pick $q^i=m^i$ and $p^i=1$ and choose $(\sfM_{{\bf q},{\bf p}})^I{}_J=\delta^I{}_J$ and $(\sfE_{{\bf q},{\bf p}})^I{}_J=(\sfA_{{\bf m}})^I{}_J\equiv \sfA^I{}_J(m)$. Furthermore the conditions \eqref{yCcond} ensures that $\tilde\sfC^{\bf m}_{IJ}\equiv \tilde\sfC_{IJ}(m)\in\mathbb{Z}$ for any $m^i\in\mathbb{Z}$, so that we can choose $(\sfN_{{\bf q},{\bf p}})^L{}_J=\delta^I{}_J$ and $\sfK_{IJ}^{{\bf q},{\bf p}}=\sfP_{IJ}^{{\bf q},{\bf p}}=\tilde\sfC_{IJ}^{\bf m}$. In this case, one can explicitly integrate out $c_I$ and set $f^I=F^I_{\text{\tiny L}}$, getting the invertible topological defect
\be\label{caldCY2} 
\begin{aligned}
\cald_{{\bf m}}(\Sigma)=&\,\exp\left(\ii\, m^i\oint_\Sigma\calj_i-\frac{\ii}{4\pi}\tilde\sfC^{\bf m}_{IJ}\oint_\Sigma A^I_{\text{\tiny L}}\wedge F^J_{\text{\tiny L}}\right)\\
&\quad~~~~~\times\int\cald b\exp\left(\frac\ii{2\pi}\oint_\Sigma b_I\wedge \left[F^I_{\text{\tiny L}}-(\sfA_{{\bf m}})^I{}_JF^J_{\text{\tiny R}}\right]\right)\,.
\end{aligned}
\ee
This coincides with the defect that can be more directly obtained  by combining  the invertible $\sfA$  and $\sfC$ interfaces appearing in \eqref{WSZ}. 

It would be important to better understand the spectrum of possible GZ topological defects in this class of models, and  their fusion rules. 
This is particularly crucial if one wants to consider periodic axions, that is, to impose the identification $\phi^i\simeq \phi^i+m^i$, $m^i\in\mathbb{Z}$, differently from what assumed so far. This periodicity  would amount to gauge the integral axionic shifts, hence summing over all possible insertions of the invertible  defects \eqref{caldCY2} in the path-integral. Since the GZ  group \eqref{CYG} is abelian, the adjoint action of the $\cald_{\bf m}$ defects  relates non-invertible  GZ topological defects  $\cald_{{\bf q},{\bf p}}$ implementing the same gluing conditions which, however, do not in general coincide.  Only linear combinations of GZ topological defects  that are invariant under this adjoint action for any $m^i\in\mathbb{Z}$ correspond to global symmetries of the theory with periodic axions. We will return to these aspects in  Section \ref{sec:gaugingSB}  from a more general point of view, leaving a more detailed study of this class of models to future work.

%%%%%%%%%%%%%%%%%%%%%%%%%%%%%%%%%%%%%%%%%%%%%%%%%%%%%%%%%%%

\subsection{A one-modulus example}

Just to get an idea of what the above formulas can more concretely look like, we consider a simple example, related to  the compactification of type IIA theory on the quintic Calabi-Yau. Namely, we focus on its low-energy  bosonic sector, corresponding to one complex field $\phi$ and two gauge fields, $A^0$ and $A^1$. In this case 
the exact {\em perturbative} prepotential takes the form  \eqref{CYprepot} with $X^I=(X^0,X^1)$, $\kappa_{111}=5$ and \cite{Candelas:1990rm}
\be 
\mathsf{y}_{00}=0 \quad,\quad \mathsf{y}_{01}=\frac{1}{12}\quad,\quad \mathsf{y}_{11}=\frac{1}{2}\,,
\ee
up to possible additional integral integral contributions to  $\mathsf{y}_{IJ}$ which do not change the physics. (One can easily check that the conditions \eqref{yCcond} are satisfied.) 
There is only one complex field  $\phi=a+ \ii s=X^1/X^0$. As in the rest of this section, we assume an agnostic viewpoint, taking $a\in\mathbb{R}$ and ignoring the fact that in string models $a$ is actually periodic. The K\"ahler potential \eqref{CYK} becomes, up to an irrelevant additive constant: 
\be 
K=-\log\left(s^3-\frac3{10}\rho\right)\,.
\ee
It is interesting to observe that a non-vanishing $\rho$ breaks the SL$(2,\mathbb{R})$ symmetry of the scalar metric associated with $\rho=0$. If $\rho\neq 0$, the only surviving Killing direction is the axionic shift $a\rightarrow a +\alpha$. (In the type IIA string theory model,  $\rho$ has the specific value $25\zeta(3)/\pi^3$ and represents a perturbative loop correction \cite{Candelas:1990rm}.)

 Let us consider rational axionic shifts $\alpha=\alpha_{\bf{q,p}}=q/p$, $q,p\in \mathbb{Z}$, $p\geq 1$. The transformation is determined by the matrices
 \be 
\sfA_{{\bf q},{ \bf p}}=\left(\begin{array}{cc}1 & 0\\
 \frac{q}p & 1
 \end{array}\right)\quad,\quad \tilde\sfC_{{\bf q},{\bf p}}=\\
 \left(\begin{array}{cc} -\frac{5q^3}{3p^3}-\frac{q^2}{2p^2}+\frac{q}{6p} & \frac{5q^2}{2p^2}+\frac{q}{2p} \\
 \frac{5q^2}{2p^2}+\frac{q}{2p} & - \frac{5q}{p}
 \end{array}\right)\,.
 \ee
 We may then choose the factorization $\sfA_{{\bf q},{\bf p}}= \mathsf{E}^{-1}_{\bf{q,p}} \mathsf{M}_{\bf{q,p}}$ with 
 \be 
 \mathsf{E}_{\bf{q,p}}=\left(\begin{array}{cc}1 & 0\\
 0 & p
 \end{array}\right)\quad,\quad
 \mathsf{M}_{\bf{q,p}}=\left(\begin{array}{cc}1 & 0\\
 q & p
 \end{array}\right)\,,
 \ee
as in \eqref{CYME}, and $\tilde\sfC_{{\bf q},{\bf p}}=\sfP_{{\bf q},{\bf p}}\sfN_{\bf{q,p}}^{-1}$, where $(\sfN_{\bf{q,p}})^I{}_J= N_{\bf q,\bf p}\delta^I{}_J$ as in \eqref{simpleNqp} with 
 \be 
 N_{{\bf q},{\bf p}}=6p^3\,.
 \ee
 This satisfies \eqref{tildeCN} for any choice of $q$ and $p$.  Consequently $\sfK^{\bf{q,p}}$ as defined after \eqref{caldCY} becomes 
 \be 
\sfK^{\bf{q,p}}=6p^3\left(\begin{array}{cc}-10 q^3+3q^2p -qp^2 & 15q^2p+3qp^2\\
 15q^2p+3qp^2 & -30qp^2
 \end{array}\right)
 \ee
 Given these quantities, one can easily write down the corresponding GZ topological defect \eqref{caldCY}. 

%%%%%%%%%%%%%%%%%%%%%%%%%%%%%%%%%%%%%%%%%%%%%%%%%%%%%%%%%%

%%%%%%%%%%%%%%%%%%%%%%%%%%%%%%%%%%%%%%%%%%%%%%%%%%%%%%%%%%
%%%%%%%%%%%%%%%%%%%%%%%%%%%%%%%%%%%%%%%%%%%%%%%%%%%%%%%%%%

%%%%%%%%%%%%%%%%%%%%%%%%%%%%%%%%%%%%%%%%%%%%%%%%%%%%%%%%%%

\section{Remarks on gauging and symmetry breaking}
\label{sec:gaugingSB}

So far  we have been completely agnostic about the UV completion of the GZ models. As emphasized in the Introduction, GZ models should generically be interpreted as effective field theories, valid only at low-energy.  In a rigid QFT context, the  GZ global symmetries constructed in this paper may or may not admit a counterpart in the UV complete theory. On the other hand, once quantum gravity is taken into account,  the GZ symmetries are expected to be either gauged or approximate at low-energy, as any other global (invertible or non-invertible) symmetry \cite{Misner:1957mt,Banks:2010zn,Harlow:2018tng}.

Various combined mechanisms could be responsible for the realization of this expectation, already at low energies. For instance, (part of) the non-invertible $\mathscr{G}_\mathbb{Q}$ symmetries may have a mixed 't  Hooft anomaly with gravity, of the kind described for instance in \cite{Witten:1995gf}, which might automatically break these symmetries once we couple a dynamical gravitational sector. While we  postpone a detailed discussion  of this possibility  to the future, in this section we would like to briefly describe a simple alternative (and/or complementary) symmetry breaking mechanism, that would work even without coupling a dynamical gravitational sector.   Suppose that the  symplectic representation \eqref{symprep} of $\mathscr{G}$ is faithful, so that  $\mathscr{G}$ can be regarded as a subgroup of ${\rm Sp}(2n,\mathbb{R})$, and let us consider what happens if we try to gauge the subgroup $\mathscr{G}_\mathbb{Z}$ of invertible  GZ symmetries (as suggested by experience with string theory models), assuming that it has no pure 't Hooft anomalies. In terms of the corresponding GZ defects $\cald_\cals$, constructed by combining the interfaces 
\eqref{WSZ} and the invertible defects   \eqref{caluS},  this gauging corresponds to summing over all possible
insertions of the defects $\cald_\cals$ with $\cals\in \mathscr{G}_\mathbb{Z}$ \cite{Gaiotto:2014kfa}.  

Consider now a {\em non}-integral  element $\calr\in \mathscr{G}_\mathbb{Q}$ and a corresponding non-invertible GZ defect $\cald_\calr$. For any $\cals\in \mathscr{G}_\mathbb{Z}$, we then have 
\be\label{adjaction} 
\cald^{-1}_\cals \times \cald_\calr \times\cald_\cals=\cald_{\cals^{-1}\calr\cals}\,,
\ee
where $\cald_{\cals^{-1}\calr\cals}$ is a GZ defect associated with the non-integral $\cals^{-1}\calr\cals\in \mathscr{G}_\mathbb{Q}$. It is now clear that, if  $\cals^{-1}\calr\cals\neq \calr$, then $\cald_\calr $ is not a well defined operator in the $\mathscr{G}_\mathbb{Z}$ gauged theory.  In this sense, the non-invertible symmetry associated with $\cald_\calr$ is broken. Notice that this breaking is directly correlated with the breaking of the magnetic and electric U(1) one-form symmetries, whose two-form currents  $F^I$ and $G_I$ transform as in \eqref{gentrb} and are, at least partially, removed from the spectrum of gauge invariant operators. This correlation follows from the realization  of the GZ defects in terms of the half-space gaugings described in Appendices \ref{app:doublegauging} and \ref{app:Cinterfaces}.

The above duality-induced symmetry breaking mechanism   may fail if $\mathscr{G}$ is abelian, or, more generically, if   $\calr\in \mathscr{G}_\mathbb{Q}$ is invariant under $\mathscr{G}_\mathbb{Z}$ conjugation. The simplest example is provided by the axion-Mawell model discussed in Section \ref{sec:warmup}, in which  $\cald_\cals\times \cald_{\calr}=\cald_{\calr}\times \cald_\cals$, for any  $\cals\in\mathscr{G}_\mathbb{Z}\simeq \mathbb{Z}$ and any $\calr\in\mathscr{G}_\mathbb{Q}\simeq \mathbb{Q}$. One can then gauge $\mathscr{G}_\mathbb{Z}$, i.e.\ assume that the axion is  periodic as in \cite{Choi:2022jqy,Cordova:2022ieu,Choi:2022fgx}, without breaking the GZ non-invertible symmetries. However, this conclusion does not a priori apply to any abelian $\mathscr{G}$. Indeed, in general we have 
\be\label{abelianD}
\cald^{ -1}_\cals \times \cald_\calr \times \cald_\cals=\cald'_{\calr}\,,
\ee
where, for each $\cals\in\mathscr{G}_\mathbb{Z}$,   $\cald'_{\calr}$ may be  different from the $\cald_\calr$. This possibility is due to the degeneracy of non-invertible GZ defects associated with the same element of $ \mathscr{G}_\mathbb{Q}$.

The survival of the non-invertible symmetry in the $\mathscr{G}_\mathbb{Z}$-gauged theory appears strictly related to the existence of a unique or of a  finite number of ``minimal" topological GZ defects in the parent ungauged theory. Indeed, we expect that the $\mathscr{G}_\mathbb{Z}$ action \eqref{abelianD} preserves the defect minimality.  So, if  
 an element $\calr\in \mathscr{G}_\mathbb{Q}$ corresponds to a finite number of minimal GZ defects of the ungauged theory, 
finite linear combination thereof may give a gauge invariant topological defect,  which would hence survive in the $\mathscr{G}_\mathbb{Z}$ gauged theory. For instance, this is what we expect to happen in the class of models studied in Section \ref{sec:CYdefects}. Clearly, properly testing these qualitative expectations would require a better understanding of the minimal defects, which possibly requires a case-by-case study and is beyond the scope of this paper.

%%%%%%%%%%%%%%%%%%%%%%%%%%%%%%%%%%%%%%%%%%%%%%%%%%%%%%%%%%
\section{Conclusions}
\label{sec:conclusion}

In this paper we have shown that the classical symmetries described by  Gaillard-Zumino in \cite{Gaillard:1981rj}, which involve continuous electromagnetic  duality rotations, admit a quantum realization in terms of non-invertible operators. Our conclusion applies to any non-gravitational GZ model and any  GZ symmetry group $\mathscr{G}$. In absence of  mixed 't Hooft anomalies between the GZ symmetries and gravity, this conclusion can be immediately extended to gravitational GZ models as well.  More precisely,  we have shown that for any rational GZ symmetry $\cals\in\mathscr{G}_\mathbb{Q}\subset \mathscr{G}$ one can construct a corresponding topological defect $\cald_\cals$. This  defect is  non-invertible, unless $\cals$ belongs to the subgroup $\mathscr{G}_\mathbb{Z}\subset \mathscr{G}_\mathbb{Q}$ of integral GZ symmetries. Our general claims have been illustrated in some concrete GZ models. Other well-known examples of GZ models are provided by theories with extended supersymmetry, such as the maximal ungauged supergravity \cite{Cremmer:1979up,deWit:1982bul}. However, we emphasize that our results are completely independent of supersymmetry.   

Starting from the general results presented in this paper one may proceed in  various possible directions. 

\begin{itemize}

\item 
First of all, we have provided just a `proof of principle' of the existence of non-invertible GZ symmetries, emphasizing that the degeneracy of non-invertible GZ defects. Namely, the fact that each GZ transformation  $\cals\in\mathscr{G}_\mathbb{Q}$ corresponds to several, and potentially infinite, non-invertible defects  $\cald_\cals$. On the other hand, for specific transformations, namely $\sfA$ and $\sfC$ transformations, we showed that  there exist `minimal' defects  $\cald_\cals$, which  then play a distinguished role and reduce the degeneration.  
It would be extremely interesting to better understand whether this is true more generically, namely if there is some notion of minimality which holds for more general defects and which implies that, for any given $\cals\in\mathscr{G}_\mathbb{Q}$, there exists a finite number of minimal defects $\cald_\cals$, or maybe even a unique one.  

\item  We have already emphasized the relevance of the electric and magnetic one-form symmetries in the construction of the GZ defects. It would be interesting to more systematically investigate the interplay of our GZ non-invertible (zero-form) symmetries with other higher-form symmetries, as done for QED \cite{Choi:2022jqy} and for axion-Maxwell \cite{Choi:2022fgx}, and to understand their overall categorical structure \cite{Copetti:2023mcq,DelZotto:2024ngj}. Finally, it would be interesting to analyze the fate of GZ-symmetries once charged fermions are added.

\item An interface $\calw_\cals$ becomes a topological defect for theories with a decoupled neutral sector at a fixed point of the $\cals$ action. This observation is particularly interesting when the neutral sector includes a moduli space of scalar fields. At these fixed points, only a (discrete) subgroup of $\mathscr{G}_{\mathbb Q}$ survives. For instance, in the $n=1$ case the symmetries surviving at $\cals$ fixed points should correspond to the non-invertible defects studied in \cite{Niro:2022ctq, Cordova:2023ent, Hasan:2024aow, Paznokas:2025auw}. It would be interesting to understand this mechanism in more detail, and to clarify the physical interplay between the choice of a vacuum in the moduli space and the $\calw_\cals$ interfaces that survive as symmetry defects at the $\cals$ fixed point. 
A related question is to what extent the spontaneous breaking of the GZ non-invertible symmetries captures the structure of the neutral sector away from the fixed points, in the spirit of \cite{Arbalestrier:2024oqg,Arbalestrier:2025poq}, or alternatively as in \cite{GarciaEtxebarria:2022jky,Karasik:2022kkq}. 
Moreover, as studied in \cite{Bashmakov:2022uek,Antinucci:2022cdi}, non-Lagrangian $\mathcal N=2$ four-dimensional class S theories exhibit non-invertible symmetries forming subgroups of the symplectic group ${\rm Sp}(2n,\mathbb Q)$, and it is natural to ask  whether they  admit any relation with the non-invertible GZ symmetries studied in this paper.  

\item It is crucial to study the anomalies of the GZ symmetries we have constructed. Such anomalies would provide obstructions to gauging subgroups of $\mathscr{G}_{\mathbb Q}$, and may be relevant in the context of gauged supergravity theories.    Even more interestingly, there may exist mixed anomalies with the gravitational sector, of the type discussed in \cite{Witten:1995gf}.   If not curable by some cancellation  mechanism, for instance as the one at work on D3-branes \cite{Bachas:1999um} or along the lines of  \cite{Seiberg:2018ntt}, these anomalies would provide a purely field-theoretic gravitational symmetry breaking mechanism  of these non-invertible symmetries.
Anomalies of non-invertible symmetries are, however, more intricate to formulate than in the standard invertible case. Recent progress in this direction has been made in \cite{Antinucci:2025fjp, Antinucci:2023ezl, Cordova:2023bja}. A particularly powerful approach is the symmetry topological field theory (SymTFT) framework introduced in \cite{Ji:2019jhk, Pulmann:2019vrw, Gaiotto:2020iye, Apruzzi:2021nmk, Freed:2022qnc}, which has recently been generalized to include non-abelian continuous symmetries \cite{Bonetti:2024cjk, Jia:2025jmn, Apruzzi:2025hvs, Bonetti:2025dvm}. The SymTFT framework provides a systematic way to analyze non-invertible defects and their anomalies, and therefore may serve as a useful alternative viewpoint on our construction of GZ non-invertible symmetries.

\item In Section \ref{sec:gaugingSB} we outlined a simple mechanism that leads, in one shot,  to the gauging of  GZ invertible symmetries and the breaking of GZ non-invertible ones. In combination with the possible mixed gravitational anomalies mentioned above, this may further contribute to
the expected absence of global symmetries in quantum gravity models, without needing the explicit incorporation of UV ingredients. This
mechanism may provide a purely bottom-up motivation for the gauging of the   $\mathscr{G}_\mathbb{Z}$ subgroup. Of course, the possibility of realizing this mechanism depends on the potential  presence of (pure) 't Hooft anomalies for $\mathscr{G}_\mathbb{Z}$, of the kind e.g.\ studied  in \cite{Hsieh:2019iba}, which would obstruct the gauging of $\mathscr{G}_\mathbb{Z}$. In addition, the efficiency of this mechanism depends on the structure of the theory and of its GZ group $\mathscr{G}$.   It would be extremely interesting to explore the connection of this observation with the alternative bottom-up arguments for the existence of U-duality groups recently given in \cite{Delgado:2024skw}, and with the related results of \cite{Baines:2025upi}, which focuses on  models with symmetric moduli spaces and study in detail their  asymptotic limits. 
On the other hand, a world-sheet realization of our symmetries, along the lines of \cite{Bachas:2012bj}, would provide a useful top-down perspective thereon, which may also clarify a potential connection with the results of \cite{Kaidi:2024wio,Heckman:2024obe}.

\end{itemize}

%%%%%%%%%%%%%%%%%%%%%%%%%%%%%%%%%%%%%%%%%%%%%%%%%%%%%%%%%%

%\vspace{1cm}

%\centerline{\large\em Acknowledgments}

%\vspace{0.5cm}

\section*{Acknowledgments}

\noindent We thank Riccardo Argurio, Costas Bachas, Guillaume Bossard, Alessandra Gnecchi, Alessandra Grieco, Gianluca Inverso, Ho-Tat Lam, Raffaele Savelli, Luigi Tizzano, Irene Valenzuela and Roberto Volpato for useful discussions. In particular, we thank Riccardo Argurio and Costas Bachas for useful comments on the draft, and Alessandra Grieco and Irene Valenzuela for collaboration on a related project. We also thank Riccardo Argurio for sharing a preliminary version of \cite{Arbalestrier:2025jsg}. This work was supported in part by the Italian MUR Departments of Excellence grant
2023-2027 “Quantum Frontiers” and by the MUR-PRIN contract 2022YZ5BA2 - “Effective Quantum Gravity”. The work of FA is partially supported by the University of Padua under the 2023 STARS Grants@Unipd programme (GENSYMSTR – Generalized Symmetries from Strings and Branes) and in part also by the grant NSF PHY-2309135 to the Kavli Institute for Theoretical Physics (KITP). 

%%%%%%%%%%%%%%%%%%%%%%%%%%%%%%%%%%%%%%%%%%%%%%%%%%%%%%%%%%%%%%%%%%%%%%%%%%%%%%%%%%%%%%%%%%%

\newpage

\appendix

\section{Two-derivative models}
\label{app:twoderiv}

In this Appendix we review the realization of the GZ conditions  described in Section \ref{sec:GZ} in two-derivative models including only $n$ vectors $A^I$ and  a set of neutral massless scalars $\phi^i$ parametrizing a moduli space $\calm$.
Important  examples of these  GZ models are provided by coset models, which are discussed in some detail in \cite{Gaillard:1981rj} and for instance appear in extended supergravities.

The most general two-derivative Lagrangian for such a field content takes the form  
\be\label{gen2derL}
\begin{aligned} 
\call&=\frac1{4\pi} \Im\caln_{IJ}(\phi)F^I\wedge *F^J+\frac1{4\pi}\Re\caln_{IJ}(\phi)F^I\wedge F^J-\frac12\calg_{ij}(\phi)\d\phi^i\wedge *\d\phi^j\\
&=\frac1{2\pi}\Re\left[\caln_{IJ}(\phi)F^I_+\wedge F^J_+\right]-\frac12\calg_{ij}(\phi)\d\phi^i\wedge *\d\phi^j\,,
\end{aligned}
\ee
where we have grouped the field dependent gauge couplings and theta terms in the complex symmetric matrix  $\caln_{IJ}(\phi)=\caln_{JI}(\phi)$, and we have introduced the imaginary self-dual field strengths
\be 
F_+^I\equiv \frac12(F^I-\ii*F^I)\quad~~~~\Rightarrow \quad~~~~ *F_+^I=\ii F_+^I\,.
\ee
The $\tau$-Maxwell theory of Section \ref{sec:tauMaxwell} is a particular case with $n=1$ and $\caln=-\tau$. 

The dual field strengths $G_I$ as defined in \eqref{G_I} are given by 
\be\label{linearG_I} 
G_I=\Im\caln_{IJ}*F^J+\Re\caln_{IJ}F^J\,.
\ee
It may also be useful to rewrite this equation in terms of the  imaginary self-dual combinations $G^+_I\equiv \frac12(G_I-\ii*G_I)$:
\be\label{G+rel} 
G^+_I=\caln_{IJ}(\phi)F^J_+\,.
\ee

The Lagrangian $ \call_\cals$ appearing  in \eqref{Lvariation2} is more easily obtained by imposing that the form of \eqref{G+rel} is preserved under a general symplectic transformation \eqref{gentrb}. Namely, we should have   $\tilde G^+_I=\tilde\caln_{IJ}(\phi)\tilde F^J_+$ for some $\tilde\caln_{IJ}(\phi)$.  Decomposing $\cals$  as in \eqref{genS} and imposing \eqref{gentrb} we get the identity
\be\label{calntr} 
\tilde\caln_{IJ}(\phi)=\big(\big[\mathsf{C}+\mathsf{D}\caln(\phi)\big]\big[\mathsf{A}+\mathsf{B}\caln(\phi)\big]^{-1}\big)_{IJ}\,.
\ee
The Lagrangian $ \call_\cals(\tilde F,\phi)$ is then obtained by replacing $\caln_{IJ}(\phi)$ and $F^I$ with  $\tilde\caln_{IJ}(\phi)$ and $\tilde F$, respectively, in \eqref{gen2derL}. Notice that the condition \eqref{Lvariation2} is trivially satisfied, because of 
\be 
\Re\left[\caln_{IJ}(\phi)F^I_+\wedge F^J_+\right]=\Re\left(F^I_+\wedge G_I^+\right)=\frac12F^I\wedge G_I\,, 
\ee
and of the analogous identity with tilded quantities. Hence
\be\label{2derLid} 
\call(F,\phi)-\frac1{4\pi}F^I\wedge G_I=\call_\cals(\tilde F,\phi)-\frac1{4\pi}\tilde F^I\wedge \tilde G_I=-\frac12\calg_{ij}(\phi)\d\phi^i\wedge *\d\phi^j\,.
\ee

So far, we have not imposed any non-trivial GZ restriction yet. Imposing \eqref{tildecall} we conclude that, if $\cals\in\mathscr{G}$, $f_\cals(\phi)$ must define an isometry of the moduli space  metric $\d s^2_\calm \equiv \calg_{ij}(\phi)\d\phi^i\d\phi^j$ and $\caln_{IJ}$ must transform  as follows:
\be\label{GZcond2der} 
\caln_{IJ}(f_\cals(\phi))=\tilde\caln_{IJ}(\phi)\,.
\ee
For instance, in the $\tau$-Maxwell theory this condition  corresponds to \eqref{adtautr}.

 In terms of the infinitesimal deformations \eqref{delTphi}, $\xi_T(\phi)$ is a Killing vector of $\d s^2_\calm$. It acts on $\caln_{IJ}$ as follows
 \be\label{infcaln} 
 \delta_T\caln_{IJ}(\phi)=\xi_T^i(\phi)\del_i\caln_{IJ}(\phi)=\left(W+Y\caln -\caln U-\caln V\caln\right)_{IJ}\,,
 \ee
 where we used the decomposition \eqref{Tsplit} of $T$. The corresponding current defined by \eqref{defJ}  is 
 \be\label{ggcurr} 
 \calj_T=-g_{ij}(\phi)\xi^i(\phi)*\d\phi^j\,.
 \ee
 By using the $\phi^i$ equations of motion 
 \be\label{genphiEoM}  E_i=\calg_{ij}D*\d\phi^j+\frac1{2\pi}\Re\left[\del_i\caln_{IJ}(\phi)F^I_+\wedge F^J_+\right]=0\,,
 \ee
 with $ D*\d\phi^i\equiv\d*\d\phi^i+\Gamma^i{}_{jk}(\phi)\d\phi^j\wedge *\d\phi^j$, one can then check the general identity \eqref{dJnon0}.

\section{Integral symplectic transformations  and interfaces}
\label{app:Spdualities}

In this Appendix we review how the integral symplectic group ${\rm Sp}(2n,\mathbb{Z})$ acts on GZ Lagrangians and we construct the corresponding interfaces \eqref{WSZ}, following \cite{Ganor:1996pe,Gaiotto:2008ak,Kapustin:2009av}. 
 The general element of ${\rm Sp}(2n,\mathbb{Z})$ can be decomposed as the product of integral powers of \eqref{Sfactor}, with $\sfA\in {\rm GL}(n,\mathbb{Z})$ and $\sfC=\sfC^{\rm t}\in{\rm Mat}(n,\mathbb{Z})$ \cite{mumford2007tata}.  Thus, we can focus on these three cases,  as  more general $\calw_\cals$ interfaces with $\cals\in {\rm Sp}(2n,\mathbb{Z})$ can be obtained by concatenating these $\sfA$, $\sfC$ and $\Omega$ interfaces and their inverses. By adopting the same notation introduced in Section \ref{sec:GZ2}, we will denote  the corresponding transformed Lagrangians  as $\call_\sfA\equiv \call_{\cals_{\mathsf{A}}}$, $\call_\sfC\equiv \call_{\cals_{\mathsf{C}}}$ and $\call_\Omega\equiv \call_{\cals_{\Omega}}$, respectively.

\subsection{{\sf A} transformations and interfaces}

An  ${\sf A}\in {\rm GL}(n,\mathbb{Z})$ transformation amounts to just the reshuffle of the U(1) gauge fields
\be\label{Amix} 
\tilde F^I=\sfA^I{}_J F^J\,.
\ee 
 The new Lagrangian $\call_\sfA$ is then simply defined by $\call_\sfA(\tilde F,\phi)=\call(F,\phi)$ and is clearly physically equivalent to $\call(F,\phi)$, provided the gauge fields are related as in \eqref{Amix}. 

The $\calw_\sfA(\Sigma)$  interface, connecting two such equivalent descriptions as in \eqref{WSpicture}, must only implement the correct gluing conditions. It is easy to see that this requirement is realized by the insertion of the interface
\be\label{Awall}  
\calw_{\mathsf{A}}(\Sigma)=\int\cald b\exp\left[\frac{\ii}{2\pi}\oint_\Sigma b_I\wedge (F^I_{\text{\tiny L}}-\mathsf{A}^I{}_JF^J_{\text{\tiny R}})\right]\,, 
\ee 
where $b_I$ are U(1) world-volume connections on $W$. Indeed, integrating out $b_I$ one gets the gluing conditions $F^I_{\text{\tiny L}}|_\Sigma=\mathsf{A}^I{}_JF^J_{\text{\tiny R}}|_\Sigma$. On the other hand, the infinitesimal variation of the gauge fields $A^I_\tL$ and $A^I_\tR$ produces a term localized on $\Sigma$, which vanishes only if   $G_{\tL\,I}|_\Sigma=-\d b_I$ and $(\sfA^{-1})^J{}_IG_{\tR\,J}|_\Sigma=-\d b_I$, hence realising the appropriate gluing condition $G_{\tL\,I}|_\Sigma=(\sfA^{-1})^J{}_IG_{\tR\,J}|_\Sigma$.

\subsection{{\sf C} transformations and interfaces}

Consider now an integral $\sfC$ transform of the Lagrangian. This is defined by 
\be\label{callC} 
\call_\sfC(F,\phi)=\call(F,\phi)+\frac{1}{4\pi}\sfC_{IJ}  F^I\wedge F^J\,.
\ee
Note that the corresponding symplectic transformation does not act on the field strengths $F^I$, in the sense that $\tilde F^I=F^I$, while it does transform the duals field strengths: $\tilde G_I=G_I+\sfC_{IJ}F^J$.  
Since we assume that our four-dimensional manifolds are spin, the last term in \eqref{callC} is always an integral multiple of $2\pi$ and hence, again, does not change the  physical properties of the theory.

The corresponding interface $\calw_\sfC(\Sigma)$ acting as in \eqref{WSpicture} can be obtained first using $\call(F,\phi)$ on both sides of $\Sigma$, and then rewriting $\call(F_\tL,\phi)$ as follows
\be 
\call(F_\tL,\phi)=\call_{\sfC}(F_\tL,\phi)- \frac{1}{4\pi}\sfC_{IJ} F^I_\tL\wedge F^J_\tL\,.
\ee
The last term can then be rewritten as $-\frac{1}{4\pi}\sfC_{IJ}\oint_\Sigma A^I_\tL\wedge F^J_\tL$. Adding a delta-function factor that imposes $F^I_\tL|_\Sigma=F^I_\tR|_\Sigma$,  we get the interface operator \eqref{WSZc}, namely
\be 
\calw_{\mathsf{C}}(\Sigma)=\exp\left(-\frac{\ii}{4\pi}\mathsf{C}_{IJ}\oint_\Sigma A^I_{\text{\tiny R}}\wedge F^J_{\text{\tiny R}}\right) \int\cald b\,\exp\left[\frac{\ii}{2\pi}\oint b_I\wedge (F^I_{\text{\tiny L}}-F^I_{\text{\tiny R}})\right]\,.
\ee
As a further check, extremizing with respect   $A^I_\tL$ and $A^I_\tR$ produces the localised contributions  $G_{\tL\,I}|_\Sigma=-\d b_I$ and $G_{\tR\,J}|_\Sigma+\sfC_{IJ}F^J_\tR|_\Sigma=-\d b_I$, hence realizing  $G_{\tL\,I}|_\Sigma=G_{\tR\,I}|_\Sigma+\sfC_{IJ}F^J_\tR|_\Sigma$.

\subsection{$\Omega$ transformations and interfaces}
\label{app:Omegaduality}

An $\Omega$ transformation corresponds to a simultaneous S-duality on all U(1) gauge fields.  Following \cite{Witten:1995gf,Gaiotto:2008ak}, we now explain why it produces a physically equivalent theory and construct the corresponding interface.

As in \cite{Witten:1995gf},  we can extend our elementary degrees of freedom without modifying the theory as follows.  Introduce two new types of fields:  a set of U(1) two-form potentials  $B^I$ and a new set  
of U(1) one-form potentials $\tilde A^I$,  with field strengths $\tilde F^I=\d \tilde A^I$. Furthermore we impose that the two-form potentials  $B^I$ gauge the $n$ electric one-form symmetries of our generic GZ model. Namely, we impose   invariance under the one-form gauge symmetry
\be\label{extgauge} 
\calb^I\rightarrow \calb^I+2\pi\Lambda^I\quad,\quad F^I \rightarrow F^I-2\pi\Lambda^I\,,
\ee
where  $\Lambda^I$ are arbitrary closed two-forms that identify integral cohomology classes,  $\Lambda^I\in H^2(X,\mathbb{Z})$. Of course, we also impose invariance under  the ordinary small and large gauge transformations  of $A^I$ and  $\tilde A^I$. The GZ Lagrangian $\call(F,\phi)$ is then replaced by 
\be\label{SdualB} 
\call(\calf,\phi)-\frac{1}{2\pi}\delta_{IJ} \calf^I\wedge \tilde F^J\,,
\ee
with 
 \be 
 \calf^I\equiv F^I+B^I\,.
 \ee

In order to show that this Lagrangian describes the same physics of the original one, let us first integrate out the gauge fields $\tilde A^I$. This generates a delta-function that imposes  $\d B^I=0$ and the cohomological condition  $B^I\in 2\pi H^2(X,\mathbb{Z})$. Given this restriction, one can  gauge-fix \eqref{extgauge} by setting $B^I\equiv 0$ (or equivalently setting $F^I\equiv0$ and renaming $B^I\rightarrow F^I$), getting back the original GZ Lagrangian $\call(F,\phi)$. Note also that the equations of motion of $B^I$ impose the relations  
\be\label{FDeq} 
\tilde F^I=2\pi\delta^{IJ}\frac{\del\call(\calf,\phi,\del\phi)}{\del\calf^J}\,.
\ee
In the gauge $B^I\equiv 0$, in which we get back the original theory, this reduces to  $\tilde F^I=\delta^{IJ}G_J$, with $G_I$ defined as in \eqref{G_I}.

In order to get the dual description, we choose to fix the one-form gauge symmetry  \eqref{extgauge} by imposing $F^I\equiv 0$, so that $B^I=\calf^I$. The equations of motion of  $B^I$ then give the relation \eqref{FDeq}.  One can in principle invert \eqref{FDeq} to get $\calf$ in terms of $\tilde F^I$ and $\phi^i$ (and $\del_\mu\phi^i$). By integrating out $B^I$ one then arrives at the S-dual Lagrangian\footnote{This step is straightforwardly justified only if the integration is Gaussian, and hence the Lagrangian $\call(F,\phi)$ is at most quadratic in $\calf$. For more general theories it is still valid at the classical level, and we implicitly assume that its validity holds at the  quantum level too, at least within the effective field theory description. Furthermore, a more  careful dualization may involve anomalous $\phi$ dependent factors as in \cite{Witten:1995gf}, originating from  a mixed 't Hooft  anomaly with the gravitational sector. }  
\be\label{Omegalag} 
\call_{\Omega}(\tilde F,\phi)=\left[\call(F,\phi)-\frac{1}{2\pi}\delta_{IJ}F^I\wedge \tilde F^J\right]_{ F^I={\calf^I(\tilde F)}}\,.
\ee
By performing a further S-duality, we get the identification $\tilde G_I=-F^I$, so that in \eqref{Omegalag} we can make the identification  $\calf^I(\tilde F)=-\delta^{IJ}\tilde G_J(\tilde F)$.

Now, the interface $\calw_\Omega(\Sigma)$ implementing an $\Omega$-transformation can be obtained by performing the above procedure only  on one  side of $\Sigma$  \cite{Gaiotto:2008ak}. In other words, it can be obtained from a  `half-space gauging' of the electric one-form symmetries. On the right  side of $\Sigma$, let us rename $A^I$ as  $A^I_\tR$ and use the original theory $\call(F_\tR,\phi)$, while on the left side we want to use \eqref{SdualB}, renaming $A^I$ and $\tilde A^I$ as $\hat A^I$ and  $A^I_\tL$, respectively.  Furthermore,  we also need  to enable the gluing condition $\calf^I|_\Sigma=F^I_\tR|_\Sigma$, which in the gauge $B^I\equiv 0$ reduces to $\hat F^I|_\Sigma=F^I_\tR|_\Sigma$. As will become clear shortly, this requires the addition of a boundary term $\frac1{2\pi}\delta_{IJ}\oint_\Sigma A^I_\tL\wedge F^J_\tR$, so that   the total action on the left-hand side  becomes
\be\label{Sdualbound} 
\int_{\calt_\tL} \call(\calf,\phi)-\frac1{2\pi}\delta_{IJ}\int_{\calt_\tL}\calf^I\wedge F^J_\tL+\frac1{2\pi}\delta_{IJ}\oint_\Sigma A^I_\tL\wedge F^J_\tR\,,
\ee
with $\calf^I\equiv \hat F^I+\calb^I$.
If we try to rerun the above argument and integrate out $A^I_\tL$, we get a delta-function that localizes the path-integral on   $B^I$ that satisfy  $\d B^I=0$, $B^I\in 2\pi H^2(\Sigma,\mathbb{Z})$ and  $\calf^I|_\Sigma=F^I_\tR|_\Sigma$.  
Imposing the gauge-fixing  $B^I\equiv  0$ we indeed  go back to the original description on the entire spacetime. 
Instead, fixing the gauge $\hat F^I\equiv 0$ and integrating out $B^I$ on the left-hand side, we obtain  the dual Lagrangian plus a boundary term 
\be\label{Sdualbound1}
\int_{\calt_\tR} \call_{\Omega}(F_\tL,\phi)+\frac1{2\pi}\delta_{IJ}\oint_\Sigma A^I_\tL\wedge F^J_\tR\,. 
\ee
The boundary term gives the $\Omega$ interface already written in \eqref{WSZa}:
\be 
\calw_\Omega(\Sigma)=\exp\left(\frac\ii{2\pi}\delta_{IJ}\oint_\Sigma A^I_\tL\wedge F^J_\tR\right)\,,
\ee
which realizes \eqref{WSpicture} with $\cals=\Omega$. This can be more directly  checked, since in presence of $\calw_\Omega(\Sigma)$  the bulk $A^I_\tL$ and $A^I_\tR$ equations of motion get the following boundary contribution: 
\be 
F^I_\tL|_\Sigma =\delta^{IJ}G_{\tR\,J}|_\Sigma \quad,\quad  G_{\tL\,I}|_\Sigma=-\delta_{IJ}F^J_\tR|_\Sigma\,.
\ee
These are indeed  the correct gluing conditions.

%%%%%%%%%%%%%%%%%%%%%%%%%%%%%%%%%%%%%%%%%%%%%%%

\section{Integral coprime matrix factorizations}
\label{app:coprime}

In this appendix we collect some definitions and properties regarding the factorization of rational matrices in terms of integral matrices. More details can be found in  \cite{Vidyasagar2011}, which works with matrices taking values in a general principal ideal domain and the corresponding field of fractions, while here we focus on $\mathbb{Z}$ and $\mathbb{Q}$, respectively.  

Any $n\times n$ rational  matrix $\sfQ\in {\rm Mat}(n,\mathbb{Q})$ can be factorized in terms of pairs $(\sfM,\sfN)$ and $(\tilde\sfM,\tilde\sfN)$ of  {\em integral} matrices, with invertible $\sfN$ and $\tilde\sfN$, as follows:
\be 
\sfQ= \sfM \sfN^{-1}=\tilde\sfN^{-1}\tilde\sfM\,.
\ee
The factorization defined by $(\sfM,\sfN)$ is  {\em right-coprime}  if there exist two integral matrices $\sfX,\sfY\in {\rm Mat}(n,\mathbb{Z})$ such that:
\be\label{Rcop}
\sfX\sfM-\sfY\sfN=\mathds{1}\,,
\ee
while the factorization defined by $(\tilde\sfM,\tilde\sfN)$ is  {\em left-coprime} if there exist two integral matrices $\tilde\sfX,\tilde\sfY\in {\rm Mat}(n,\mathbb{Z})$ such that:
\be\label{Lcop}
\tilde\sfM\tilde\sfX-\tilde\sfN\tilde\sfY=\mathds{1}\,.
\ee
Furthermore, any rational matrix $\sfQ\in {\rm Mat}(n,\mathbb{Q})$ admits both a   right-coprime and  left-coprime factorization. 

A right-coprime factorization $\sfQ= \sfM \sfN^{-1}$ is also {\em weakly}  right-coprime \cite{quadrat2003fractional}, that is, such that
\be\label{Rcoprime} 
{\bm\alpha}\in \mathbb{Q}^n\quad\text{and}\quad \sfM {\bm\alpha},\sfN{\bm\alpha}\in \mathbb{Z}^n\quad\Leftrightarrow\quad {\bm\alpha}\in \mathbb{Z}^n\,.
\ee
Similarly, a left-coprime factorization $\sfP= \tilde\sfN^{-1}\tilde\sfM $ is also  weakly  left-coprime, namely:
\be\label{lcoprime} 
{\bm\alpha}\in \mathbb{Q}^n\quad\text{and}\quad \tilde\sfM^{\rm t} {\bm\alpha},\tilde\sfN^{\rm t}{\bm\alpha} \in \mathbb{Z}^n\quad\Leftrightarrow\quad {\bm\alpha}\in \mathbb{Z}^n\,.
\ee
Take for instance a right-coprime factorization $\sfQ= \sfM \sfN^{-1}$.  Assuming  $\sfM {\bm\alpha},\sfN{\bm\alpha}\in \mathbb{Z}^n$, the identity \eqref{Rcop}  implies that ${\bm\alpha}=(\sfX\sfM-\sfY\sfN){\bm\alpha}\in\mathbb{Z}$. Therefore, the factorization is also weakly right-coprime.

Coprime factorizations are, in a certain sense, the `minimal' possible factorizations. In order to explain this point, we can focus on right-coprime factorizations, as the argument for left-coprime ones is completely analogous. 
Suppose that $\sfQ= \sfM \sfN^{-1}$  is a right-coprime factorization and take another (not necessarily right-coprime) factorization $\sfQ= \sfM' \sfN'^{-1}$. Then 
\be 
\sfM'=\sfM\sfT\quad,\quad \sfN'=\sfN\sfT\,,
\ee
with $\sfT=\sfN^{-1}\sfN'$, $\det\sfT\neq 0$. Take  now any integral vector ${\bm m}\in \mathbb{Z}^n$. Since $\sfM'$ and $\sfN'$ are integral matrices, $\sfM'{\bm m}=\sfM\sfT{\bm m}\in \mathbb{Z}^n$ and $\sfN'{\bm m}=\sfN\sfT{\bm m}\in \mathbb{Z}^n$. By \eqref{Rcoprime} we then deduce that $\sfT{\bm m}\in \mathbb{Z}^n$. Since  ${\bm m}\in \mathbb{Z}^n$ is arbitrary, we conclude that 
\be
\sfT\in {\rm Mat}(n,\mathbb{Z})\,.
\ee 
If $\sfM'$ and $\sfN'$ are right-coprime too, then we can apply the same  argument to argue that $\sfT^{-1}\in {\rm Mat}(n,\mathbb{Z})$, and hence $|\det\sfT|=1$, that is 
\be 
\sfQ=\sfM'\sfN'^{-1} \quad\text{right-coprime}\quad \Leftrightarrow \quad \sfT\in{\rm GL}(n,\mathbb{Z})\,.
\ee

Similar conclusions hold for left-coprime factorizations $\sfQ=\tilde\sfN^{-1}\tilde\sfM=\tilde\sfN'^{-1}\tilde\sfM'$. If $\sfQ=\tilde\sfN^{-1}\tilde\sfM$ is left-coprime, then $\tilde\sfM'=\sfT\tilde\sfM$ and $\tilde\sfN'=\sfT\tilde\sfN$ with $\sfT\in {\rm Mat}(n,\mathbb{Z})$. Furthermore, $\sfQ=\tilde\sfN'^{-1}\tilde\sfM'$ is left-coprime too if (and only if) $\sfT\in {\rm GL}(n,\mathbb{Z})$. 

These results imply that  right- or left-coprime factorizations are unique up to unimodular transformations. Notice that, in order to show this property, we have only used  the weak coprimitivity conditions \eqref{Rcoprime} and \eqref{lcoprime}.
So, the above argument  also proves that any weakly right/left-coprime factorization is related to any right/left-coprime factorization  by an unimodular transformations. Since right/left-coprime factorizations of rational matrices always exist, we conclude that any  weakly right/left-coprime factorization is in fact right/left-coprime.   Hence,  one could alternatively use \eqref{Rcoprime} and \eqref{lcoprime} to characterize  coprime factorizations of rational matrices.

%%%%%%%%%%%%%%%%%%%%%%%%%%%%%%%%%%%%%%%%%%%%%%%

\section{{\sf A} interfaces from half-space gauging}
\label{app:doublegauging}

In this appendix we discuss how the $\sfA$ interfaces introduced in Section \ref{sec:GZAdefects} can be obtained from a half-space gauging procedure, along the lines of \cite{Choi:2021kmx}. In general, any $\sfA$ matrix admits an integral factorization as in \eqref{AMEdec}. We will  separately consider  the cases $\sfE=\mathds{1}$ and $\sfM=\mathds{1}$ in \eqref{AMEdec}, which will correspond to magnetic and electric half-space gaugings, respectively.  From \eqref{WAfact}, it follows that a more general $\sfA$ interface can be obtained by an ordered  sequence of half-space gaugings. We emphasize that, in general, these half-space gaugings cannot be performed simultaneously, because of the mixed anomaly between electric and magnetic one-form symmetries \cite{Gaiotto:2008ak}.

\subsection{Interfaces from magnetic one-form half-space gauging}
\label{app:Mhalfgaug}

We first assume that $\sfE=\mathds{1}$, i.e.\ $\sfA=\sfM$. In this case the $\sfA$ interface can be obtained by  following the half-space gauging prescription outlined in   \cite{Choi:2021kmx}. 

We first observe that new theory is obtained from the gauging of the finite  subgroup of the magnetic U(1)$^n$ one-form symmetry group of the theory. The new Lagrangian can be written as
\be\label{LMB}
\call(F,\phi)+ \frac1{2\pi} F^I\wedge B_I-\frac1{2\pi}\sfM^I{}_J\tilde F^J \wedge B_I\,,
\ee
where we have introduced the U(1) two-form potentials  $B_I$ and field strengths $\tilde F^I=\d \tilde A^I$.
The last term imposes $\sfM^J{}_I\d B_J=0$ and the cohomological condition $\sfM^J{}_I B_J\in  2\pi H^2(X,\mathbb{Z})$. This means that, adopting the notation introduced in Section \ref{sec:GZdefects2}, the two-form potentials $B_I$ identify an element of $H^2(X,\Gamma^*_\sfM)$. Hence \eqref{LMB} implements a gauging of the finite $\Gamma^{*(1)}_\sfM$ one-form symmetry group. By integrating out $B_I$ we get the condition $F^I=\sfM^I{}_J \tilde F^J$, and the Lagrangian reduces to
$\tilde\call(\tilde F,\phi)=\call(\sfM \tilde F,\phi)$,
which coincides with $\call_{\sfA^{-1}}(\tilde F,\phi)$, since $\sfA=\sfM$ and hence $\sfA^{-1}=\sfM^{-1}$ -- see \eqref{LSA}.

Let us now introduce an oriented three-dimensional surface $\Sigma$ and assume that it divides the spacetime $X$ in two parts, $X_\tL$ ad $X_\tR$. We would like to repeat the above gauging only on $X_\tR$. First of all, let us rename $A^I\rightarrow \hat A^I$ and $\tilde A^I\rightarrow A^I_\tR$ on $X_\tR$, and $A^I\rightarrow A^I_\tL$ on $X_\tL$. We also impose the boundary conditions
\begin{equation}
    B_I|_\Sigma=0, \qquad \hat A^I|_\Sigma=A^I_\tL|_\Sigma\, ,
\end{equation} 
enforcing the second one by adding an appropriate boundary term. The resulting action is
\be 
\int_{X_\tL}\call(F_\tL,\phi)+\frac1{2\pi}\oint_\Sigma b_I\wedge\left(F^I_\tL-\hat F^I\right)+\frac1{2\pi}\int_{X_\tL}B_I\wedge\left(\hat F^I-\sfM^I{}_JF^J_\tR\right)+\int_{X_\tR}\call(\hat F,\phi)\,.
\ee
This action describes a half-space gauging of the finite $\Gamma^{*(1)}_\sfM$ one-form symmetry group. By integrating out $B_I$ as above, we then get
\be 
\int_{X_\tL}\call( F_\tL,\phi)+\frac1{2\pi}\oint_\Sigma b_I\wedge\left( F^I_\tL-\sfM^I{}_J F^J_\tR\right)+\int_{X_\tR}\call_{\sfA^{-1}}(F_\tR,\phi)\,,
\ee
with $\sfA=\sfM$. We have then  obtained an $\sfA$ interface \eqref{WA} that separates $\call$ on the left from $\call_{\sfA^{-1}}$ on the right. Had we started with $\call_\sfA$ instead of $\call$, one would have obtained the same interface, but now separating $\call_\sfA$ on the left from $\call$ on the right.

\subsection{{\sf A} interfaces from electric one-form half-space gauging}
\label{app:Ehalfgaug}

 Let us consider now the case $\sfM=\mathds{1}$ in \eqref{AMEdec}, so that $\sfA=\sfE^{-1}$. In this case one must follow a modification of the  half-space gauging procedure outlined in \cite{Choi:2021kmx}.  We start from the Lagrangian
\be\label{EmodL}
\call(\calf,\phi)+\frac{1}{2\pi}\sfE^I{}_J F'_I\wedge \calf^J\,,\quad \text{with}\quad\calf^I\equiv F^I+B^I\,,
\ee
where $B^I$ are two-form gauge potentials and $  F'_I=\d A'_I$ are U(1) field strengths. \eqref{EmodL} may be considered as a variant of \eqref{SdualB} and combines 
 the electric $U(1)$ one-form gauging $F^I\rightarrow \calf^I=F^I+B^I$, together with a BF-term which restricts this one-form gauging to a finite subgroup. Indeed, integrating out $  A'_I$ one gets a delta-function that localizes the $B^I$ path integral over the solutions of    
\be\label{EBcoho} 
\sfE^I{}_J\d B^J=0\quad,\quad \sfE^I{}_J[B^J]\in 2\pi H^2(X,\mathbb{Z})\,.
\ee 
In other words, using the notation of Section \ref{sec:GZAdefects}, the two-form potentials $B^I$ are flat and identify a class of  $H^2(X,\Gamma_\sfE)$. Hence \eqref{EmodL} implements the gauging of the finite  $\Gamma^{(1)}_\sfE$ subgroup  of the ${\rm U}(1)^n$ electric one-form symmetry group.\footnote{From this perspective, \eqref{SdualB} implements a gauging under a trivial $\Gamma^{(1)}_\sfE$ one-form symmetry group.} We cannot  impose the gauge-fixing condition $B^I\equiv 0$, as  in Appendix \ref{app:Omegaduality}. However, we can still impose the gauge-fixing condition $F^I\equiv 0$ and solve \eqref{EBcoho} by setting $B^I=(\sfE^{-1})^I{}_J\tilde F^J$, for some U(1) gauge fields $\tilde A^I$. This leads  to a new Lagrangian $\tilde\call$ such that
\be 
\tilde\call(\tilde F,\phi)=\call(\sfE^{-1} \tilde F,\phi)\,.
\ee
That is, $\tilde\call$  is just $\call_{\sfA^{-1}}$ as defined in \eqref{LSA} with $\sfA^{-1}=\sfE$. This shows that $\call_{\sfA^{-1}}$ with $\sfA^{-1}=\sfE$ can be obtained from an electric  $\Gamma^{(1)}_\sfE$ one-form gauging. 

As in Appendix \ref{app:Mhalfgaug}, we would now like to obtain the corresponding $\calw_\sfA(\Sigma)$ by performing a half-space gauging,  on right side of $\Sigma$, renaming   $A^I\rightarrow \hat A^I$ and $\tilde A^I\rightarrow A^I_\tR$ on $X_\tR$, and $A^I\rightarrow A^I_\tL$ on $X_\tL$.  We  implement the half-space gauging  by using $\call(F_\tL,\phi)$ on $X_\tL$, and \eqref{EmodL} with $\calf^I=\hat F^I+B^I$ on $X_\tR$. However, differently from what we did in Appendix \ref{app:Mhalfgaug}  following \cite{Choi:2021kmx}, here we do not impose the $B^I|_\Sigma=0$ and $\hat A^I|_\Sigma=A^I_\tL|_\Sigma$ separately. Rather, together with \eqref{EBcoho} the following boundary condition is satisfied
\begin{equation}
    \calf^I|_\Sigma=F^I_\tL|_\Sigma\, .
\end{equation}  
We realize this condition by introducing new boundary U(1) one-form potentials $b_I$ and $c^I$ and adding  the boundary terms 
\be\label{EmodL1}
\frac{1}{2\pi}\sfE^I{}_J\oint_\Sigma b_I\wedge F^J_\tL-\frac1{2\pi}\oint_\Sigma c^I\wedge \left(\d b_I+F'_I\right)\,. 
\ee
Integrating out $A'_I$ one now  gets, in addition to \eqref{EBcoho}, the condition 
\be 
\d c^I=\sfE^I{}_J\calf^J|_\Sigma\,.
\ee 
Implementing these constraints, the second term in \eqref{EmodL} and \eqref{EmodL1} together reduce to 
\be\label{EmodL2}
\frac1{2\pi}\sfE^I{}_J\oint_\Sigma b_I \wedge (F^J_\tL-\calf^J)\,.
\ee
This enforces the required boundary conditions. 

 Finally, as in absence of boundaries, we can  fix the gauge $\hat F^I=0$  and solve \eqref{EBcoho} by setting $B^I=(\sfE^{-1})^I{}_JF^J_\tR$. On $X_\tR$ we get, as above,  the Lagrangian $\call_{\sfA^{-1}}(F_\tR,\phi)$ with $\sfA^{-1}=\sfE$, while \eqref{EmodL2} becomes
\be\label{EmodL2p}
\frac1{2\pi}\oint_\Sigma b_I \wedge (\sfE^I{}_JF^J_\tL-F^I_\tR)\,.
\ee
We have thus  obtained  our interface \eqref{WA} with $\sfM=\mathds{1}$, i.e.\ $\sfA=\sfE^{-1}$, separating $\call$ on the left from $\call_{\sfA^{-1}}$ on the right. Repeating the argument by starting with $\call_{\sfA}$ rather than $\call$, one obtains the same interface, but now separating $\call_{\sfA}$ on the left and  $\call$ on the right. 

Finally we observe that, since  ${\calw}^{(\sfE,\mathbbm{1})}={\overline\calw}^{(\mathbbm{1},\sfE)}$, we can alternatively obtain it by performing on {\em left} of $\Sigma$ a {\em magnetic} half-space gauging of the same kind described in Appendix  \ref{app:Mhalfgaug}, using  $\sfE^I{}_J$ instead of $\sfM^I{}_J$. Similarly, ${\calw}^{(\mathbbm{1},\sfM)}={\overline\calw}^{(\sfM,\mathbbm{1})}$
can also be obtained by performing on left of $\Sigma$ an  electric half-space gauging of the same kind described in this appendix, using  $\sfM^I{}_J$ instead of $\sfE^I{}_J$.

%%%%%%%%%%%%%%%%%%%%%%%%%%%%%%%%%%%%%%%%%%%%%%%

\section{Bulk origin of minimal TQFTs and {\sf C} interfaces}
 \label{app:Cinterfaces}

 In this appendix we show how the  $\sfC$ interfaces of Section  \ref{sec:GZCdefects} can be obtained from a half-space gauging procedure. 
 In order to do that, we will first provide a bulk formulation of the minimal $\cala_{\sfC}$  and non-minimal $\calz_{\sfP,\sfN}$ TQFTs supported by these interfaces.

\subsection{Bulk realization of the minimal TQFT}
\label{app:bulkmin}

In Section \ref{sec:minimalC} we have discussed how one can associate a minimal TQFT, denoted as $\cala_\Sigma^{(\sfN,\sfP)}$, to any choice of $\sfC_{IJ}$ mod $\mathbb{Z}$. In this appendix  we discuss how $\cala_\Sigma^{(\sfN,\sfP)}$ can be realized on the boundary of a four-dimensional topological theory. In the case $n=1$, this theory has already been discussed in \cite{Kapustin:2014gua}, Appendix B of \cite{Gaiotto:2014kfa} and Appendix E of  \cite{Hsin:2018vcg}.

 Let us pick a right-coprime integral factorization \eqref{Cfact} and a four-dimensional space $Y$ such that $\Sigma=\del Y$. Consider the continuum description of \eqref{Banomalypol}. We promote $\calb^I$ to dynamical two-form potentials $B^I$ obeying Dirichlet boundary conditions $B^I|_\Sigma=0$. One gets the action 
 \be\label{4dmin} 
-\frac1{4\pi}(\sfN^{\rm t}\sfP)_{IJ}\int_Y B^I\wedge  B^J+\frac1{2\pi}N^I{}_J\int_Y  B^J\wedge  \tilde F_I\,,
 \ee
 where $\tilde F_I=\d\tilde A_I$ are U(1) two-form field strengths. 
 This bulk action is invariant mod $2\pi\mathbb{Z}$ under the one-form and zero-form gauge transformations 
 \be\label{Bcgauge} 
 B^I\rightarrow B^I+\d\Lambda^I\quad,\quad \tilde A_I\rightarrow \tilde A_I+\sfP_{IJ}\Lambda^I+\d\lambda_I\quad,\quad  \Lambda^I|_\Sigma=0\,.
 \ee
The second term in \eqref{4dmin} constrains $B^I$ to be $\Gamma^{(1)}_\sfN$ two-form potentials: 
\be\label{minBNcond} 
\d B^I=0\quad,\quad N^I{}_J[B^J]\in 2\pi H^2(Y,\Sigma,\mathbb{Z})\,.
\ee
Hence the potentials $B^J$ determine a class in  $H^2(Y,\Sigma,\Gamma_\sfN)$.  The $B_I$ equations of motion are
\be\label{BBulkeq} 
\sfN^J{}_I\left(\sfP_{JK}B^K-\tilde F_J\right)=0\,.
\ee
Notice that the boundary condition $B^I|_\Sigma=0$ may be regarded as boundary contribution to the $\tilde A_I$ equations of motion, and implies that $N^J{}_I\tilde F_J|_{\Sigma}=0$. 

One can construct the following gauge invariant operators
\begin{subequations}\label{VWminop}
\begin{align}
U_{\bf m}(C,\gamma)&=\exp\left(\ii m_I\int_C B^I \right)\quad,\quad \del C=\gamma\subset \Sigma=\del Y\,,\label{VWminopa}\\ 
 V_{\bf n}(C,\gamma)&=\exp\left(\ii n^I\int_\gamma \tilde A_I -\ii n^I\sfP_{IJ}\int_C B^J \right)\quad,\quad \del C=\gamma\subset Y\,,\label{VWminopb} \\
L_{\bf n}(\gamma)&=\exp\left(\ii n^I\oint_\gamma \tilde A_I\right)\quad,\quad~~~~\gamma\subset \Sigma=\del Y\,.\label{Lnbulkmin}
\end{align}
\end{subequations} 
Note that the $L_{\bf n}(\gamma)$ is  invariant under the gauge transformations \eqref{Bcgauge} since $\Lambda^I|_\Sigma=0$. 

Let us first assume that $\Sigma=\del Y=\emptyset$, so that the lines $L_{\bf n}$ disappear from the spectrum. First observe, the operators \eqref{VWminopb} that depend on the open surface must be considered as trivial, since they can only contribute to correlators through contact terms.  Hence the operators \eqref{VWminopb} can be non-trivial only if they do not depend on the surface $C$, and hence can be considered as   genuine line operators. This happens only if $n^I\sfP_{IJ}\int_C B^J\in 2\pi\mathbb{Z}$ for any $C$ and $B^I$. Since $\frac1{2\pi}N^I{}_JB^J$ defines an integral two-cochain and $(\sfP\sfN^{-1})_{IJ}=(\sfP\sfN^{-1})_{JI}$, this happens precisely if  $(\sfP\sfN^{-1})_{IJ}n^J=0$. Setting $\alpha^I\equiv (\sfN^{-1})^I{}_Jn^J$, we  have $\sfN^I{}_J\alpha^J=n^I\in\mathbb{Z}$ and $\sfP_{IJ}\alpha^J\in\mathbb{Z}$. But since $\sfN$ and $\sfP$ are right-coprime, we must have  $\alpha^I\equiv m^I \in\mathbb{Z}$ and $n^I=\sfN^I{}_Jm^J$. It follows that only the operators $V_{\sfN {\bf m}}$ can be  genuine line operators. But because of the equations of motion \eqref{BBulkeq}, $V_{\sfN {\bf m}}=1$ and hence all the  $V_{\bf n}$ operators are trivial. 

Consider now the    operators \eqref{VWminopa}. Since $\sfP$ and $\sfN$ are right-coprime, $\sfX\sfP-\sfY\sfN=\mathds{1}$ for some integral matrices $\sfX$ and $\sfY$. Using \eqref{minBNcond}, this implies that
\be 
U_{\bf m}(C)=\exp\left(\ii m_I\sfX^{IJ}\sfP_{JK}\int_CB^K\right)\,.
\ee
But this in turn implies that $U_{\bf m}(C)$ can be split in the product of  open surface operators $V_{\bf n}$, with $n^I=m_J\sfX^{JI}$. Since  $V_{\bf n}=1$ for all $n^I$, we see that the operators  $U_{\bf m}$  are trivial too. We therefore conclude that if $\sfP$ and $\sfN$ are right-coprime, and $\del X=\emptyset$, the topological theory \eqref{4dmin}  is trivial.

Now let us go back to the four-dimensional space with boundary $\del X=\Sigma$. In addition to the new boundary line operators \eqref{Lnbulkmin}, we can also consider $U_{\bf m}(C,\gamma)$ with $\del C=\gamma\subset \Sigma$. But the   above arguments on  the triviality of the bulk theory imply that these are genuine boundary line operators too. So, we can simply denote them as $U_{\bf m}(\gamma)$. In fact, from  $V_{\bf n}(C,\gamma)=1$ it follows that, if we pick $m_I=n^J\sfP_{JI}$,  $U_{{\bf m}}$ can be identified with     $L_{\bf n}$:
\be\label{ULidentity} 
L_{\bf n}(\gamma)=U_{\sfP^{\rm t}{\bf n}}(\gamma)\,.
\ee
Observe also that \eqref{minBNcond} implies that   $U_{\sfN^{\rm t}{\bf k}}=1$, for any ${\bf k}\in V^*_\mathbb{Z}$. 
Together with \eqref{ULidentity}, this also implies that $U_{\sfN {\bf k}}=1$ for any $\tilde{\bf k}\in V_\mathbb{Z}$,  since  $\sfP^{\rm t}\sfN \tilde{\bf k}=\sfN^{\rm t}\sfP\tilde{\bf k}$. Hence the 
$U_{\bf m}$ and $L_{\bf n}$ are labeled by the elements of  $\Gamma^*_\sfN$ and $\Gamma_\sfN$, respectively. Furthermore, using again the right-coprimitivity of $\sfP$ and $\sfN$, we can also invert \eqref{ULidentity} and express the $U_{\bf m}$ lines in terms of the $L_{\bf n}$ lines. Indeed,  
 any ${\bf m}\in V^*_\mathbb{Z}$ can be rewritten as $\sfP^{\rm t}\sfX^{\rm t}{\bf m}-\sfN^{\rm t}\sfY^{\rm t}{\bf m}$ for some integral matrices $\sfX$ and $\sfY$, so that $U_{\bf m}=U_{\sfP^{\rm t}\sfX^{\rm t}{\bf m}}$. From \eqref{ULidentity} it then follows that 
\be 
U_{\bf m}(\gamma)=L_{\sfX^{\rm t}{\bf m}}\,.
\ee
Hence, we can either use $L_{\bf n}$ or $U_{\bf m}$ as generators of all the boundary lines. 

Finally, as in \cite{Kapustin:2014gua,Gaiotto:2014kfa}, it may be convenient to  relax the condition $B^I|_\Sigma$ and restore the one-form gauge symmetry on the boundary  by adding  boundary Stueckelberg fields $a^I$, such that 
\be\label{Bcgauge2} 
a^I\rightarrow a^I+\Lambda^I|_{\Sigma}\,,
\ee
under \eqref{Bcgauge}.
We can then make the full system gauge invariant by adding the boundary action
\be 
\frac{1}{4\pi}(\sfN^{\rm t}\sfP)_{IJ}\oint_\Sigma a^I\wedge f^J-\frac1{2\pi}\sfN^I{}_J\oint_\Sigma a^J\wedge \tilde F_I\,. 
\ee
The boundary lines can now be written in a fully gauge invariant way
\be 
L_{\bf n}(\gamma)=\exp\left[\ii n^I\oint_\gamma \left(\tilde A_I-\sfP_{IJ}a^J\right)\right]\quad,\quad U_{\bf m}(\gamma)=\exp\left[\ii m_I\left(\int_C B^I-\oint_\gamma a^I\right)\right]\,. 
\ee
This description makes it easy to compute the brading of the these lines. The insertion of an $L_{\bf n}(\gamma)$ modifies the $a^I$ equations of motion into 
\be 
\sfN^J{}_I\left(\tilde F_J|_\Sigma-\sfP_{JK}f^K\right)=-n^J\sfP_{JI}\delta_2(\gamma)\,.
\ee
By using this equation and $\sfC=\sfP\sfN^{-1}$, one can check that the braiding rule of the $L_{\bf n}$ lines is given by \eqref{genbraid}.
This shows that the lines $L_{\bf n}$ -- or, equivalently, the lines $U_{\bf m}$ -- provide a realization of $\cala_\Sigma^{(\sfN,\sfP)}$. This, of course, holds also for the partially gauge-fixed description with $B^I|_\Sigma=0$.  

\subsection{Coupling to background gauge fields}
\label{app:bulkminback}

We now consider the coupling of the this minimal TQFT  to background $\Gamma^{(1)}_\sfN$ gauge fields defined on $\Sigma$. If we use the description \eqref{Lnbulkmin} of the topological lines, this corresponds to adding to \eqref{4dmin} the boundary term
\be\label{bulkcoup1} 
\frac1{2\pi}\sfN^I{}_J\int_\Sigma \tilde A_I\wedge \calb^J\,,
\ee
where $\calb^I$ denote the $\Gamma^{(1)}_\sfN$  background gauge fields, i.e. such that $\frac1{2\pi}\sfN^I{}_J\calb^J$ identify an integral uplift of an element  of  $H^2(\Sigma,\Gamma_\sfN)$. 
Hence we get 
\be\label{4dmin2} 
-\frac1{4\pi}(\sfN^{\rm t}\sfP)_{IJ}\int_Y B^I\wedge  B^J+\frac1{2\pi}\sfN^I{}_J\int_Y  B^J\wedge  \tilde F_I+\frac1{2\pi}\sfN^I{}_J\int_\Sigma \tilde A_I\wedge \calb^J\,,
 \ee
whose partition function gives $\cala_\Sigma^{(\sfN,\sfP)}[\calb]$. 
Integrating out $\tilde A_I$ localizes the   $B^I$ integration to a discrete sum over   (flat)  $\Gamma^{(1)}_\sfN$ bulk gauge fields $B^I$, that is such that $\frac1{2\pi}\sfN^I{}_JB^J\in H^2(Y,\mathbb{Z})$, with $B^I|_\Sigma=-\calb^I$. We can then write $ B^I$ as $ B^I=\Omega^I-\hat\calb^I$, where $\hat\calb^I$ are  $\Gamma^{(1)}_\sfN$  bulk  gauge fields that  provide any  given bulk extension of the background $\calb^I$, and  $\Omega^I$   are dynamical  $\Gamma^{(1)}_\sfN$  bulk  gauge fields with 
$\Omega^I|_\Sigma=0$. In other words,  $\omega^I\equiv \frac1{2\pi}\sfN^I{}_J\Omega^J$ represent the integral uplift of  the relative cohomology classes $\bm{\omega} \in H^2(Y,\Sigma;\Gamma_\sfN)$, which hence  identify the physically inequivalent choices of $\Omega^I$. The action  \eqref{4dmin2} reduces to
\be\label{4dmin3} 
-\frac1{4\pi}(\sfN^{\rm t}\sfP)_{IJ}\int_Y (\Omega^I-\hat\calb^I)\wedge  (\Omega^J-\hat\calb^J)\,.
 \ee
The associated partition function obtained by summing over $ \bm\omega\in H^2(Y,\Sigma;\Gamma_\sfN)$ defines the minimal TQFT coupled to $\calb^I$, $\cala^{(\sfN,\sfP)}_\Sigma[\calb]$. In this form the anomaly \eqref{infanomaly} is manifest.

One  can also define the coupling to background fields through the  line operators \eqref{VWminopa}. Since they are defined in terms of periods of $B^I$ over relative bulk cycles $C$, that is, such that $\del C\subset\Sigma$, a natural coupling to background potentials is given by a bulk term
\be\label{BBcoupling} 
\frac{1}{2\pi}\sfN^I{}_J\int_Y B^J\wedge \tilde\calb_I\,,
\ee
where we are already restricting ourselves to flat $B^I$ potentials, such that $B^I|_\Sigma=0$ and $\sfN^I{}_JB^J\in 2\pi H^2(Y,\mathbb{Z})$. Hence, they identify  elements of $ H^2(Y,\Sigma;\Gamma_\sfN)$ and, adopting the notation introduced  after \eqref{4dmin2},  we can rename them as  $B^I=\Omega^I$.

On the other hand, in \eqref{BBcoupling} $\tilde\calb_I$ are flat $\Gamma^{*(1)}_\sfN$  two-form potentials, whose  boundary restriction $\tilde\calb_I|_\Sigma$  defines an alternative choice of three-dimensional  background potentials for the $\Gamma^{(1)}_\sfN$ one-form symmetry.  
In Appendix \ref{app:bulkmin} we discussed the equivalence between the $L_{\bf n}$ and $U_{\bf m}$ line operators. In the present context, the same arguments  show that the two choices  of $\Gamma^{(1)}_\sfN$  background gauge fields provides by $\calb^I$ and $\tilde\calb_I|_\Sigma$ are related by 
\be\label{BBrel} \tilde\calb_I|_\Sigma=\sfP_{IJ}\calb^J\quad \text{mod}\quad 2\pi H^2(\Sigma,\mathbb{Z})\,,
\ee 
 which  corresponds to a proper identity in $H^2(\Sigma,\Gamma^*_\sfN)$.\footnote{In this sense, by using the B\'ezout identity $\sfX\sfP-\sfY\sfN=\mathds{1}$, \eqref{BBrel} is invertible into the identity  $\calb^I=\sfX^{IJ}\tilde\calb_J|_\Sigma$ modulo $2\pi H^2(\Sigma,\mathbb{Z})$.}   The relation \eqref{BBrel} can be extended to the identity  $\tilde\calb_I=\sfP_{JI}\hat\calb^J$ mod $2\pi H^2(Y,\mathbb{Z})$,   involving the corresponding bulk extensions. By using it in \eqref{BBcoupling},  we get the bulk theory 
\be\label{4dmin3a}
\begin{aligned}
&-\frac1{4\pi}(\sfN^{\rm t}\sfP)_{IJ}\int_Y \Omega^I\wedge  \Omega^J+\frac{1}{2\pi}(\sfN^{\rm t}\sfP)_{IJ}\int_Y \Omega^I\wedge \hat\calb^J\\
&=
\frac1{4\pi}(\sfN^{\rm t}\sfP)_{IJ}\int_Y \hat\calb^I\wedge \hat\calb^J-\frac1{4\pi}(\sfN^{\rm t}\sfP)_{IJ}\int_Y (\Omega^I-\hat\calb^I)\wedge  (\Omega^J-\hat\calb^J)\,.
\end{aligned}
\ee
Comparing  \eqref{4dmin3a} to \eqref{4dmin3}, we conclude that the corresponding path integral is given by
\be\label{YBpathint}
\quad \cala^{(\sfN,\sfP)}_\Sigma[\calb]\,\exp\left(\frac\ii{4\pi}(\sfN^{\rm t}\sfP)_{IJ}\int_Y \hat\calb^I\wedge \hat\calb^J\right)\,.
\ee
The bulk term precisely cancels the anomaly of  $\cala^{(\sfN,\sfP)}_\Sigma[\calb]$, and hence the full system is anomaly free.

 \subsection{{\sf C} interfaces from half-space gauging}
 \label{app:Chalfgaug}

Let us now consider a GZ Lagrangian $\call(F,\phi)$ defined on a spacetime $X$, and an orientable submanifold $\Sigma$. Furthermore we assume that $\Sigma$ splits $X$ into two halves $X_\tL$ and $X_\tR$, with $\del X_\tL=\Sigma=-\del X_\tR$. Let us now  couple this theory a topological sector defined by \eqref{4dmin3a}, with $Y=X_\tL$ and $\hat\calb^I=(\sfN^{-1})^I{}_J F^J$. That is, we add to $\int_{X_\tL}\call(F,\phi)$ the terms
\be\label{Chalf-space gauging} 
-\frac1{4\pi}(\sfN^{\rm t}\sfP)_{IJ}\int_Y \Omega^I\wedge  \Omega^J+\frac{1}{2\pi}\sfP_{IJ}\int_Y F^I\wedge\Omega^J\,,
 \ee
 with $\Omega^I$ defined as in Appendix \ref{app:bulkminback}.
This procedure can be regarded as a half-space gauging with  discrete torsion, restricted to $X_\tL$, of the magnetic $\Gamma^{(1)}_\sfN$ one-form symmetry group of the GZ models. The  gauging is defined by the `charges' $\sfP_{IJ}$ and the discrete two-form potentials  satisfy the boundary conditions $\Omega^I|_\Sigma=0$. Applying \eqref{YBpathint} we can immediately conclude that this half-space gauging has the effect of inserting the terms 
\be
\cala^{(\sfN,\sfP)}_\Sigma[\sfN^{-1}F]\exp\left(\frac\ii{4\pi}\sfC_{IJ}\int_{X_\tL}F^I\wedge F^J\right)\,,
\ee
into the path integral. 
In other words, the net effect is to perform an $\cals_\sfC$ transformation of the theory on $X_\tL$, and insert the TQFT $\cala^{(\sfN,\sfP)}_\Sigma[\sfN^{-1}F]$. Therefore, using \eqref{calicala}, we get precisely the minimal realization of the $\sfC$ interface \eqref{WC2}.
This generalizes the result of  \cite{Choi:2022jqy}. While  the rationale is basically the same, our half-space gauging procedure  differs by some interesting details from the procedure  followed in \cite{Choi:2022jqy}.

%%%%%%%%%%%%%%%%%%%%%%%%%%%%%%%%%%%%%%%%%%%%%%%
\if{

\section{Some fusions of GZ topological defects}
\label{app:fusions}

\subsection{GZ defects in Calabi-Yau models}

Consider the ordered fusion $\cald_{{\bf q},{\bf p}}\cald_{\bf m}$ of the GZ topological defects \eqref{caldCY} and \eqref{caldCY2}, by using \eqref{CYME} and  \eqref{simpleNqp}. The intermediate background field strengths $F^I$ become a world-volume field strengths  $\tilde f^I=\d\tilde a^I$, and we get 
\be\label{caldCYapp}
\begin{aligned}
\cald_{{\bf q},{\bf p}}\cald_{\bf m}(\Sigma)=&\,\exp\left[\ii(\alpha^i_{{\bf q},{\bf p
}}+m^i)\oint_\Sigma\calj_i\right]\times \\
&\int\cald a\cald c\,\exp\left(-\frac{\ii}{4\pi}\,\sfK_{IJ}^{{\bf q},{\bf p}}\oint_\Sigma  a^I\wedge f^J+\frac\ii{2\pi}\oint_\Sigma c_I\wedge \left[(\sfN_{{\bf q},{\bf p}})^I{}_J f^J-F^I_{\text{\tiny L}}\right]\right)\\
&\int\cald\tilde a \cald b\exp\left(\frac\ii{2\pi}\oint_\Sigma b_I\wedge \left[(\sfM_{{\bf q},{\bf p}})^I{}_JF^J_{\text{\tiny L}}-(\sfE_{{\bf q},{\bf p}})^I{}_J\tilde f^J\right]\right)\\
&\exp\left(-\frac{\ii}{4\pi}\tilde\sfC^{\bf m}_{IJ}\oint_\Sigma \tilde a^I\wedge \tilde f^J\right)\int\cald \tilde b \exp\left(\frac\ii{2\pi}\oint_\Sigma \tilde b_I\wedge\left[\tilde f^I-(\sfA_{\bf m})^I{}_JF^J_{\text{\tiny R}}\right]\right)
\,,
\end{aligned}
\ee
where $(\sfA_{\bf m})^I{}_L\equiv \sfA^I{}_J(m)$, with $\sfA^I{}_J(\alpha)$ as in \eqref{pertsympl0a}.  By integrating out $\tilde b_I$ we get 
\be\label{caldCYapp2}
\begin{aligned}
\cald_{{\bf q},{\bf p}}\cald_{\bf m}(\Sigma)=&\,\exp\left[\ii(\alpha^i_{{\bf q},{\bf p
}}+m^i)\oint_\Sigma\calj_i\right]\exp\left[-\frac{\ii}{4\pi}(\sfA^{\rm t}_{\bf m}\tilde\sfC^{\bf m}\sfA_{\bf m})_{IJ}\oint_\Sigma A^I_{\text{\tiny R}}\wedge F^J_{\text{\tiny R}}\right]\times \\
&\int\cald a\cald c\,\exp\left(-\frac{\ii}{4\pi}\,\sfK_{IJ}^{{\bf q},{\bf p}}\oint_\Sigma  a^I\wedge f^J+\frac\ii{2\pi}\oint_\Sigma c_I\wedge \left[(\sfN_{{\bf q},{\bf p}})^I{}_J f^J-F^I_{\text{\tiny L}}\right]\right)\\
&\int\cald b\exp\left(\frac\ii{2\pi}\oint_\Sigma b_I\wedge \left[(\sfM_{{\bf q},{\bf p}})^I{}_JF^J_{\text{\tiny L}}-(\sfE_{{\bf q},{\bf p}}\sfA_{\bf m})^I{}_JF^J_{\text{\tiny R}}\right]\right)
\,.
\end{aligned}
\ee

The path-integral over $b_I$ acts as a delta-function which imposes  
\be\label{ppcondCY} 
 (\sfM_{{\bf q},{\bf p}})^I{}_JF^J_{\text{\tiny L}}|_\Sigma=(\sfE_{{\bf q},{\bf p}}\,\sfA_{\bf m})^I{}_JF^J_{\text{\tiny R}}|_\Sigma\,.
\ee
The condition \eqref{ppcondCY} is not always solvable and implies that the path-integral over $b_I$ acts as a projector over the  bulk gauge fields summed in the path-integration. This projector vanishes if the bulk left and right line bundles do not satisfy the boundary topological condition 
\be\label{linebbulk} 
(\sfM_{{\bf q},{\bf p}})^I{}_J[F^J_{\text{\tiny L}}]=(\sfE_{{\bf q},{\bf p}}\,\sfA_{\bf m})^I{}_J [F^J_{\text{\tiny R}}]\quad,\quad  \text{in}\,\  H^2(\Sigma,\mathbb{Z})\,.
\ee
If instead it is satisfied, then the projector imposes the gluing condition 
\be\label{FAAF} 
F^I_{\text{\tiny R}}|_\Sigma=(\sfA^{-1}_{\bf m}\,\sfA^{-1}_{{\bf q},{\bf p}})^I{}_JF^J_{\text{\tiny L}}|_\Sigma+\omega^I\,,
\ee
where $(\sfA_{{\bf q},{\bf p}})^I{}_J\equiv \sfA^I{}_J(\alpha_{{\bf q},{\bf p}})=(\sfM^{-1}_{{\bf q},{\bf p}}\sfE_{{\bf q},{\bf p}})^I{}_J$ and   $\omega^I\in 2\pi H^2(\Sigma,\mathbb{Z})_{\rm tor}$ is a torsional class such that $(\sfE_{{\bf q},{\bf p}})^I{}_J \omega^J=0$. By using the explicit form \eqref{CYME} of $\sfE_{{\bf q},{\bf p}}$, this is equivalent to imposing  
\be 
\omega^0=0\quad, \quad p^i\omega^i=0\quad,\quad i=1,\ldots n-1.
\ee

Repeating the computation by reversing the defect ordering, we get
 \be\label{caldCYapp3}
\begin{aligned}
\cald_{\bf m}\cald_{{\bf q},{\bf p}}(\Sigma)=&\,\exp\left[\ii(\alpha^i_{{\bf q},{\bf p
}}+m^i)\oint_\Sigma\calj_i\right]\exp\left(-\frac{\ii}{4\pi}\tilde\sfC^{\bf m}_{IJ}\oint_\Sigma A^I_{\text{\tiny L}}\wedge F^J_{\text{\tiny L}}\right)\times 
\\
&\int\cald a\cald c\,\exp\left(-\frac{\ii}{4\pi}\,\sfK_{IJ}^{{\bf q},{\bf p}}\oint_\Sigma  a^I\wedge f^J+\frac\ii{2\pi}\oint_\Sigma c_I\wedge \left[(\sfN_{{\bf q},{\bf p}})^I{}_J f^J-(\sfA^{-1}_{\bf m})^I{}_JF^J_{\text{\tiny L}}\right]\right)\\
&\int\cald b\exp\left(\frac\ii{2\pi}\oint_\Sigma b_I\wedge \left[(\sfM_{{\bf q},{\bf p}}\sfA^{-1}_{\bf m})^I{}_JF^J_{\text{\tiny L}}-(\sfE_{{\bf q},{\bf p}})^I{}_J F^J_{\text{\tiny R}}\right]\right)
\,.
\end{aligned}
\ee
 Now the path-integral over $b_I$ acts as a delta-function which imposes  
\be\label{ppcondCY2} 
 (\sfM_{{\bf q},{\bf p}}\sfA^{-1}_{\bf m})^I{}_JF^J_{\text{\tiny L}}|_\Sigma=(\sfE_{{\bf q},{\bf p}})^I{}_JF^J_{\text{\tiny R}}|_\Sigma\,.
\ee
This implies that the path-integral acts as a projector over the bulk line bundles fulfilling the boundary topological condition
\be\label{linebbulk2} 
(\sfM_{{\bf q},{\bf p}}\sfA^{-1}_{\bf m})^I{}_J[F^J_{\text{\tiny L}}]=(\sfE_{{\bf q},{\bf p}})^I{}_J[ F^J_{\text{\tiny R}}]\quad,\quad  \text{in}\,\   H^2(\Sigma,\mathbb{Z})\,.
\ee
Once this is satisfied, \eqref{ppcondCY2} can be solved by setting 
\be\label{FAAF2} 
F^I_{\text{\tiny L}}|_\Sigma=(\sfA_{\bf m}\sfA_{{\bf q},{\bf p}})^I{}_JF^J_{\text{\tiny R}}|_\Sigma+(\sfA_{\bf m})^I{}_J\tilde\omega^J\,,
\ee
where $\tilde\omega^I\in 2\pi H^2(\Sigma,\mathbb{Z})_{\rm tor}$ and $(\sfA^{-1}_{\bf m}\sfM_{{\bf q},{\bf p}})^I{}_J \tilde\omega^J=0$. By using the explicit form of $\sfA_{\bf m}$ and $\sfM_{{\bf q},{\bf p}}$, this condition reduces to $\tilde\omega^0=0$ and $p^i\tilde\omega^i=0$, which coincide with the conditions for $\omega^I$ appearing in \eqref{FAAF}. Hence  \eqref{FAAF} and \eqref{FAAF2} coincide too. We could also solve \eqref{ppcondCY} and \eqref{ppcondCY2} expressing $F^J_{\text{\tiny R}}|_\Sigma$ in terms of $F^J_{\text{\tiny L}}|_\Sigma$, obtaining the same condition. 

Consider the effect of integrating out $c_I$. By first considering \eqref{caldCYapp2}, we get a delta-function which localizes the path integral over the configurations satisfying
\be\label{NfFL0} 
N_{{\bf q},{\bf p}} f^I=F^I_{\text{\tiny L}}|_\Sigma\,.
\ee
Again we get a projector. 
It is non-vanishing only if $F^I_{\text{\tiny L}}|_\Sigma$ satisfies the cohomological constraint  
\be\label{NfFL} 
\frac1{2\pi}\left[F^I_{\text{\tiny L}}|_\Sigma\right]\in H^2(\Sigma,N_{{\bf q},{\bf p}}\mathbb{Z})\,.
\ee
 Once this constraint is fulfilled, \eqref{NfFL0} fixes  $f^I$ in terms of $F^I_{\text{\tiny L}}|_\Sigma$. Up to the projection imposing \eqref{NfFL} and \eqref{linebbulk},  \eqref{caldCYapp2} reduces to $\exp\left[\ii(\alpha^i_{{\bf q},{\bf p
}}+m^i)\oint_\Sigma\calj_i\right]$ times 
 \be
 \exp\left[\ii(\alpha^i_{{\bf q},{\bf p
}}+m^i)\oint_\Sigma\calj_i-\frac{\ii}{4\pi}(\sfA^{\rm t}_{\bf m}\sfC^{\bf m})_{IJ}\oint_\Sigma A^I_{\text{\tiny R}}\wedge F^J_{\text{\tiny R}}-\frac{\ii}{4\pi}(\sfC^{{\bf q},{\bf p}}\sfA^{-1}_{{\bf q},{\bf p}})_{IJ}\oint_\Sigma A^I_{\text{\tiny L}}\wedge F^J_{\text{\tiny L}}\right]\,.
 \ee
 Note that the second Chern-Simons term is well-defined  because the line bulndles are restricted as in \eqref{linebbulk}. By using \eqref{FAAF} we can rewrite it as 
 \be\label{finCYD1}
\caln\exp\left[\ii(\alpha^i_{{\bf q},{\bf p
}}+m^i)\oint_\Sigma\calj_i-\frac{\ii}{4\pi}\left(\sfA^{-1{\rm t}}_{{\bf q},{\bf p}}\sfC^{\bf m}\sfA^{-1}_{{\bf m}}\sfA^{-1}_{{\bf q},{\bf p}}+\sfC^{{\bf q},{\bf p}}\sfA^{-1}_{{\bf q},{\bf p}}\right)_{IJ}\oint_\Sigma A^I_{\text{\tiny L}}\wedge F^J_{\text{\tiny L}}\right]\,,
 \ee
 where $\caln$ contains the dependence on the torsion elements $\omega^I=\d\alpha^I$ appearing in \eqref{FAAF}:
 \be 
\caln=\exp\left[-\frac{\ii}{4\pi}(\sfA^{\rm t}_{\bf m}\sfC^{\bf m})_{IJ}\oint_\Sigma \alpha^I\wedge \omega^J-\frac{\ii}{4\pi}\left(\sfA^{-1{\rm t}}_{{\bf q},{\bf p}}\sfC^{\bf m}+\sfA^{\rm t}_{\bf m}\sfC^{\bf m}\sfA^{-1}_{\bf m}\sfA^{-1}_{{\bf q},{\bf p}}\right)_{IJ}\oint_\Sigma \alpha^I\wedge F^J_{\text{\tiny L}}\right]\,.
 \ee

Consider now \eqref{caldCYapp3}.  By integrating out $c_I$ we get 
 \be\label{NfFL1} 
N_{{\bf q},{\bf p}} f^I=(\sfA^{-1}_{\bf m})^I{}_JF^J_{\text{\tiny L}}|_\Sigma\,.
\ee
 Since $\sfA_{\bf m}$ is unimodular, this gives the  same topological condition \eqref{NfFL} for $F^I_{\text{\tiny L}}|_\Sigma$ and again, if  \eqref{NfFL} is satisfied, fixes  $f^I$ in terms of $F^I_{\text{\tiny L}}|_\Sigma$. Hence, up to the overall projectors that impose \eqref{NfFL} and \eqref{linebbulk}, \eqref{caldCYapp3} reduces to 
 \be\label{finCYD2}
 \exp\left[\ii(\alpha^i_{{\bf q},{\bf p
}}+m^i)\oint_\Sigma\calj_i-\frac{\ii}{4\pi}(\sfC^{\bf m}\sfA^{-1}_{\bf m}+\sfA^{-1{\rm t}}_{{\bf m}}\sfC^{{\bf q},{\bf p}}\sfA^{-1}_{{\bf q},{\bf p}}\sfA^{-1}_{{\bf m}})_{IJ}\oint_\Sigma A^I_{\text{\tiny L}}\wedge F^J_{\text{\tiny L}}\right]\,.
 \ee
By writing the first

 In order to prove that \eqref{finCYD1} and \eqref{finCYD2} coincide, up to the torsion factor, we need to use the fact that the symplectic matrices \eqref{pertsympl0} commute. In particular, this implies  that $\sfA_{{\bf q},{\bf p}}$ and $\sfA_{{\bf m}}$ commute and 
 \be 
\sfC^{{\bf q},{\bf p}}\sfA_{{\bf m}}+\sfA^{-1{\rm t}}_{{\bf q},{\bf p}}\sfC^{{\bf m}}=\sfC^{{\bf m}}\sfA_{{\bf q},{\bf p}}+\sfA^{-1{\rm t}}_{{\bf m}}\sfC^{{\bf q},{\bf p}}\,.
 \ee
Multiplying both sides by $\sfA^{-1}_{{\bf m}}\sfA^{-1}_{{\bf q},{\bf p}}$ from the right, we get
 \be 
\sfC^{{\bf q},{\bf p}}\sfA^{-1}_{{\bf q},{\bf p}}+\sfA^{-1{\rm t}}_{{\bf q},{\bf p}}\sfC^{{\bf m}}\sfA^{-1}_{{\bf m}}\sfA^{-1}_{{\bf q},{\bf p}}=\sfC^{{\bf m}}\sfA^{-1}_{{\bf m}}+\sfA^{-1{\rm t}}_{{\bf m}}\sfC^{{\bf q},{\bf p}}\sfA^{-1}_{{\bf q},{\bf p}}\sfA^{-1}_{{\bf m}}\,.
 \ee
 This identity shows that \eqref{finCYD1} and \eqref{finCYD2} coincide, up to the  
}
\fi

%%%%%%%%%%%%%%%%%%%%%%%%%%%%%%%%%%%%%%%%%%%%%%%

%%%%%%%%%%%%%%%%%%%%%%%%%%%%%%%%%%%%%%%%%%%%%%%%%%%%%%%%%%
%%%%%%%%%%%%%%%%%%%%%%%%%%%%%%%%%%%%%%%%%%%%%%%%%%%%%%%%%%

%%%%%%%%%%%%%%%%%%%%%%%%%%%%%%%%%%%%%%%%%%%%%%%%%%%%%%%%%%
\newpage
\bibliographystyle{jhep}
\bibliography{references}

\end{document}